\documentclass[a4paper,11pt]{article}
\pdfoutput=1 
\usepackage{jheppub}
\usepackage{amsmath}
\usepackage{amsfonts}
\usepackage{amssymb}
\usepackage{slashed}
\usepackage{bm}
\usepackage{graphicx}%
\usepackage[T1]{fontenc} 
\newcommand{\dhd}{{\textstyle d}
\lower.03ex\hbox{\kern-0.38em$^{\scriptstyle-}$}\kern-0.05em{}}
\newcommand{\dbar}{{\textstyle \delta}
\lower.03ex\hbox{\kern-0.38em$^{\scriptstyle-}$}\kern-0.05em{}}
\newcommand{\half}{{1\over 2}}

\newcommand{\barz}{{\bar z}}

\newcommand{\blambda}{{\bar \lambda}}
\newcommand{\bfi}{{\bar \phi}}

\newcommand{\bLa}{{\bar \Lambda}}

\newcommand{\calf}{{\cal F}}

\newcommand{\caln}{{\cal N}}  
\newcommand{\calo}{{\cal O}}

\newcommand{\calu}{{\cal U}} 
\newcommand{\calv}{{\cal V}} 
\newcommand{\calw}{{\cal W}}

\newcommand{\tile}{{\tilde e}}

\newcommand{\tilk}{{\tilde k}} 
\newcommand{\tiln}{{\tilde n}} 

\newcommand{\tils}{{\tilde s}}

\newcommand{\tilx}{{\tilde x}} 
\newcommand{\tilz}{{\tilde z}}

\newcommand{\tilU}{{\tilde U}}
\newcommand{\tilV}{{\tilde V}}

\newcommand{\tibeta}{\tilde {\beta}}

\newcommand{\tiLa}{\tilde {\Lambda}}

\newcommand{\talfa}{\tilde {\alpha}} 

\newcommand{\talef}{\tilde {\aleph}} 
\newcommand{\taleph}{\tilde {\aleph}}

\newcommand{\bre}{\breve {e}}

\newcommand{\brek}{\breve {k}} 
\newcommand{\bres}{\breve {s}} 
\newcommand{\brex}{\breve {x}} 
 
\newcommand{\brez}{\breve {z}} 

\newcommand{\che}{{\check e}}
\newcommand{\chek}{{\check k}}
\newcommand{\ches}{{\check s}}
\newcommand{\chex}{\check {x}} 
\newcommand{\chez}{{\check z}}

\begin{document}

\pdfoutput=1.  
\abstract{The structure constants of  twist-two operators with spin $j$ in the BFKL limit $g^2\rightarrow 0, j\rightarrow 1$ and ${g^2\over j-1}\sim 1$ are
found from the calculation of the  three-point correlator of twist-two light-ray operators in the triple Regge limit. 
 It is well known that the anomalous dimensions of twist-two operators in this limit 
are determined by the BFKL intercept.
Similarly, the obtained structure constants are determined by an analytic function of three BFKL intercepts. 

}.   
\arxivnumber{}
\affiliation{ Physics Department, Old Dominion University, Norfolk, VA 23529, USA \\and \\Thomas Jefferson National Accelerator Facility, Newport News, VA 23606, USA}
\emailAdd{balitsky@jlab.org}.

\title{\boldmath Structure constants of twist-two light-ray operators in the triple Regge limit}
\author{I. Balitsky}
\preprint{JLAB-THY-19-2682}
\maketitle

\flushbottom

\section{Introduction\label{aba:sec1}}
The approximate conformal invariance of pQCD makes it very useful in practical calculations. 
Any leading-order pQCD result which does not have explicit $\beta$-function can be obtained from 
conformally invariant amplitudes. Moreover, the results obtained
in a close conformal ``neighbor'' of QCD, the ${\cal N}=4$ SYM theory, can be used as a starting point of QCD calculation. Typically, 
the result in ${\cal N}=4$ theory gives the most complicated part of pQCD result, i.e. the one with maximal transcendentality.  
This is explicitly confirmed in many cases,  for example in  the calculation of anomalous dimensions of twist-two 
operators \cite{Moch:2004pa,Vogt:2004mw} and  cusp anomalous dimension
at the  three-loop level \cite{Grozin:2014hna,Grozin:2015kna}.  
Actually, it is worthwhile to start a pQCD calculation from the corresponding analysis in ${\cal N}=4$ SYM. The Lagrangian
might seem more complicated but the result will be obtained in a more streamlined and controlled way and it will 
give the most transcendental part of the QCD result. 

It is well known that the all correlation functions (correlators)  in a conformal theory are 
in principle determined if one knows the anomalous dimensions of all primary operators
operators and the corresponding structure constants determined by three-point correlators.  
The important class of local operators is represented by the so-called twist-two operators 
encountered in many phenomenological applications in
 QCD starting from the famous case of  deep inelastic scattering.  
As to anomalous dimensions of twist-two operators in ${\cal N}=4$ SYM, 
there was a considerable progress in recent years due to the development of QCS method \cite{Gromov:2013pga,Gromov:2014caa}
resulting in analytic expressions at large $N_c$ up to 7th order of perturbation theory \cite{Marboe:2016igj}
and very accurate numerical calculations at any coupling constant up to a strong coupling limit \cite{Gromov:2014bva}.  

In contrast, the study of structure constants of twist-two operators is not at the same level yet. For arbitrary spins, the structure 
constants of three twist-two operators
are explicitly known only at the tree level \cite{Kazakov:2012ar,Sobko:2013ema}. There are  calculations of the structure constants of two protected operators and a twist-two operator, the most recent 
performed using the hexagon approach \cite{Basso:2015zoa} up to the four-loop \cite{Eden:2016aqo} and five-loop \cite{Chicherin:2018avq} level.
However, for the correlator of three non-protected operators the hexagon approach gives only general prescription for calculations  and to get
explicit results further development of hexagon method is necessary. There is also a related QCS calculation of three-cusp Wilson loop 
similar to correlator of one protected and two non-trivial operators \cite{Cavaglia:2018lxi}, but at this stage it is not clear whether such result
can be used to get the correlator of three twist-two operators.

In this circumstances, it is very useful to find examples of explicit calculation of twist-two structure constants, especially in the approximations which go beyond the 
leading orders of perturbation theory. One of the most interesting examples is the structure constants of twist-two operators in the so-called  BFKL limit when the Lorentz spin
of the twist-two operator tends to one: $\omega=j-1\rightarrow 0$, coupling constant $g^2={g_{{}_{\rm YM}}^2\over 16\pi^2}N_c$ is small but the ratio ${g^2\over \omega}$ is fixed. This limit is closely related to the high-energy behavior of amplitudes, roughly speaking ${g^2\over\omega}\sim g^2\ln E $ where $E$ is the energy. The problem of high-energy behavior of amplitudes has a long story
  starting with Heisenberg-Froissart bound $\ln^2\!E$ for total cross section 
  which has not been constructively explained in any (field or string) theory in more than
  50 years. The most popular idea is to reduce the gauge theory at high 
  energies to 2+1 effective theory which can be solved (exactly or 
  by computer simulations). Unfortunately, despite the multitude 
  of attempts, the Lagrangian for 2+1 QCD or  ${\cal N}=4$ SYM at high energies is not written yet. 
  In this context, the complementary approach of conformal bootstrap may be helpful.
  One may start with twist-two operators in the BFKL limit
 and  use  knowledge of anomalous dimensions and structure constants
  of these operators  to construct the high-energy amplitudes.
  Of course, the high-energy behavior of amplitudes is not completely determined by twist-two operators, 
for example the BFKL equation involves twist two as well as all higher twists. Still, the effective 
conformal theory of twist-two operators at small $\omega$'s appears to be a good place to start.

Since the conformal twist-two operator in ${\cal N}=4$ SYM looks like
\begin{equation}
\calo^j(x)~=~{\rm Tr}F_{+i}D_+^{j-2}F_+^{~i}~+~{\rm gluinos + scalars}
\label{fla1}
\end{equation}
the point $\omega=j-1\rightarrow 0$ is an unorthodox point corresponding to the non-local operator ${\rm Tr}F_{+i}D_+^{\omega-1}F_+^{~i}$.  The explicit form  of this non-local operator is a so-called light-ray operator - a bilocal operator with the  light-like gauge link. Such light-ray operators are extensively studied in QCD since matrix elements of those operators define parton distribution densities for forward case and so-called GPDs in the off-forward case (see the book \cite{Collins:2011zzd} for a review). For ${\cal N}=4$ SYM, the supersymmetric generalization of QCD light-ray operators \cite{Balitsky:1987bk} is presented in Ref. \cite{Balitsky:2013npa}  following the corresponding work on the supermultiplet of twist-two local operators \cite{Belitsky:2003sh}.

The anomalous dimensions of twist-two operators in the BFKL limit can be obtained from Regge asymptotics of the four-point correlators resulting in the equation
 $\omega=\aleph(\Delta)$ 
where $\Delta$ is the dimension of the operator and $\aleph(\Delta)$ is the famous Pomeron intercept. In QCD, it is  known only up to the NLO order \cite{Fadin:1998py}, but 
in ${\cal N}=4$ SYM it is studied well beyond that: there are explicit perturbative expressions at the  NNLO level  \cite{Gromov:2015vua,Velizhanin:2015xsa,Caron-Huot:2016tzz}, numerical estimates at few extra orders \cite{Gromov:2015vua} 
and several terms for the large-coupling expansion around graviton point $j=2$ \cite{Costa:2012cb,Kotikov:2013xu,Brower:2014wha,Gromov:2014bva}. 

Thus, the theory of anomalous dimensions of twist-two operators in the BFKL limit seems to be well developed and it would
be very interesting to bring the study of structure constants up to the same level.
The most direct way to find the structure constants in the BFKL limit is to compute the correlation function of the corresponding three light-ray operators. 
This was done in Refs. \cite{Balitsky:2015tca,Balitsky:2015oux} using the non-linear evolution equation for color dipoles 
\cite{Balitsky:1995ub,Kovchegov:1999ua,Kovchegov:1999yj}
and the result is that the structure constant is determined by so-called three-pomeron vertex \cite{Korchemsky:1997fy} projected onto Lipatov's eigenfunctions of the BFKL kernel \cite{Lipatov:1985uk}. 
However, by this method it is possible to obtain structure constants only at $\omega_1=\omega_2+\omega_3$ and generalization to
arbitrary  $\omega$'s requires the analysis of perturbative diagrams in the triple Regge limit. It should be noted that the triple Regge limit is a somewhat novel regime of resummation in perturbation theory.  
 Roughly speaking,
 it describes the interaction of three particles going with the speed near speed of light along $x$, $y$, and $z$ directions.
 Such limit was not studied in QCD (or any other QFT) except for Ref. \cite{White:1999zp} devoted to possible anomaly 
coming from three pomerons interacting  by quark exchange (in our LLA calculations  quark exchanges are neglected since they are subleading at high energies).

 In this paper, following the logic of earlier papers \cite{Balitsky:2013npa,Balitsky:2015tca,Balitsky:2015oux},  I calculate the correlator of three light-ray operators (\ref{fla1}) in the 
triple BFKL limit $g^2,\omega_i\rightarrow 0$ and ${g^2\over\omega_i}\sim 1$ in the leading logarithmic approximation (LLA).  
The three light rays are collinear to three linearly independent light-like vectors $n_1$, $n_2$, and $n_3$. 
To simplify the complicated spin structure of a general correlator of  three light-ray operators, I place  these operators on 
the same line in the direction orthogonal to all $n_i$, and integrate each light-ray operator along the total translation in the corresponding $n_i$ direction. 
As demonstrated in Ref. \cite{Balitsky:2015oux}, the resulting correlator has only one tensor
structure and computing the coefficient in front of that structure is the aim of this paper. Moreover, since it is well known that 
in the LLA-Regge limit the contributions of gluino and scalar fields are sub-leading, 
the obtained result for three-point correlator will be valid in QCD as well.

The paper is organized as follows. In Sect. \ref{sect:rych} I recall the generic structure of 3-point correlator for local twist-2 operators
and present the form of the correlator of three ``forward'' local operators 
integrated over the total translation in corresponding light-like directions. This formula is generalized to correlator of three twist-two light-ray operators in Sect. \ref{sect:LRs}. 
In Sect. \ref{sect:wframes} I define ``Wilson frame'' operators and in Sect. \ref{sect:2wframes} remind 
the calculation of two-point correlator of these 
operators in the BFKL limit. In Sect. \ref{sect:3wframes} which is central to this paper, I calculate the correlator of three Wilson frames
in the BFKL limit and present the result for the structure constant. 
In the Conclusions section I discuss the obtained result and its relation to the result of Ref. \cite{Balitsky:2015tca}.
The Appendix contains derivations
of technical results used in Sect. \ref{sect:3wframes}.

\section{3-point correlators and structure constants of ``forward'' operators \label{sect:rych}}

The general structure of 3-point correlators of local operators with spin was found in Ref. \cite{Costa:2011mg} to be 
\footnote{To save space, throughout the paper we use notation 
$\langle \calo_1(x_1)...\calo_n(x_n)\rangle \equiv \langle T\{\calo_1(x_1)...\calo_n(x_n)\}\rangle$.}
\begin{align}
\langle\mathcal{O}^{j_1}_{n_1}(x) \mathcal{O}^{j_2}_{n_2}(y)\mathcal{O}^{j_3}_{n_3}(z)\rangle
=\sum_{m_{12}, m_{13}, m_{23}\ge 0} \lambda_{m_{12},m_{23},m_{13}}
 \left[\begin{array}{ccc}\Delta_1 & \Delta_2 & \Delta_3 \\ j_1 & j_2 & j_3 \\ m_{23} & m_{13} & m_{12}\end{array}\right]
 \nonumber
\end{align}
where $\calo^{l}_{n}(x)$ is a spin-$l$ operator with indices contracted with light-like vector $n$, the square brackets represent some tensor structures 
and the sum over $m_{ij}$ goes over positive integers satisfying certain inequalities. Following Refs.  \cite{Balitsky:2015tca} 
and \cite{Balitsky:2015oux}
I consider the correlator of
three ``forward'' operators integrated over corresponding light-like lines
\begin{equation}
\!\int\! dv_1dv_2dv_3~\langle \calo^{j_1}_{n_1}(v_1n_1+z_{1_t})\calo^{j_2}_{n_2}(v_2n_2+z_{2_t})\calo^{j_3}_{n_3}(v_3n_3+z_{3_t})\rangle
\end{equation}
where the transverse separations $z_{i_t}$ are orthogonal to all $n_i$. It has been demonstrated in Ref. \cite{Balitsky:2015oux} 
that after such integration all tensor structures collapse to one and we get:
\begin{eqnarray}
&&\hspace{-1mm}
\!\int\! dv_1dv_2~\langle \calo^{j}_{n_1}(v_1n_1+z_{1_t})\calo^{j'}_{n_2}(v_2n_2+z_{2_t})\rangle~=~\delta(j-j'){C(\Delta,j)s_{12}^{j-1}\over |z_{12_t}^2|^{\Delta-1}}\mu^{-2\gamma},
\nonumber\\
&&\hspace{-1mm}
\!\int\! dv_1dv_2dv_3~\langle \calo^{j_1}_{n_1}(v_1n_1+z_{1_t})\calo^{j_2}_{n_2}(v_2n_2+z_{2_t})\calo^{j_3}_{n_3}(v_3n_3+z_{3_t})\rangle
\nonumber\\
&&\hspace{-1mm}
=~C(\Delta_i,j_i){s_{12}^{j_1+j_2-j_3-1\over 2}\over |z_{12_t}|^{\Delta_1+\Delta_2-\Delta_3-1}}
{s_{13}^{j_1+j_3-j_2-1\over 2}\over |z_{13_t}|^{\Delta_1+\Delta_3-\Delta_2-1}}{s_{23}^{j_2+j_3-j_1-1\over 2}\over |z_{23_t}|^{\Delta_2+\Delta_3-\Delta_1-1}}
\mu^{-\gamma_1-\gamma_2-\gamma_3}
\label{defc}
\end{eqnarray}
where $\mu$ is the normalization point, $s_{ij}\equiv -2n_i\cdot n_j$, $z_{ij}\equiv z_i-z_j$ and $\Delta_i$ are dimensions (canonical $d_i$  plus anomalous $\gamma_i$) 
of operators $\calo_i$.

As was mentioned in the Introduction,
the most interesting operators for possible phenomenological applications are the twist-two operators. The supermultiplet
of twist-2 operators in $\caln=4$ SYM was explicitly constructed in Ref. \cite{Belitsky:2003sh}. In our case of ``forward'' operators
it reads
\begin{eqnarray}
&&\hspace{-11mm}
S_1^j~=~\calo_g^j+{1\over 2}\calo_\lambda^j-{1\over 2}\calo_\phi^j,~~~~~~~~~
S_2^j~=~\calo_g^j-{1\over 2(j-1)}\calo_\lambda^j+{j+1\over 6(j-1)}\calo_\phi^j
\nonumber\\
&&\hspace{-11mm}
S_3^j~=~\calo_g^j-{j+2\over j-1}\calo_\lambda^j-{(j+1)(j+2)\over 2j(j-1)}\calo_\phi^j
\label{sjs}
\end{eqnarray}
where
\footnote{We use metric $g^{\mu\nu}=(-1,1,1,1)$ and the covariant derivative is $\nabla_\mu=(\partial_\mu-ig_{{}_{\rm YM}}[A_\mu$,).}
\begin{eqnarray}
&&\hspace{-11mm}
\calo_\phi^j(x_t)~=~\int\! du~\bfi_{AB}^a\nabla_n^j\phi^{ABa}(un+x_t),~~~
\nonumber\\
&&\hspace{-11mm}
\calo_\lambda^j(x_t)~=~\int\! du~i\blambda_A^a\nabla_n^{j-1}\sigma_n \lambda_A^a(un+x_t)
\nonumber\\
&&\hspace{-11mm}
\calo_g^j(x_t)~=~\int\! du~F_{ni}^a\nabla_n^{j-2}F_n^{ai}(un+x_t),~~~
\label{lops}
\end{eqnarray}

The operators (\ref{sjs}) are  multiplicatively renormalized operators with anomalous dimensions 
\begin{equation}
\gamma^{S_1}_j(\alpha_s)~=~4[\psi(j-1)+\gamma_E]+O(\alpha_s^2), 
~~~~~\gamma^{S_2}_j~=~\gamma^{S_1}_{j+2},~~~~~\gamma^{S_3}_j~=~\gamma^{S_1}_{j+4}
\label{anomdim2}
\end{equation}

As mentioned in the Introduction, the goal is to calculate the structure constant $C$ in Eq. (\ref{defc}) in 
the ``triple BFKL limit'' $g^2\rightarrow 0, \omega_i=j_i-1\rightarrow 0$ but 
${g^2\over\omega_i}\sim 1$. 
 However,  at $\omega\rightarrow 0$ these gluon operators are no longer local. Instead, they 
are represented by so-called light-ray operators discussed in the next Section.

\section{Light-ray operators in the BFKL limit \label{sect:LRs}}

\subsection{Light-ray operators as an analytic continuation of local operators}
Light-ray (LR) operators are defined as bilocal operators with light-like separation and gauge links providing gauge invariance. For example, the gluon light-ray twist-two
operator is defined as
\begin{equation}
F^a_{\alpha\xi}(x)[x,y]^{ab}F_\beta^{b\xi}(y),~~~~~~(x-y)^2=0
\label{gluonlr}
\end{equation}
where the gauge link $[x,y]$ is defined as
\begin{equation}
[x,y]~\equiv~{\rm Pexp} \big\{-ig_{{}_{\rm YM}}\!\int_0^1\! du~(x-y)_\mu A^\mu(ux+(1-u)y)\big\}
\end{equation}
These operators represent the sum of local operators of twist two convoluted with light-like vector $x-y$. They possess extra UV divergencies in addition to usual self-energy 
and vertex UV divergencies so they are defined with a set of counterterms and the dependence of this counterterms on the UV cutoff  defines the evolution equations
for light-ray operators.

The LR operator (\ref{gluonlr}) can be interpreted as an analytic continuation of a local operator to non-integer number 
of covariant derivatives. Indeed, if we can represent $F^a_{n\xi}\nabla_n^{j-2}F_n^{a\xi}(0)$ as
\begin{eqnarray}
&&\hspace{-1mm}
F^a_{n\xi}\nabla_n^{j-2}F_n^{a\xi}(0)~=~{\Gamma(j-1)\over 2\pi i}\!\int_C\! du ~u^{1-j}F^a_{n\xi}e^{-u\nabla_n}F_n^{a\xi}(0)
\nonumber\\
&&\hspace{27mm}
=~{\Gamma(j-1)\over 2\pi i}\!\int_C\! du ~u^{1-j}F^a_{n\xi}(ux)[ux,0]^{ab}F_n^{b\xi}(0)
\end{eqnarray}
where $C$ is the contour of integration in Hankel's formula for gamma-function.
\footnote{The path of integration starts at $\infty+i0$ at the real axis, goes to $\epsilon+i0$,
circles the origin in the counterclockwise direction with radius $\epsilon$ to the point $\epsilon-i0$, and returns to the point $\infty-i0$.}
At non-integer $j$ this formula can be simplified to
\begin{eqnarray}
&&\hspace{-1mm}
F^a_{n\xi}\nabla_n^{j-2}F_n^{a\xi}(0)~=~{1\over\Gamma(2-j)}\!\int_0^\infty \! du ~u^{1-j}F^a_{n\xi}(ux)[ux,0]^{ab}F_n^{b\xi}(0)
\nonumber\\
&&\hspace{-1mm}
\Rightarrow~\!\int_0^\infty \! du ~u^{1-j}F^a_{n\xi}(ux)[ux,0]^{ab}F_n^{b\xi}(0)~=~\Gamma(2-j)F^a_{n\xi}\nabla_n^{j-2}F_n^{a\xi}(0)
\label{operff}
\end{eqnarray}
At $j=-\half +i\varsigma$ this light-ray operator realizes the principal series irreducible representation of \(sl(2|4)\) with 
conformal spin \(J=j+1=\frac{1}{2}+i\varsigma\).  
Since it is well-defined at $J=\half +i\varsigma$ it can be uniquely analytically continued to 
the whole complex plane of $J$ and the continuation to integer $J=k+1$ gives local operator as a residue in the pole at $j=k$.

The generalization of supermultiplet of twist-two operators (\ref{sjs})  to the case of complex spin \(j\)  was constructed in  \cite{Balitsky:2013npa}. 
We defined ``forward'' parity-even light-ray operators as 
\begin{eqnarray}
&&\hspace{-1mm}
\calf_n^j(x_t)~=~\int_0^\infty\!\! dl~l^{1-j}\calf_n(l,x_t),~~\Lambda_n^j(x_t)=~\int_0^\infty\!\! dl~l^{-j}\Lambda_n(l,x_t)
\nonumber\\
&&\hspace{-1mm}
\Phi_n^j(x_t)=~\int_0^\infty\!\! dl~l^{-1-j}\Phi_n(l,x_t)
\label{lrays}
\end{eqnarray}
where
\begin{eqnarray}
&&\hspace{-1mm}
\calf_n(l,x_t)~=~\!\int\!dv~F_{n\xi}^a(ln+vn +x_t)[l+v,v]^{ab}F_n^{b \xi}(vn+x_t)
\nonumber\\
&&\hspace{-1mm}
\Lambda_n(l,x_t)~=~{i\over 2}\!\int\!dv 
\big[-\blambda_A^a(ln+vn+x_t)[l+v,v]_x^{ab}\sigma_n\lambda_A^b(vn+x_t)
\nonumber\\
&&\hspace{34mm}
+~\blambda_A^a(vn+x_t)[v,l+v]^{ab}\sigma_n\lambda_A^b(ln+vn+x_t)\big]
\nonumber\\
&&\hspace{-1mm}\Phi_n(l,x_t)~=~\!\int\! dv~\phi_I^a(ln+vn +x_t)[l+v,v]_x^{ab}\phi_I^b(vn+x_t)
\label{forwlrs}
\end{eqnarray}
where $[u,v]_x$ is a shorthand notation for $[un+x_t,vn+x_t]$. The corresponding renorm-invariant light-ray operators are given by 
 \cite{Balitsky:2013npa}
\begin{eqnarray}
&&\hspace{-11mm}
S_1^j~=~\calf_j-{j-1\over 2}\Lambda_j-{1\over 2}j(j-1)\Phi_j,~~~~~~~~~
S_2^j~=~\calf_j+{1\over 2}\Lambda_j-{j+1\over 6}\Phi_j
\nonumber\\
&&\hspace{-11mm}
S_3^j~=~\calf_j+(j+2)\Lambda_j-{(j+1)(j+2)\over 2}\Phi_j
\label{sjlrs}
\end{eqnarray}
where the difference between the coefficients here and in Eq. (\ref{sjs}) is due to Eq. (\ref{operff}).

It is demonstrated in Ref. \cite{BALITSKY:2014zza} 
that analytic continuation of anomalous dimensions of local operators Eq. (\ref{sjs}) to non-integer $j$ 
by  integrals of DGLAP kernels gives the anomalous dimensions of light-ray operators (\ref{sjlrs}).  Consequently,
the anomalous dimensions of light-ray operators (\ref{sjlrs}) are related by the same Eq. (\ref{anomdim2}) as local operators
(\ref{sjs}).

Since supersymmetric light-ray operators $S^j$ are analytic continuation of local operators, one should expect the same formulas 
as (\ref{defc}) for correlators of local operators:
\begin{eqnarray}
&&\hspace{-1mm}
\langle S^j(z_{1t})S^{j'}(z_{2t})\rangle~=~\delta(j-j'){C(j,\Delta)s_{12}^{j-1}\over (z_{12t}^2)^{\Delta-1}}\mu^{-2\gamma}
\label{2pointCF}
\end{eqnarray}
and
\begin{eqnarray}
&&\hspace{-1mm}
\langle S_{n_1}^{j_1}(z_{1_t})S_{n_2}^{j_2}(z_{2_t})S_{n_3}^{j_3}(z_{3_t})\rangle
\nonumber\\
&&\hspace{-1mm}
=~C(\Delta_i,j_i){s_{12}^{j_1+j_2-j_3-1\over 2}\over |z_{12_t}|^{\Delta_1+\Delta_2-\Delta_3-1}}
{s_{13}^{j_1+j_3-j_2-1\over 2}\over |z_{13_t}|^{\Delta_1+\Delta_3-\Delta_2-1}}{s_{23}^{j_2+j_3-j_1-1\over 2}\over |z_{23_t}|^{\Delta_2+\Delta_3-\Delta_1-1}}
\mu^{-\gamma_1-\gamma_2-\gamma_3}
\label{defclr}
\end{eqnarray}
Note that from Eq. (\ref{operff}) we see that the canonical dimension of light-ray operators (\ref{lrays}) is $j+2$. 

As mentioned above, the goal of this paper is to find structure constants of operators $S_1^j$ in the BFKL limit 
$g^2\rightarrow 0, \omega_i=j_i-1\rightarrow 0,{g^2\over\omega_i}\sim 1$. The important observation is that 
at small $\omega$  it is sufficient to study 
the correlator of three gluon operators $\int\! du~F_{ni}^a\nabla_n^{\omega-1}F_n^{ai}(un+x_t)$. Indeed, solving Eqs. (\ref{sjlrs}) we see that
\begin{eqnarray}
&&\hspace{-1mm}
\calf_j~=~{S_1^j+\omega\Big({5+\omega\over 6}S_1^j+\big(6+{13\over 2}\omega+{3\over 2}\omega^2\big)S_2^j-{5+4\omega\over 6}S_3^j\Big)
\over 1+6\omega+6\omega^2+{3\over 2}\omega^3}
\label{fis1}
\end{eqnarray}
so at small $\omega=j-1$ the operator $S_1^j$ is approximately equal to gluon operator $\calf^j$.

\subsection{Correlators of the light-ray operators in the BFKL limit: what to expect \label{general}}
As we noted above, the BFKL limit for light-ray operators (\ref{lrays}) is $\omega=j-1\rightarrow 0$, $g^2\rightarrow 0$ but ${g^2\over\omega}\sim 1$. 
From Eq. (\ref{fis1}) we see that in this limit only gluon light-rays survive so hereafter we will identify $S_1^\omega$ from Eq. (\ref{sjs}) with $\calf^\omega$.
It is well known that anomalous dimension of light-ray operators $\calf^\omega$ in the BFKL limit  is given by the  solution of equation $\omega=\aleph(\Delta)$ 
where $\Delta$ is the dimension of the operator and $\aleph(\Delta)$ is the famous Pomeron intercept 
\begin{equation}
\omega~=~\taleph(\gamma+\omega)~~~\Rightarrow~~~\gamma=\gamma^\ast(\omega,g^2)
\label{gammast}
\end{equation}
where $\gamma+\omega=\Delta-3$ and $\taleph(\gamma)$ is the pomeron intercept 
\begin{equation}
\taleph(\gamma,g^2)~=~4g^2\big[2\psi(1)-\psi\big(-{\gamma\over 2}\big)-\psi\big(1+{\gamma\over 2}\big)\big]~+~O(g^4)
\label{talef}
\end{equation}
At present, two more terms in the perturbative expansion of the intercept are known \cite{Gromov:2015vua,Velizhanin:2015xsa,Caron-Huot:2016tzz}. 

The coefficient $C(\omega,\Delta)$ in the BFKL limit was calculated in Ref. \cite{Balitsky:2013npa}
(see also Eq. (\ref{corr2result}) below)
\begin{eqnarray}
&&\hspace{-1mm}
C(j,\Delta)~=~~16g^2N_c^2
{2^{1-2\xi^\ast}\pi\over{\xi^\ast}^2\sin\pi\xi^\ast \Gamma^2\big(1-{\xi^\ast\over 2}\big)\Gamma^2\big({1\over 2}+{\xi^\ast\over 2}\big)\aleph'(\xi^\ast)}
\label{C2frames}
\end{eqnarray}
where  $\xi^\ast=\gamma^\ast+\omega=\Delta-3$. 

As I mentioned in the Introduction, the goal of this paper is to obtain the structure constant $C(\Delta_i,j_i)$ in the triple BFKL limit $g^2\rightarrow 0$, $\omega_1\sim\omega_2\sim\omega_3\rightarrow 0$, but ${g^2\over \omega}\sim 1$. 
It will be demonstrated that the structure constant as a function of $g^2$ and $\omega_i=j_i-1$ has the form
\begin{eqnarray}
&&\hspace{-1mm}
C(j_i,\Delta_i,g^2)~
\label{Cgeneral}\\
&&\hspace{-1mm}
=~{iN_c^2\omega_1\omega_2\omega_3
F\big[\gamma^\ast(\omega_1,g^2),\gamma^\ast(\omega_2,g^2),\gamma^\ast(\omega_3,g^2)\big]
\over \pi^3(\omega_1+\omega_2-\omega_3) (\omega_1+\omega_3-\omega_2) (\omega_2+\omega_3-\omega_1)}
\Big[1+O(g^2)+O(\omega_i)+O\big({1\over N_c^2}\big)\Big]
\nonumber
\end{eqnarray}
where function $F(\gamma_1,\gamma_2,\gamma_3)$ is given by a certain integral over two-dimensional coordinates  represented as a quartic Mellin-Barnes integral in Appendix \ref{sect:lambda}. It should be noted that the singularities $(\omega_i-\omega_j-\omega_k)^{-1}$ are of general nature and come from the boost invariance of the correlator (\ref{defclr}) in the limit $n_j\rightarrow n_k$ \cite{Balitsky:2015tca}, see the discussion in Appendix \ref{sect:BKase}.

For the calculation of structure constants I use the method
developed in Refs. \cite{Balitsky:2015tca} 
and \cite{Balitsky:2015oux}
based on calculation of correlators of ``Wilson frames'' operators which are basically the light-ray operators with point-splitting UV regularization. 
It is explained in the next Section. 

\section{Wilson frames \label{sect:wframes}}

\begin{figure}[htb]
\begin{center}
\includegraphics[width=55mm]{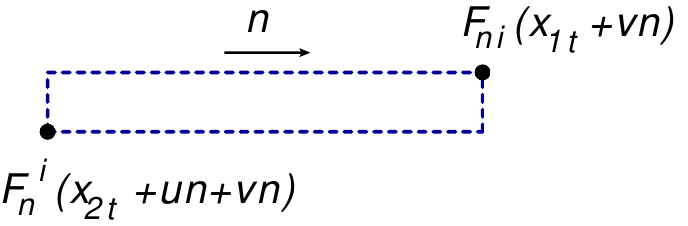}
\end{center}
\caption{Gluon ``Wilson frame'' operator\label{fig:wframe}}
\end{figure}

As we demonstrated in Ref. \cite{Balitsky:2013npa}, one cannot study correlators of LR operators in the BFKL approximation since the contribtions would be singular. 
Instead, one should consider the ``Wilson frame'' - LR operator with the point splitting in the transverse direction, see e.g. Fig. \ref{fig:wframe} for the gluon operator. 
We need the ``forward'' Wilson frame integrated over total translation in the corresponding light-like direction
%
%
\begin{eqnarray}
&&\hspace{-1mm}
\calf_n^j~(x_{1t},x_{2t})~\equiv~\int_0^\infty\!\! du~u^{1-j}\calf_n(u;x_{1t},x_{2t}),
\nonumber\\
&&\hspace{-1mm}
 \calf_n(u;x_{1t},x_{2t})~\equiv~\int\! dv~ 2{\rm Tr}\big\{F_{n\xi}(x_{1t}+un+vn)[un+vn+x_{1t},vn+x_{1t}]
\nonumber\\
&&\hspace{24mm}
\times~[vn+x_{1t},vn+x_{2t}]F_n^{~\xi}(x_{2t}+vn)[x_{2t}+vn,x_{2t}+un+vn]\big\}.  
\label{wframe}
\end{eqnarray}
As $x_{1t}\rightarrow x_{2t}$ the Wilson-frame operator  $\calf_n(l;x_{1t},x_{2t})$ reduces to LR operator $\calf(l,x_{1t})$ defined in Eq. (\ref{forwlrs}).
Moreover, it is intuitively clear that the point splitting $x_{12_t}$ serves as an UV cutoff for the light-ray operator in this limit, at least in the leading log approximation.

One can define also gluino and scalar ``Wilson frames'' by similar formulas and write down combinations but, as we mentioned above, we do not need their
explicit form since at small $\omega$'s everything is determined by gluon operators $\calf^j$.
Thus,  we  define Wilson-frame operators (\ref{wframe}) stretched in $n_1$, $n_2$ or $n_3$ directions and calculate their correlator at small $\omega_i$

\begin{figure}[htb]
\begin{center}
\includegraphics[width=111mm]{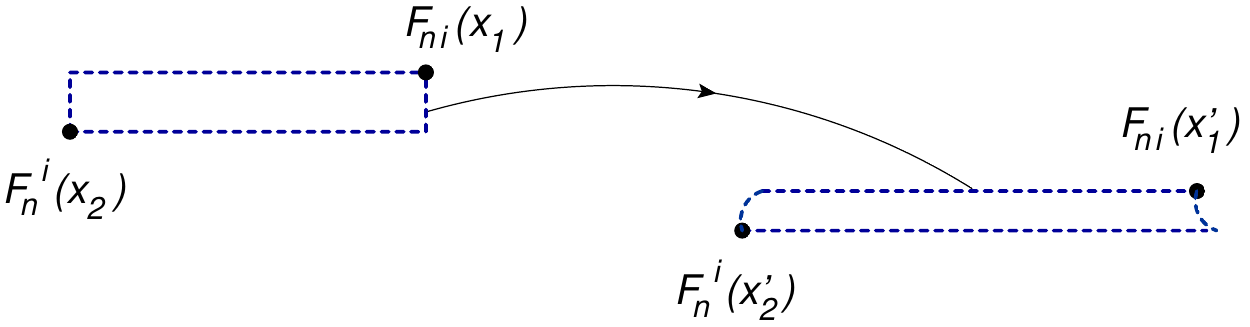}
\end{center}
\caption{Transformation of Wilson frame under inversion $x_\mu\rightarrow x_\mu/x^2$\label{fig:conf}}
\end{figure}

It should be emphasized that narrow Wilson-frame operators are approximately conformally invariant: if one makes the inversion around the point $(0,0,0,a_t)$  
one gets the long and narrow Wilson frame with somewhat distorted ends, see Fig. \ref{fig:conf}.
However, since we are calculating the correlators of Wilson frames in the leading BFKL approximation,
the logarithmic integrals are determined by the whole range of integration over $u$ and small corrections 
at the fringes can be neglected in the leading-log approximation. Thus, one should expect the conformal 
formulas for the two- and three- point correlators of Wilson frames in the limit of small width of frames 
of the same form as Eqs. (\ref{2pointCF}) and (\ref{defclr}).
\begin{eqnarray}
&&\hspace{-1mm}
\big\langle \calf_{n_1}^{j}\big(z_t+{w_t\over 2},z_t-{w_t\over 2}\big)\calf_{n_2}^{j'}\big(z'_t+{w'_t\over 2},z'_t-{w'_t\over 2}\big)\big\rangle
~\stackrel{w_t,w'_t\rightarrow 0}{=}~\delta(\nu-\nu'){C(\nu,\Delta)\over |z_t-z'_t|^{2\Delta(j)}}|w_tw'_t|^{\gamma(j)}
\nonumber\\
\label{corr2s}
\end{eqnarray}
and
\begin{eqnarray}
&&\hspace{-1mm}
\big\langle \calf_{n_1}^{j_1}\big(z_{1_t}+{w_{1_t}\over 2},z_{1_t}-{w_{1_t}\over 2}\big) \calf_{n_2}^{j_2}\big(z_{2_t}+{w_{2_t}\over 2},z_{2_t}-{w_{2_t}\over 2}\big)
 \calf_{n_3}^{j_3}\big(z_{3_t}+{w_{3_t}\over 2},z_{3_t}-{w_{3_t}\over 2}\big)\big\rangle
\label{corr3fs}\\
&&\hspace{-1mm}
\stackrel{w_{i_t}\rightarrow 0}{=}~C(\Delta_i,j_i){s_{12}^{j_1+j_2-j_3-1\over 2}|w_{1_t}|^{\gamma(j_1)}\over |z_{12_t}|^{\Delta(j_1)+\Delta(j_2)-\Delta(j_3)-1}}
{s_{13}^{j_1+j_3-j_2-1\over 2}|w_{2_t}|^{\gamma(j_2)}\over |z_{13_t}|^{\Delta(j_1)+\Delta(j_3)-\Delta(j_2)-1}}
{s_{23}^{j_2+j_3-j_1-1\over 2}|w_{3_t}|^{\gamma(j_3)}\over |z_{23_t}|^{\Delta(j_2)+\Delta(j_3)-\Delta(j_1)-1}}
\nonumber
\end{eqnarray}
with point-splitting distances $w_t$ serving as UV cutoffs similar to cutoff $\mu$ for the light-ray operators
in Eqs. (\ref{2pointCF}) and (\ref{defclr}).

Our goal is the three-point formula (\ref{corr3fs}) but first I remind the derivation of the BFKL asymptotics of two-point correlator (\ref{corr2s}) obtained in Ref.  \cite{Balitsky:2013npa} which will serve as a building block for three-frame calculation.

\section{Correlator of two Wilson frames in the BFKL limit \label{sect:2wframes}}
The CF of two Wilson-frame operators in Regge kinematics is calculated in the same way as four-point correlator of local operators 
$\langle T\{\calo(x_1)\calo(x_2)\calo(y_1)\calo(y_2)\}\rangle$ in the Regge limit $x_{1n_1},y_{1n_2}\rightarrow \infty$,  $x_{2n_1},y_{2n_2}\rightarrow -\infty$ and the 
rest of coordinates fixed. (Hereafter I use the notation $x_n\equiv x\cdot n$). Let me remind
the essential steps of such calculation (see e.g. Ref. \cite{Balitsky:2009yp}). 

\subsection{Rapidity factorization for 4-point correlators in the Regge limit.}

Let us consider the correlator of four scalar operators 
\footnote{For definiteness, one may think about Konishi operator $\calo=\phi^a_I\phi^a_I$.} 
\begin{eqnarray}
&&\hspace{-3mm}
 A(x_1,x_2,x_3,x_4)~\equiv~\mu^{-4}(\mu^4x_{12}^2x_{34}^2)^{2+\gamma_o}\langle T\{\calo(x_1)\calo(x_2)\calo(x_3)\calo(x_4)\}\rangle~
 \nonumber\\
 &&\hspace{-1mm}
 x_1=u_1n_1+x_{1_\perp}, ~~~x_2=v_1n_1+x_{2_\perp},~~~ x_3=u_2n_2+x_{1_\perp}, ~~~x_4=v_2n_2+x_{4_\perp}
 \label{ada}
\end{eqnarray}
where $\gamma_o$ is the  anomalous dimension of $\calo$. 
In the Regge limit $s_{12}=-2n_1\cdot n_2\rightarrow \infty$ and $x_{i_\perp}$ fixed. The amplitude (\ref{ada})
is a function of two conformal ratios which can be chosen in the Regge limit as
\begin{eqnarray}
&&\hspace{-5mm}
R~=~{x_{13}^2x_{24}^2\over x_{12}^2x_{34}^2}~\simeq~
{u_1u_2v_1v_2s_{12}^2\over x_{12_t}^2x_{34_t}^2},~~~~~~
 \nonumber\\
 &&\hspace{-1mm}
r~=~R\Big[1-{x_{14}^2x_{23}^2\over  x_{13}^2x_{24}^2}+{1\over R}\Big]^2
~\simeq~{(u_1u_2x_{34_\perp}^2+v_1v_2x_{12_\perp}^2-u_1v_2x_{23_\perp}^2-v_1u_2x_{14_\perp}^2)^2
\over u_1u_2v_1v_2x_{12_t}^2x_{34_\perp}^2}
\label{Rr}
\end{eqnarray}
so that $R$ increases with ``energy'' $s_{12}=-2n_1\cdot n_2$ while $r$ is energy-independent.
\footnote{To avoid confusion, we reserve the notation $a_t$ for the component of the vector 
$a$ orthogonal to three light-like vectors $n_1,n_2,n_3$ and use the notation 
$a_\perp$ when we discuss components orthogonal to the 
two light-like vectors $n_1$ and $n_2$.} This corresponds to the momentum-space definition of Regge limit 
$s/m_\perp^2\gg 1$ where $m_\perp^2$ is a characteristic mass scale of the process, in our case the scale
of inverse characteristic transverse distances.

In general, the calculation of particle scattering in the Regge limit is based on the rapidity factorization  
of the amplitude into the product of ``projectile impact factor'' with rapidities close 
to those of the projectile particle, ``target impact factor'' with rapidities close
to the those of the target, and scattering of color dipoles encompassing the rapidities 
in the region between projectile and the target. 
\begin{figure}[htb]
\begin{center}
\includegraphics[width=88mm]{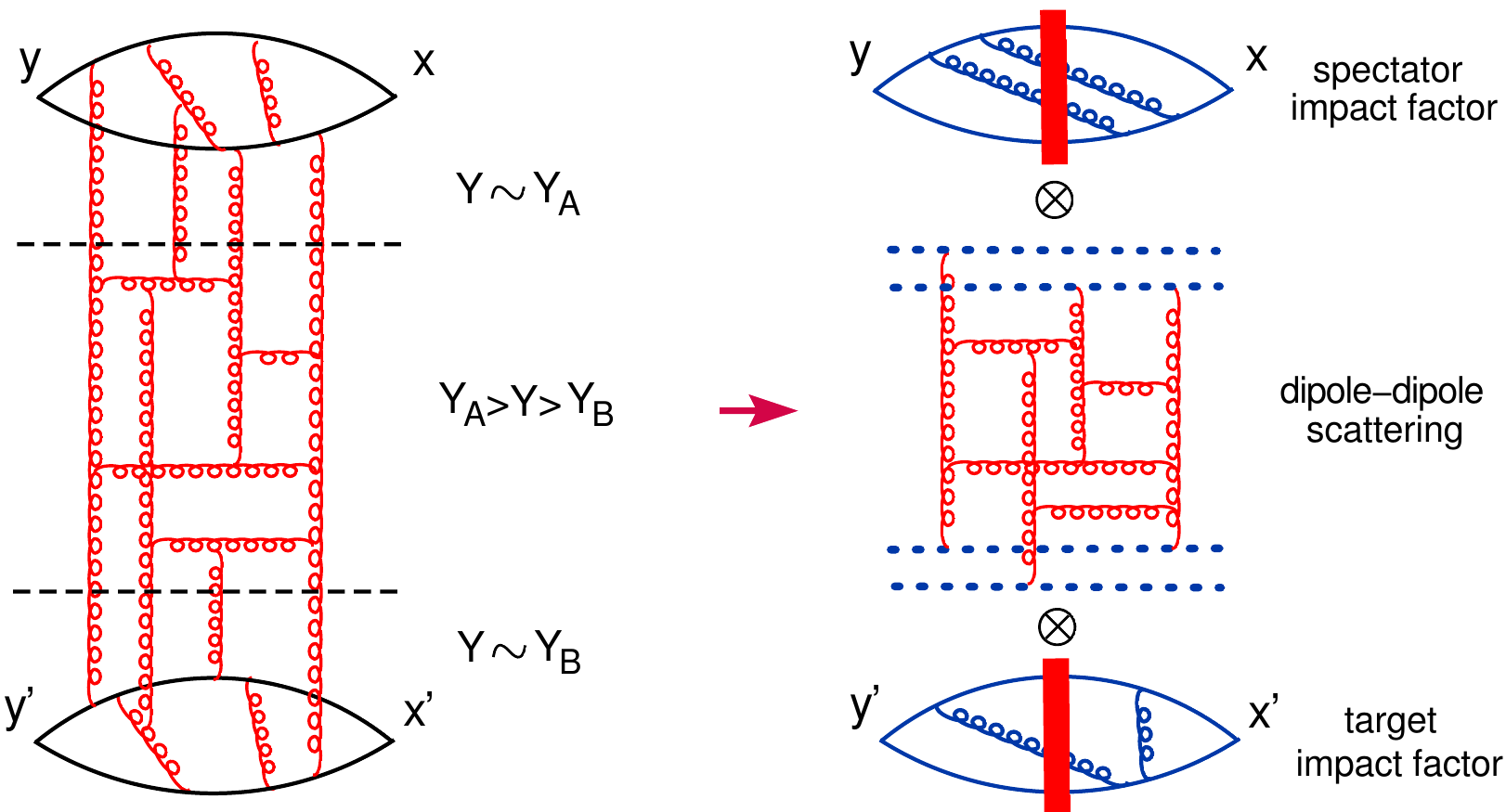}
\end{center}
\caption{Rapidity factorization for 4-point correlator in the Regge limit\label{fig:4pointCF}}
\end{figure}
Technically, one expands 
the $T\{\calo(x_1)\calo(x_2)\}$
 in the set of Wilson-line operators with the first being so-called ``color dipole'' 
$\calu(z_{1_\perp},z_{2_\perp})$
\begin{eqnarray}
&&\hspace{-1mm}
T\{\calo(x_1)\calo(x_2)\}~=~\int\! d^2z_{1_\perp}d^2z_{2_\perp} ~I(x_1,x_2;z_{1_\perp},z_{2_\perp} )
 {\rm Tr} \{U(z_{1_\perp})U^\dagger(z_{2_\perp})\}+...
 \nonumber\\
 &&\hspace{-1mm}
 \calu(z_{1_\perp},z_{2_\perp})~\equiv~1-{1\over N_c}{\rm Tr} \{U(z_{1_\perp})U^\dagger(z_{2_\perp})\}
\label{dipole}
\end{eqnarray}
where integration goes over $z_\perp$ orthogonal to both $n_1$ and $n_2$, the Wilson line $U$ is defined as
\begin{equation}
U(z_\perp)~\equiv~[\infty n_1+z_\perp,-\infty n_1+z_\perp]
\end{equation}
and dots stand for higher orders of perturbation theory and more Wilson lines. 
The rapidities\footnote{The definition of rapidity for the particle with momentum $k=\alpha n_1+\beta n_2+k_\perp$ is 
$Y\equiv\half\ln{\alpha\over \beta}$.}
 inside the color dipole should be cut from above by characteristic rapidities in the integrals forming the impact factor. 
To ensure conformal invariance of the rapidity factorization, one should expand in `` composite conformal dipoles'' introduced in Ref. \cite{Balitsky:2009xg}.
\begin{equation}
T\{\calo(x_1)\calo(x_2)\}~=~\int\! d^2z_{1_\perp}d^2z_{2_\perp} ~I(x_1,x_2;z_{1_\perp},z_{2_\perp} )
{\cal U}_{\rm conf}^{Y_A}(z_1,z_2)+...
\label{5.5}
\end{equation}
where 
\begin{equation}
{\cal U}_{\rm conf}^Y(z_1,z_2)~=~1-{1\over N_c}{\rm Tr} \{U(z_{1_\perp})U^\dagger(z_{2_\perp})\}~+~\alpha_s\times
~ ({\rm four~Wilson ~lines~ correction})
\label{conformaldipole}
\end{equation}
is a conformal composite dipole
and 
$Y_A=\half\ln{4x_{1n_2}x_{2n_2}\over s_{12}x_{12_t}^2}$ 
is the conformally invariant rapidity cutoff. 
The explicit form of the 4-lines correction is presented in Ref. \cite{Balitsky:2009xg} , but we do not need it for the leading BFKL logs. 

Since we are interested in Regge asymptotics, it is sufficient to consider highest eigenvalue of BFKL intercept with spin 0. 
Defining a projection of the conformal dipole (\ref{conformaldipole}) on Lipatov's eigenfunctions
\cite{Lipatov:1985uk} with spin 0, we get
\begin{eqnarray}
&&\hspace{-1mm}
 {\cal U}_{\rm conf}^Y(\nu,z_0)~\equiv~{1\over \pi^2}\!\int\! {d^2z_1d^2z_2\over z_{12}^4}~
\Big({z_{12}^2\over z_{10}^2z_{20}^2}\Big)^{\half -i\nu}~ {\cal U}_{\rm conf}^Y(z_1,z_2)
\nonumber\\
&&\hspace{-1mm}
 {\cal U}_{\rm conf}^Y(z_1,z_2)~=~\!\int_{-\infty}^\infty\!d\nu~{\nu^2\over \pi^2}\!\int\! d^2z_0
 \Big({z_{12}^2\over z_{10}^2z_{20}^2}\Big)^{\half+i\nu} {\cal U}_{\rm conf}^Y(\nu,z_0)~~+~{\rm higher~spins}
\label{uconf}
\end{eqnarray}
and therefore one can rewrite Eq. (\ref{5.5}) as follows
\begin{equation}
T\{\calo(x_1)\calo(x_2)\}~=~\int\! d^2z_0\!\int\!d\nu ~I_A(\nu )
\bigg[{x_{12}^2x_{1n_2}x_{2n_2}\over {(x_1-z_0)_\perp^2\over  x_{1n_2}}-{(x_2-z_0)_\perp^2\over  x_{2n_2}}}\bigg]^{\half+i\nu}
 {\cal U}_{\rm conf}^{Y_A}(\nu,z_0).
\end{equation}

Repeating the same expansion for the ``target'' we get
\begin{equation}
T\{\calo(x_3)\calo(x_4)\}~=~\int\! d^2z_0\!\int\!d\nu ~I_A(\nu )
\bigg[{x_{34}^2x_{3n_1}x_{4n_1}\over {(x_3-z_0)_\perp^2\over  x_{3n_1}}-{(x_4-z_0)_\perp^2\over  x_{4n_1}}}\bigg]^{\half+i\nu}
 {\cal V}_{\rm conf}^{Y_B}(\nu,z_0).
\end{equation}
where $Y_B=\half\ln{4x_{3n_1}x_{4n_1}\over s_{12}x_{34_\perp}^2}$ and the conformal dipole ${\cal V}_{\rm conf}^{Y_B}(\nu,z_0)$ 
is defined as
\begin{eqnarray}
&&\hspace{-1mm}
  {\cal V}_{\rm conf}^{Y_B}(\nu,z_0)~\equiv~{1\over \pi^2}\!\int\! {d^2z_3d^2z_4\over z_{34}^4}~
\Big({z_{34}^2\over z_{30}^2z_{40}^2}\Big)^{\half -i\nu}~  {\cal V}_{\rm conf}^{Y_B}(z_3,z_4)
\nonumber\\
&&\hspace{-1mm}
\calv_{\rm conf}^{Y_B}(z_{3_\perp},z_{4_\perp})~\equiv~
1-{1\over N_c}{\rm Tr} \{V(z_{3_\perp})V^\dagger(z_{4_\perp})\}_{\rm conf}^{Y_B}
\label{vconf}
\end{eqnarray}
where Wilson lines $V$ are ordered along $n_2$ direction
\begin{equation}
V(z_\perp)~\equiv~[\infty n_2+z_\perp,-\infty n_2+z_\perp]
\end{equation}

Now the 4-point CF can be represented as an integral of the product of two impact factors  $I_A(\nu),~I_B(\nu')$
and the amplitude of scattering of two color dipoles. In the leading BFKL approximation this amplitude has the form
($\alpha_s\equiv {g_{{}_{\rm YM}}^2\over 4\pi}$)
\begin{eqnarray}
&&\hspace{-1mm}
\langle {\cal U}_{\rm conf}^{Y_A}(\nu,z_0) {\cal V}_{\rm conf}^{Y_B}(\nu',z'_0)\rangle~
=~-{\alpha_s^2(N_c^2-1)\over 4N_c^2}
e^{\aleph(\nu,g^2)\ln Y_AY_B}
\label{confdipscat}\\
&&\hspace{-1mm}
\times~{16\pi^2\over \nu^2(1+4\nu^2)^2}
\Big[\delta(z_0-z'_0)\delta(\nu+\nu')+{2^{1-4i\nu}\delta(\nu-\nu')\over \pi|z_0-z'_0|^{2-4i\nu}}
{\Gamma\big({1\over 2}+i\nu\big)\Gamma(1-i\nu)\over\Gamma(i\nu)\Gamma\big(\half-i\nu\big)}\Big].
\nonumber
 \end{eqnarray}
where 
\begin{equation}
\aleph(\nu,g^2)~=~4g^2\big[2\psi(1)-\psi\big(\half+i\nu\big)-\psi\big(\half-i\nu\big)\big]~+~O(g^4)
\end{equation}
is the pomeron intercept (\ref{talef}). 

As I mentioned in the Introduction, in QCD only the $\alpha_s^2$ correction is known \cite{Fadin:1998py}
while in ${\cal N}=4$ SYM the $g^6$ term is known analytically \cite{Gromov:2015vua,Velizhanin:2015xsa,Caron-Huot:2016tzz} 
and many more can be calculated numerically
\cite{Gromov:2015vua} using Quantum Spectral Curve method \cite{Alfimov:2014bwa}.

Assembling the result for the 4-point CF(\ref{ada}) one gets the result in the form  of general formula \cite{Cornalba:2007fs}  for correlators in the ``Regge + large $N_c$''  limit
\begin{eqnarray}
&&\hspace{-3mm}
A(x_i)~\stackrel{s_{12}\rightarrow\infty}{=}~
{i\over 2}\!\int\! d\nu~f_+(\aleph(g^2,\nu))
F(g^2,\nu)
\Omega(r,\nu)R^{\aleph(g^2,\nu)/2}
\label{cornalba}
\end{eqnarray}
 where $f_+(\aleph)={e^{i\pi\aleph}-1\over \sin\pi\aleph}$ is a signature factor and
\begin{equation}
\Omega(r,\nu)~=~{\nu\over 2\pi^2}{\sin 2\nu\rho\over\sinh\rho},~~~~\cosh\rho={\sqrt{r}\over 2}
\end{equation}
is a solution of the Laplace equation in $H_3$ hyperboloid $(\partial^2_{H_3}+\nu^2+1)\Omega(r,\nu)=0$.
The dynamics is described by the pomeron intercept $\aleph(g^2,\nu)$ and the ``pomeron residue''
$F(g^2,\nu)$.  The formula (\ref{cornalba}) was proved in \cite{Cornalba:2007fs} (see also \cite{Costa:2012cb}) by considering the 
leading Regge pole in a conformal theory.
Also, it was demonstrated up to the NLO level that the structure (\ref{cornalba}) is reproduced by 
the high-energy OPE in Wilson lines \cite{Balitsky:1995ub,Balitsky:1998ya,Balitsky:2001gj}. 

\subsection{Correlator of two Wilson frames in the Regge limit \label{sect:2frames}}

The Regge limit for CF of two Wilson-frame operators means that longitudinal length of frame is much greater than the transverse separation between the frames
and the width of frames is even less.  As we mentioned, at small frame widths the frames are approximately 
conformally invariant so one may expect that the general formula (\ref{cornalba}) is applicable. At 
 $x_{12}^2,x_{34}^2\rightarrow 0$ one gets
\begin{eqnarray}
&&\hspace{-3mm}
r~\rightarrow~{(u_1-v_1)^2(u_2-v_2)^2x_{13_\perp}^4\over u_1v_1u_2v_2x_{12_\perp}^2x_{34_\perp}^2}
\label{rsmall}
\end{eqnarray}
and 
\begin{equation}
\Omega(r,\nu)~\rightarrow~{\nu\over 2\pi^2 i}\big(r^{-\half +i\nu}-r^{-\half -i\nu}\big)
\label{omegasmall}
\end{equation}
Moreover, if we consider ``forward'' correlation function
\begin{eqnarray}
&&\hspace{-5mm}
 A(l,l'; x_{1_\perp},x_{2_\perp},x_{3_\perp},x_{4_\perp})~ 
 \label{fcf}\\
&&\hspace{-5mm}
\equiv~\int_0^\infty\! dv_1dv_2~
 A(ln_1+v_1n_1+x_{1_\perp},v_1n_1+x_{2_\perp},l'n_2+v_2n_2+x_{3_\perp},v_2n_2+x_{4_\perp})
\nonumber
\end{eqnarray}
the Eq. (\ref{cornalba}) reduces to
\begin{eqnarray}
&&\hspace{-7mm} 
A(l,l'; x_{1_\perp},x_{2_\perp},x_{3_\perp},x_{4_\perp})
~\stackrel{x_{12_\perp}^2,x_{34_\perp}^2\rightarrow 0}{=}~ll'i\!\int\!{d\nu}~\Phi(\nu,g^2)
\label{ada2}\\
&&\hspace{37mm}
\times~
\Big({x_{12_\perp}^2x_{34\perp}^2\over x_{13_\perp}^4}\Big)^{\half+i\nu}
\Big({l^2{l'}^2\over x_{12_\perp}^2x_{34\perp}^2}\Big)^{\aleph(\nu,g^2)/2}f_+(\aleph)   ~ 
\nonumber
\end{eqnarray}

As noted in Sect. \ref{sect:wframes}, at small widths Wilson frames are approximately conformally invariant so 
we need to obtain the representation of Eq. (\ref{ada2}) type for the correlator
\begin{equation}
\hspace{-1mm}
\langle \calf_{n_1}\big(l;z_t+{w_t\over 2},z_t-{w_t\over 2}\big)\calf_{n_2}\big(l';z'_t+{w'_t\over 2},z'_t-{w'_t\over 2}\big)\rangle
\label{corr2frams}
\end{equation} 
at $l,l'\rightarrow\infty$ (which corresponds to  $j\rightarrow 1\Leftrightarrow \omega\rightarrow 0$ after integration over $l,l'$).
In Ref.  \cite{Balitsky:2013npa} we performed calculation of CF of two Wilson-frame operators
\begin{equation}
\hspace{-1mm}
\langle \calf_{n_1}\big(l;z_t+{w_t\over 2},z_t-{w_t\over 2}\big)\calf_{n_2}\big(l';z'_t+{w'_t\over 2},z'_t-{w'_t\over 2}\big)\rangle
\label{corr2frames}
\end{equation} 
in Regge kinematics  in the same way as four-point correlator of local operators. 
In this Section I'll reproduce that calculation in a slightly different way useful for considering 3-frame correlator 
in the next Section.

We introduce some ``rapidity divide'' $Y_0$ between $Y_A$ and $Y_B$ and integrate between $Y_A$ and $Y_0$  
and between $Y_0$ and $Y_B$ in the leading BFKL approximation. After that, we need to convolute the results with
the leading order dipole-dipole scattering amplitude.
\begin{figure}[htb]
\begin{center}
\includegraphics[width=66mm]{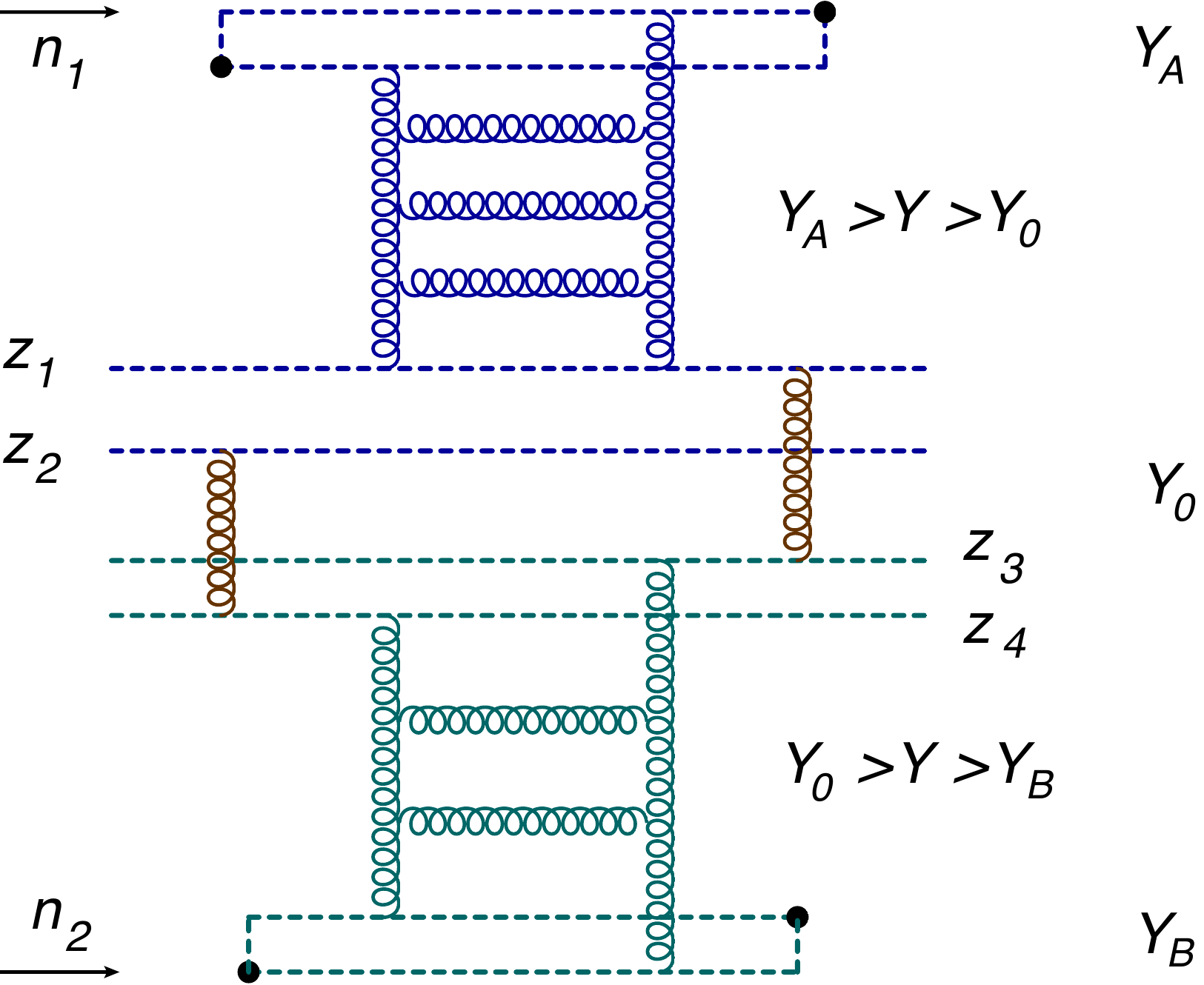}
\end{center}
\caption{Correlator of two Wilson frames in the Regge limit\label{fig:2wframes}}
\end{figure}

The first step is the expansion of Wilson frame in color dipoles.
The impact factor for Wilson frame, i.e. the coefficient of expansion of ``Wilson frame'' in color dipoles was calculated in Ref.  \cite{Balitsky:2013npa}
\begin{eqnarray}
&&\hspace{-1mm}
\calf_n(l;x_\perp,y_\perp)~=~{N_c^2\over 4\pi^3l}\!\int\! d^2z_\perp~
\big\{1-\calu(x_\perp,z_\perp)-\calu(z_\perp,y_\perp)
\label{IF1}\\
&&\hspace{7mm}
+~\calu(x_\perp,z_\perp)\calu(z_\perp,y_\perp)~+~O\big({1\over N_c}\big)\big\}^{Y_A}
\Big[{2\over (x-z)_\perp^2(y-z)_\perp^2}-{4(x-z,y-z)_\perp^2)\over (x-z)_\perp^4(y-z)_\perp^4}\Big]
\nonumber
\end{eqnarray}
where the rapidity cutoff is 
\begin{equation}
Y_A=\half\ln {l^2s_{12}\over (x-y)_\perp^2}
\label{cutoffA}
\end{equation}
 by analogy with four-point correlator. \footnote{Strictly speaking, by analogy with four-point correlator we get $\half\ln {u(u+l)s_{12}\over (x-y)_\perp^2}$
with additional intergation over $u$. However,  in Ref. \cite{Balitsky:2013npa} it was demonstrated that in the leading
log approximation this cutoff can be replaced by (\ref{cutoffA}). 
}
As explained in Ref. \cite{Balitsky:2013npa}, in order to calculate  the correlator of two Wilson frames 
we need to take into account
only the linear terms in Eq. (\ref{IF1}) so we can neglect the last quadratic term. To get the evolution of dipoles 
in Eq. (\ref{IF1}) from $Y_A$ to $Y_0$ we project onto Lipatov's eigenfunctions, i.e..  rewrite in terms 
of conformal dipoles \label{confdipolenu} and evolve these conformal dipoles in the leading BFKL order.

The projection of Eq. (\ref{IF1}) on Lipatov's eigenfunctions with spin 0 reads
\begin{eqnarray}
&&\hspace{-1mm}
\calf_n(l;x_\perp,y_\perp)
\label{projection}\\
&&\hspace{-1mm}
=~-{N_c^2\over 4\pi^3l}\!\int\!d\nu~{\nu^2\over \pi^2}\!\int\!  d^2z_0~ {\cal U}_{\rm conf}^a(\nu,z_0)\!\int\! d^2z_\perp
\Big[{2\over (x-z)_\perp^2(y-z)_\perp^2}-{4(x-z,y-z)_\perp^2)\over (x-z)_\perp^4(y-z)_\perp^4}\Big]
\nonumber\\
&&\hspace{22mm}
\times~\Big[
\Big({(x-z)_\perp^2\over (x-z_0)_\perp^2(z-z_0)_\perp^2}\Big)^{\half+i\nu} 
+\Big({(y-z)_\perp^2\over (y-z_0)_\perp^2(z-z_0)_\perp^2}\Big)^{\half+i\nu} \Big]
\nonumber\\
&&\hspace{-1mm}  
=~{N_c\over\pi^2l}\!\int\!d\nu~{\nu^2\over \pi^2}{\half+i\nu\over\half-i\nu}\!\int\! d^2z_0~
\Big[{2(x-z_0,y-z_0)_\perp^2\over (x-z_0)_\perp^2(y-z_0)_\perp^2}-1\Big]{ [(x-y)_\perp^2]^{-{1\over 2}+i\nu}
 {\cal U}_{\rm conf}(\nu,z_0)^{Y_A}   
\over [(x-z_0)_\perp^2]^{{1\over 2}+i\nu} [(y-z_0)_\perp^2]^{{1\over 2}+i\nu}}
\nonumber
\end{eqnarray}
where we used Eq. (4.11) from Ref. \cite{Balitsky:2013npa}
 to get the last line.

Moreover, in
the limit of narrow Wilson frame $(x-y)_\perp\rightarrow 0$ the integral in the r.h.s. of the above equation can be simplified.
Using
Eqs. (C.4) and (C.6) from Ref. \cite{Balitsky:2013npa} one easily obtains
\begin{eqnarray}
&&\hspace{-1mm}
\!\int\! d^2z_0~
\Big[{2(x-z_0,y-z_0)_\perp^2\over (x-z_0)_\perp^2(y-z_0)_\perp^2}-1\Big]
{[(x-y)_\perp^2]^{{1\over 2}+i\nu}
\over [(x-z_0)_\perp^2]^{{1\over 2}+i\nu} [(y-z_0)_\perp^2]^{{1\over 2}+i\nu} }
\Big({z_{12}^2\over z_{10}^2z_{20}^2}\Big)^{\half -i\nu}~
\nonumber\\
&&\hspace{-1mm}
\stackrel{x\rightarrow y}{=}~{i\pi\over 2\nu}{\big(\half-i\nu\big)^2\over\big(\half+i\nu\big)^2}
2^{4i\nu}{\Gamma\big(\half-i\nu\big)\Gamma(i\nu)\over\Gamma\big(\half+i\nu\big)\Gamma(-i\nu)}
\Big({|x-y|^2z_{12}^2\over (x-z_1)^2(x-z_2)^2}\Big)^{\half -i\nu}
\nonumber\\
&&\hspace{33mm}
-~{i\pi\over 2\nu}2^{-4i\nu}{\Gamma\big(\half+i\nu\big)\Gamma(-i\nu)\over\Gamma\big(\half-i\nu\big)\Gamma(i\nu)}
\Big({|x-y|^2z_{12}^2\over (x-z_1)^2(x-z_2)^2}\Big)^{\half +i\nu}
\label{fla5.17}
\end{eqnarray}
Recalling the definition (\ref{uconf}) of ${\cal U}_{\rm conf}$ and substituting Eq. (\ref{fla5.17}) in Eq. (\ref{projection}) 
one gets
\begin{eqnarray}
&&\hspace{-1mm}
\calf_n\big(l;z_t+{w_t\over 2},z_t-{w_t\over 2}\big)~=~{iN_c^2\over\pi^3l}
\int\! d\nu~\nu
{ 2^{-4i\nu}\Gamma\big({3\over 2}+i\nu\big)\Gamma(1-i\nu)\over \Gamma\big({3\over 2}-i\nu\big)\Gamma(1+i\nu)}
(w_t^2)^{-\half+i\nu}\calu^{Y_A}_{\rm conf}(z_t,-\nu)
\nonumber\\
\label{ffif}
\end{eqnarray}

The BFKL evolution of a conformal dipole reads
\begin{equation}
\calu_{\rm conf}^{Y_A}(\nu,z_0)~=~e^{(Y_A-Y_0)\aleph(\nu)}\calu_{\rm conf}^{Y_0}(\nu,z_0)
\end{equation}
so the result of integration over rapidities in the region $Y_A>Y>Y_0$ is
\begin{eqnarray}
&&\hspace{-1mm}
\calf_{n_1}\big(l;z_t+{w_t\over 2},z_t-{w_t\over 2}\big)~
\nonumber\\
&&\hspace{-1mm}
=~{iN_c^2\over\pi^3l}
\int\! d\nu~\nu
{ 2^{-4i\nu}\Gamma\big({3\over 2}+i\nu\big)\Gamma(1-i\nu)\over \Gamma\big({3\over 2}-i\nu\big)\Gamma(1+i\nu)}
(w_t^2)^{-\half+i\nu}e^{(Y_A-Y_0)\aleph(\nu,g^2)}\calu^{Y_0}_{\rm conf}(z_t,-\nu)
\label{evolproj}
\end{eqnarray}
where $Y_A=\ln l+\half\ln{ s_{12}\over w_t^2}$.

Repeating the same procedure for the bottom part of the diagram in Fig. \ref{fig:2wframes} one obtains
the result of integration over rapidities $Y_0>Y$ in the form
\footnote{The difference in signs of $Y_A$ and  $Y_B$ in Eqs. (\ref{evolproj}) and (\ref{evoltarg}) is due to the fact that replacement 
$n_1\leftrightarrow n_2$ should be accompanied by changing the sigh of the rapidity: $\ln{\beta\over\alpha}=-\ln{\alpha\over\beta}$.}
\begin{eqnarray}
&&\hspace{-1mm}
\calf_{n_2}\big(l';z'_t+{w'_t\over 2},z'_t-{w'_t\over 2}\big)~
\label{evoltarg}\\
&&\hspace{-1mm}
=~{iN_c^2\over\pi^3l'}
\int\! d\nu'~\nu'
{ 2^{-4i\nu'}\Gamma\big({3\over 2}+i\nu'\big)\Gamma(1-i\nu')\over \Gamma\big({3\over 2}-i\nu'\big)\Gamma(1+i\nu')}
({w'}_t^2)^{-\half+i\nu'}e^{(Y_0+Y_B)\aleph(\nu',g^2)}\calv^{Y_0}_{\rm conf}(z'_t,-\nu')
\nonumber
\end{eqnarray}
where $Y_B=\ln l'+\half\ln{s_{12}\over {w'}_t^2}$.

Using now the result for scattering of color dipoles in the leading perturbative order
\footnote{As usual, we stop the evolution of color dipoles from upper and lower parts
of the diagram in Fig.  \ref{fig:2wframes} at the points $Y_0+\delta$ and  $Y_0-\delta$.
The small $\delta$ is such that the relative energy $s_\delta=m_\perp^2e^{2\delta}$ is greater than 
the characteristic transverse scale $m_\perp^2$ but $g^2\ln{s_\delta\over m_\perp^2}=2g^2\delta\ll 1$.
In this case, one does not need to include evolution between  $Y_0+\delta$ and  $Y_0-\delta$ but can
still use the three-level formula
$$
\langle\calu(z_{1_\perp},z_{2_\perp}) \calv(z_{3_\perp},z_{4_\perp})\rangle~=~
-(1-{1\over N_c^2}){\alpha_s^2\over 8}\ln^2{ z_{13_\perp}^2z_{24_\perp}^2\over z_{14_\perp}^2z_{23_\perp}^2}
$$
which translates to Eq. (\ref{dipdiplo}) after projection on spin-0 eigenfunctions. \label{futnout}}
\begin{eqnarray}
&&\hspace{-1mm}
\langle {\cal U}_{\rm conf}^{Y_0}(-\nu,z_0) {\cal V}_{\rm conf}^{Y_0}(-\nu',z'_0)\rangle~
\label{dipdiplo}\\
&&\hspace{-1mm}
=~-\alpha_s^2
{4\pi^2\big(1-{1\over N_c^2}\big)\over \nu^2(1+4\nu^2)^2}
\Big[\delta(z_0-z'_0)\delta(\nu+\nu')-i\nu{2^{1+4i\nu}\delta(\nu-\nu')\over \pi|z_0-z'_0|^{2+4i\nu}}
{\Gamma\big({1\over 2}-i\nu\big)\Gamma(1+i\nu)\over\Gamma(1-i\nu)\Gamma\big(\half+i\nu\big)}\Big].
\nonumber
 \end{eqnarray}
we get the result for correlator of two Wilson frames in the form of Eq. (\ref{ada2}) type
\begin{eqnarray}
&&\hspace{-1mm}
\langle \calf_{n_1}\big(l;z_t+{w_t\over 2},z_t-{w_t\over 2}\big)\calf_{n_2}\big(l';z'_t+{w'_t\over 2},z'_t-{w'_t\over 2}\big)\rangle
\nonumber\\
&&\hspace{-1mm}
=~-i{g^2N_c^2\over\pi^3ll'}
\int\! d\nu~{2^{3-4i\nu}\nu\over  \big({1\over 2}+i\nu\big) \big({1\over 2}-i\nu\big)^3}
{ \Gamma\big({3\over 2}+i\nu\big)\Gamma(1-i\nu)\over \Gamma\big({3\over 2}-i\nu\big)\Gamma(1+i\nu)}
{(w_t^2{w'}_t^2)^{-\half+i\nu}e^{(Y_A+Y_B)\aleph(\nu,g^2)}\over ([z_t-z'_t)^2]^{1+2i\nu}}
\nonumber\\
&&\hspace{-1mm}
=~-i{g^2N_c^2\over\pi^3ll'}
\int\! d\nu~{2^{3-4i\nu}\nu(ll's_{12})^{\aleph(\nu)}\over  \big({1\over 2}+i\nu\big) \big({1\over 2}-i\nu\big)^3}
{ \Gamma\big({3\over 2}+i\nu\big)\Gamma(1-i\nu)\over \Gamma\big({3\over 2}-i\nu\big)\Gamma(1+i\nu)}
{(w_t^2{w'}_t^2)^{-\half+i\nu-\half\aleph(\nu,g^2)}\over [(z_t-z'_t)^2]^{1+2i\nu}}
\label{dvafa}
\end{eqnarray}
Note that the ``rapidity divide'' $Y_0$ disappeared from the result. Moreover, the scattering amplitude
(\ref{dvafa}) depends only on product of $l$ and $l'$ which is a reflection of boost invariance of the original
amplitude (\ref{wframe}): it is easy to see that if one makes boost $n_1\rightarrow \lambda n_1$ 
and $n_2\rightarrow {1\over \lambda} n_2$ the correlator  (\ref{wframe}) does not change. Now we
shall see that this property leads to the $\delta$-function in the correlator (\ref{corr2s}). 

Indeed,  the integral over $l$ and $l'$ have the form 
\begin{eqnarray}
&&\hspace{-1mm}
\langle \calf^j_{n_1}\big(z_t+{w_t\over 2},z_t-{w_t\over 2}\big)\calf^{j'}_{n_2}\big(z'_t+{w'_t\over 2},z'_t-{w'_t\over 2}\big)\rangle
~=~-i{g^2N_c^2\over\pi^3}\!\int_0^\infty\! dl dl' ~l^{-j}{l'}^{-j'}
\label{dvafaj1}\\
&&\hspace{-1mm}
\times~\theta\Big(ll'-{(z_t-z'_t)^2\over s_{12}}\Big)
\int\! d\nu~{\nu 2^{3-4i\nu}(ll's_{12})^{\aleph(\nu,g^2)}\over  \big({1\over 2}+i\nu\big) \big({1\over 2}-i\nu\big)^3}
{ \Gamma\big({3\over 2}+i\nu\big)\Gamma(1-i\nu)\over \Gamma\big({3\over 2}-i\nu\big)\Gamma(1+i\nu)}
{(w_t^2{w'}_t^2)^{-\half+i\nu-\half\aleph(\nu,g^2)}\over [(z_t-z'_t)^2]^{1+2i\nu}}
\nonumber
\end{eqnarray}
where the factor $\theta\big(ll'-{(z_t-z'_t)^2\over s_{12}}\big)$ comes from the restriction that the longitudinal  size of two 
Wilson frames should be greater than the relative transverse separation.
\footnote{This the  $s\gg m_\perp^2$ requirement for applicability of BFKL approximation recast in the coordinate-space language,
 see the discussion in Ref. \cite{Balitsky:2013npa}.}

Performing the integration over $l$ and $l'$ one obtains
where $j=\half+i\varsigma$ and  $j'=\half+i\varsigma'$.
\begin{eqnarray}
&&\hspace{-1mm}
\langle \calf^j_{n_1}\big(z_t+{w_t\over 2},z_t-{w_t\over 2}\big)\calf^{j'}_{n_2}\big(z'_t+{w'_t\over 2},z'_t-{w'_t\over 2}\big)\rangle
~=~-i\delta(\varsigma-\varsigma'){g^2N_c^2\over\pi^2}
\!\int_{-\infty}^\infty\! d\nu ~{1\over \omega-\aleph(\nu,g^2)}
\nonumber\\
&&\hspace{-1mm}
\times~
{\nu  2^{4-4i\nu}s_{12}^{\omega}\over  \big({1\over 2}+i\nu\big) \big({1\over 2}-i\nu\big)^3}
{ \Gamma\big({3\over 2}+i\nu\big)\Gamma(1-i\nu)\over \Gamma\big({3\over 2}-i\nu\big)\Gamma(1+i\nu)}
{(w_t^2{w'}_t^2)^{-\half+i\nu-\half\aleph(\nu,g^2)}\over [(z_t-z'_t)^2]^{j+2i\nu-\aleph(\nu,g^2)}}
\label{dvafaj}
\end{eqnarray}

Next, we analytically continue this formula to small $\omega=j-1$.
To estimate this integral at small $\omega$'s it is convenient to rewrite it in the variable $\xi=2i\nu-1$.  
\begin{eqnarray}
&&\hspace{-1mm}
\langle \calf^j_{n_1}\big(z_t+{w_t\over 2},z_t-{w_t\over 2}\big)\calf^{j'}_{n_2}\big(z'_t+{w'_t\over 2},z'_t-{w'_t\over 2}\big)\rangle
~=~\delta(\varsigma-\varsigma')g^4N_c^2
\!\int_{-1-i\infty}^{-1+i\infty}\! {d\xi\over 2\pi i} ~{1\over \omega-\taleph(\xi,g^2)}
\nonumber\\
&&\hspace{-1mm}
\times~{2^{5-2\xi}\pi s_{12}^\omega\over\xi^2\sin\pi\xi \Gamma^2\big(1-{\xi\over 2}\big)\Gamma^2\big({1\over 2}+{\xi\over 2}\big)}
{(w_t^2{w'}_t^2)^{\xi-\taleph(\xi,g^2)\over 2}\over [(z_t-z'_t)^2]^{2+\omega+\xi-\taleph(\xi,g^2)}}
\label{dvafaj2}
\end{eqnarray}
The notation here is 
\begin{equation}
\talef(\xi,g^2)~\equiv~\aleph\big(-i{1+\xi\over 2},g^2\big)~=~4g^2\big[2\psi(1)-\psi\big(-{\xi\over 2}\big)-\psi\big(1+{\xi\over 2}\big)\big]
~+~O(g^4)
\end{equation}
 and we often omit the $g^2$ 
dependence to avoid cluttering of the formulas.  

At small $w_t^2,{w'}_t^2$ we can close the contour of 
$\xi$ integration on the residues in the right half-plane. The two leading poles are located at $\xi^\ast=\taleph^{-1}(\omega)$ and $\xi=0$. Let us consider them in turn. 
Taking residue at $\xi^\ast=\taleph^{-1}(\omega,g^2)$ we get
\begin{eqnarray}
&&\hspace{-1mm}
\langle \calf^j_{n_1}\big(z_t+{w_t\over 2},z_t-{w_t\over 2}\big)\calf^{j'}_{n_2}\big(z'_t+{w'_t\over 2},z'_t-{w'_t\over 2}\big)\rangle
\nonumber\\
&&\hspace{-1mm}
=~\delta(\varsigma-\varsigma')g^2N_c^2
{2^{5-2\xi^\ast}\pi s_{12}^\omega\over{\xi^\ast}^2\sin\pi\xi^\ast \Gamma^2\big(1-{\xi^\ast\over 2}\big)\Gamma^2\big({1\over 2}+{\xi^\ast\over 2}\big)\aleph'(\xi^\ast)}
{(w_t^2{w'}_t^2)^{\xi^\ast-\omega\over 2}\over [(z_t-z'_t)^2]^{2+\xi^\ast}}
\label{corr2resultat}
\end{eqnarray}
Comparing this equation to general form of two-point correlator of light-ray operators (\ref{2pointCF})
we see that $\xi^\ast-\omega$  can be identified with anomalous dimension $\gamma$ so we finally get  \cite{Balitsky:2013npa}
\begin{eqnarray}
&&\hspace{-1mm}
\langle \calf^j_{n_1}\big(z_t+{w_t\over 2},z_t-{w_t\over 2}\big)\calf^{j'}_{n_2}\big(z'_t+{w'_t\over 2},z'_t-{w'_t\over 2}\big)\rangle
\nonumber\\
&&\hspace{-1mm}
=~
\delta(\varsigma-\varsigma')g^2N_c^2
{2^{5-2\xi^\ast}\pi s_{12}^\omega\over{\xi^\ast}^2\sin\pi\xi^\ast \Gamma^2\big(1-{\xi^\ast\over 2}\big)\Gamma^2\big({1\over 2}+{\xi^\ast\over 2}\big)\aleph'(\xi^\ast)}
{(w_t^2{w'}_t^2)^{\gamma^\ast\over 2}\over [(z_t-z'_t)^2]^{2+\omega+\gamma^\ast}}
\label{corr2result}
\end{eqnarray}
where $\gamma^\ast$ is a solution of the equation (\ref{gammast}) and $\xi^\ast=\gamma^\ast+\omega=\Delta-3$.  

Note that this formula is actually at the NLO  level: 
in the leading log approximation we just get $\omega=\aleph(\gamma^\ast)$ and $[(z-z')_t^2]^{-2-\gamma_\ast}$ in the r.h.s. of Eq. (\ref{corr2result}). The reason that we got the NLO equation 
(\ref{gammast}) is that we used $Y_A=\ln l+\half\ln s_{12}m_\perp^2-\half\ln m_\perp^2w_t^2$ where the last term exceeds
the LLA accuracy. 
As demonstrated in Ref. \cite{Balitsky:2013npa}, we can do this using the exact formula for the 4-point correlator (\ref{cornalba}). Unfortunately, for the 6-point correlator there is no such formula so we cannot promote our LO BFKL calculation to the NLO level.

At $g^2\ll \omega\ll1$  we get $\xi_\ast\simeq -8{g^2\over\omega}$ (recall that $\aleph(\xi)\simeq -8{g^2\over\xi}$ at small $\xi$) and therefore the result  (\ref{corr2result}) takes the form
\begin{eqnarray}
&&\hspace{-1mm}
\langle \calf^j_{n_1}\big(z_t+{w_t\over 2},z_t-{w_t\over 2}\big)\calf^{j'}_{n_2}\big(z'_t+{w'_t\over 2},z'_t-{w'_t\over 2}\big)\rangle
\nonumber\\
&&\hspace{-1mm}
\simeq~ 
-\delta(\omega-\omega'){N_c^2\omega s_{12}^\omega\over 2\pi[(z_t-z'_t)^2]^{2+\omega}}\Big({(z_t-z'_t)^2\over |w_tw'_t|}\Big)^{{8g^2\over\omega}+\omega}
\label{dvafapole}
\end{eqnarray}
which agrees with Eq. (4.17) from Ref. \cite{Balitsky:2013npa}.  

\begin{figure}[htb]
\begin{center}
\includegraphics[width=77mm]{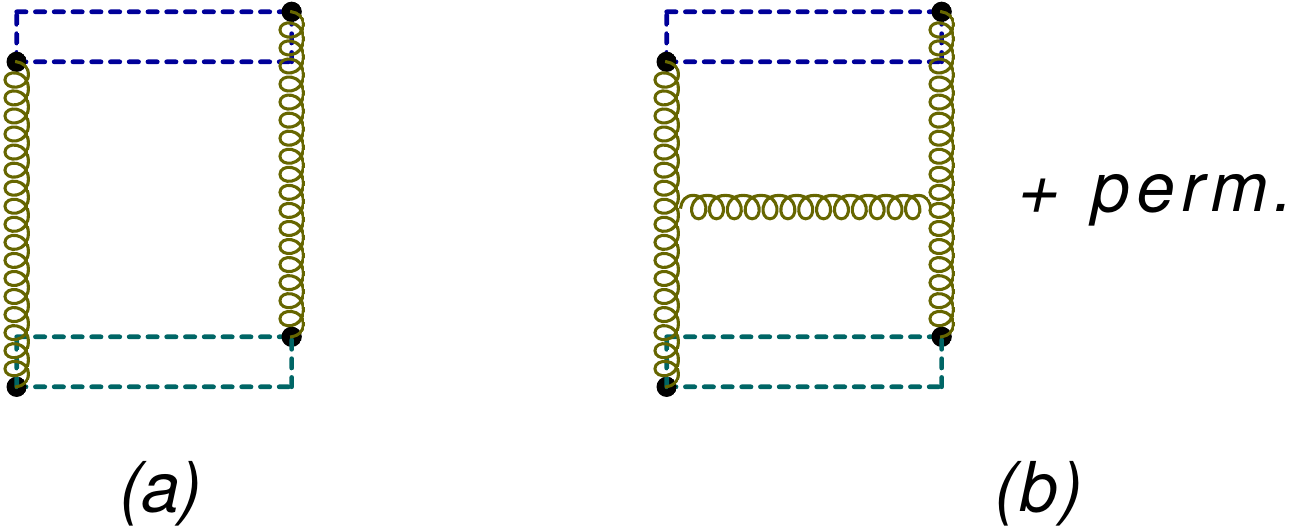}
\end{center}
\caption{First (a) and second (b) order diagrams for the correlator of two Wilson frames \label{fig:2wframesLO}}
\end{figure}

Let us consider now the pole at $\xi=0$. At small $\xi$ 
\begin{equation}
{1\over \omega-\taleph(\xi)}\simeq {\xi\over 8g^2}\Big(1-{\xi\omega\over 8g^2}+...\Big)
\end{equation}
so the residue at $\xi=0$ yields
\begin{equation}
\delta(\omega-\omega'){\omega s_{12}^\omega N_c^2\over2\pi [(z_t-z'_t)^2]^{2+\omega}}
\Big[1+{8g^2\over \omega}\ln{(z_t-z'_t)^2\over |w_tw'_t|}\Big]
\label{ostatok}
\end{equation}
Thus,  the result for diagrams in Fig. \ref{fig:2wframes} is a sum of Eq. (\ref{corr2result}) and Eq. (\ref{ostatok}). However, there are two low-order diagrams 
shown in Fig. \ref{fig:2wframesLO} that are not included in this result since the formula (\ref{IF1}) is correct starting from the second order of perturbation theory.
These diagrams should cancel the contribution of the $\xi=0$ pole (\ref{ostatok}) so the final result (\ref{corr2result}) has
proper conformal behavior. The tree-level diagram in Fig. \ref{fig:2wframesLO}a is calculated in the appendix A and the result (\ref{corr2LRs}) 
is minus the first term in the square brackets in Eq. (\ref{ostatok}). Similarly, the contribution of  diagrams in Fig. \ref{fig:2wframesLO}b should cancel the second term so
the contribution of all diagrams (in Fig.\ref{fig:2wframesLO} and Fig. \ref{fig:2wframes}) is given by Eq. (\ref{corr2result}).

\section{Correlator of three Wilson frames in the triple Regge limit \label{sect:3wframes}}
\subsection{Triple BFKL evolution}

\begin{figure}[htb]
\begin{center}
\includegraphics[width=77mm]{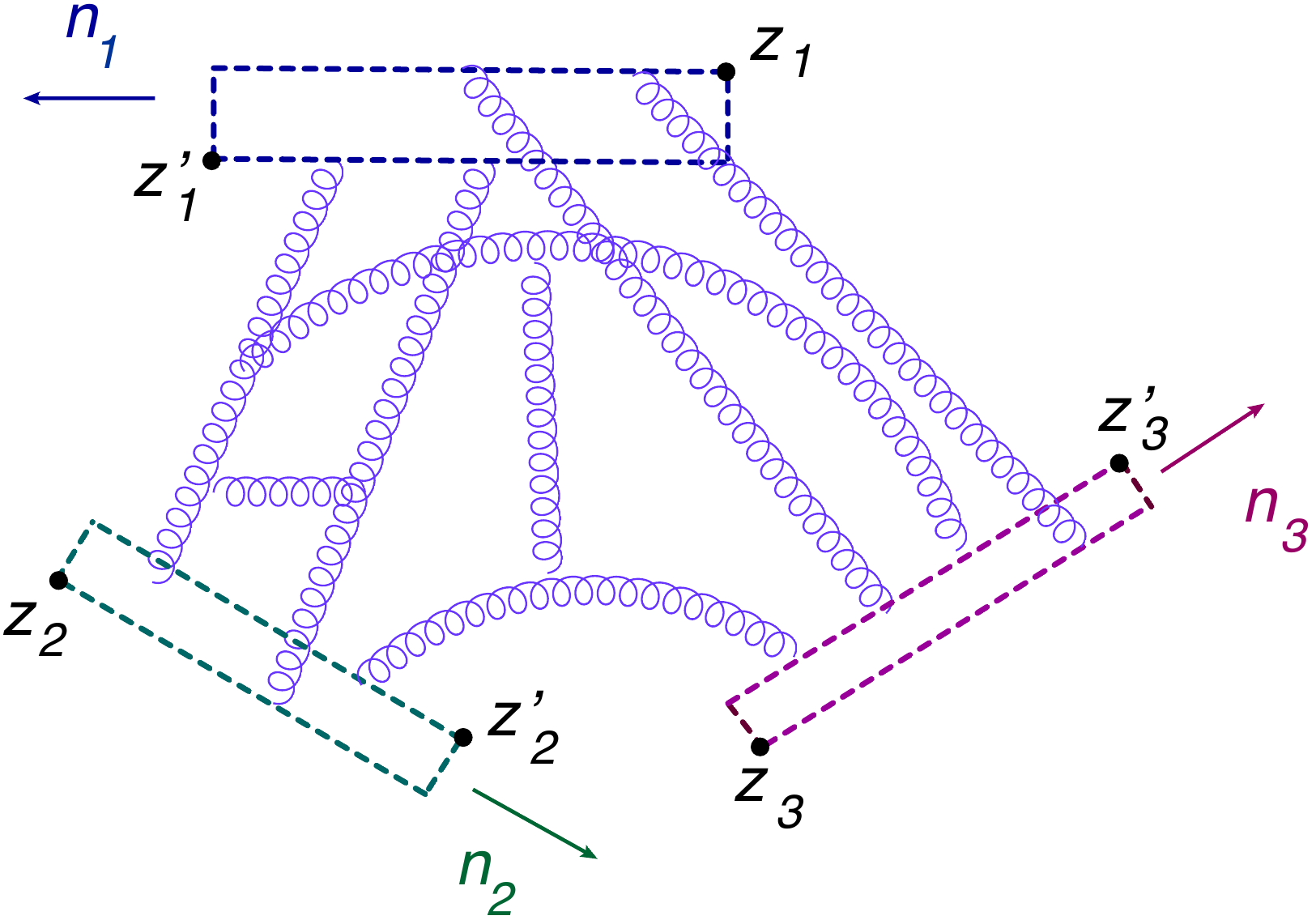}
\end{center}
\caption{Correlator of three Wison frames \label{fig:3wframesa}}
\end{figure}

To get the structure constant in Eq. (\ref{defclr}) at $\omega_i\rightarrow 0$ we will consider the correlator of three gluon Wilson frames
aligned along $n_1$, $n_2$, and $n_3$ directions:
\begin{eqnarray}
&&\hspace{-1mm}
\big\langle \calf_{n_1}^{j_1}\big(z_{1_t}+{w_{1_t}\over 2},z_{1_t}-{w_{1_t}\over 2}\big)\calf_{n_2}^{j_2}\big(z_{2_t}+{w_{2_t}\over 2},z_{2_t}-{w_{2_t}\over 2}\big)
\calf_{n_3}^{j_3}\big(z_{3_t}+{w_{3_t}\over 2},z_{3_t}-{w_{3_t}\over 2}\big)\big\rangle
\label{general1}
\end{eqnarray}
A typical diagram is shown in Fig. \ref{fig:3wframesa}.

As usual, we assume that longitudinal lengths of frames are much greater
than the transverse separations between the frames and those separations are much greater than widths of the frames. 
The form of the three-point correlators of light-ray operators  (\ref{defclr}) suggests that this correlator is determined by three BFKL 
evolutions. It will be demonstrated in this Section.

\begin{figure}[htb]
\hspace{-11mm}
\includegraphics[width=161mm]{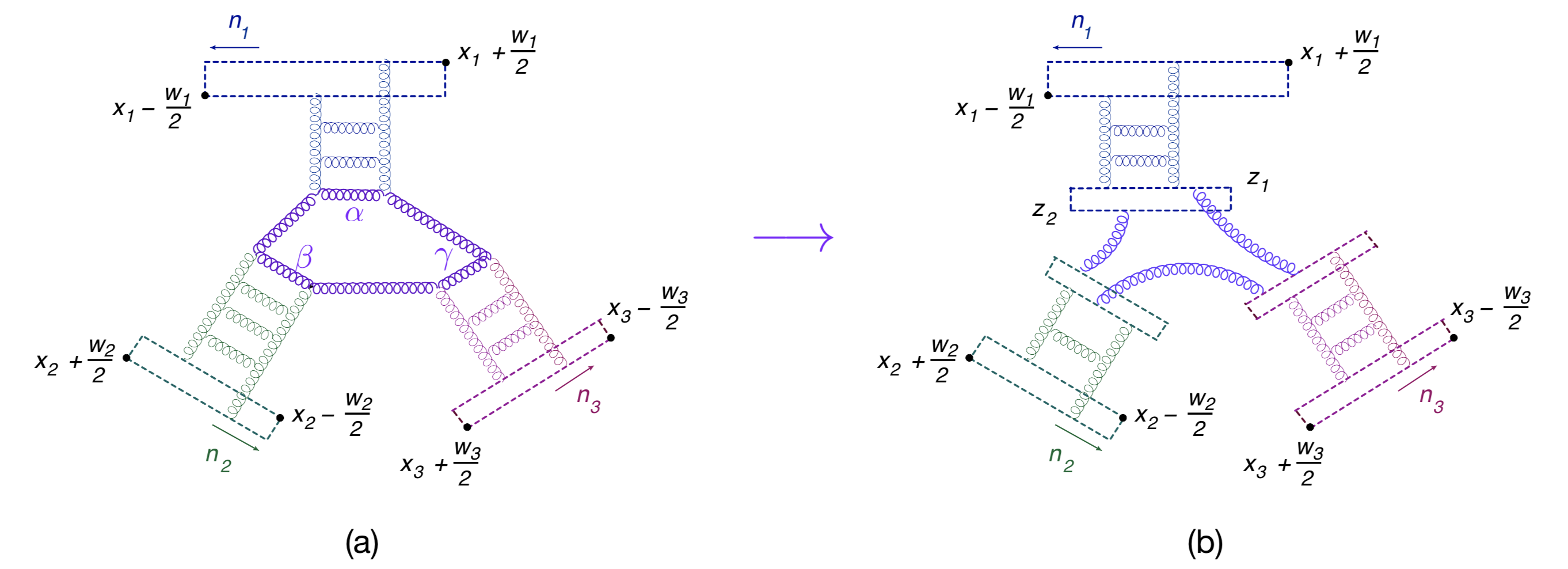}
\caption{Triple BFKL evolution.}
\label{fig:3bfklevols}
\end{figure}

The method to obtain BFKL asymptotics of a scattering amplitude  by evolution of Wilson lines is the following.
In a typical  amplitude like shown in Fig. \ref{fig:3bfklevols}a we separate the (gluon) fields according to their rapidity,
using the fact that particles with different rapidities perceive each other as Wilson lines, and study the  evolution
of these Wilson lines with respect to rapidity cutoff. 
Since we have now three light-like directions, it is convenient to introduce ``triple Sudakov variables''
\begin{equation}
k=\alpha n_1+\beta n_2+\gamma n_3+k_t,~~~~d^4k~=~{\sqrt{s_{12}s_{13}s_{23}}\over 2}d\alpha d\beta d\gamma dk_t
\label{tridakov}
\end{equation}
and consider factorization in all three of them.  
\footnote{As defined in Sect. \ref{sect:rych}, $n_i$ are light-like vectors with $s_{ij}=-2n_i\cdot n_j$ and $k_t$ is orthogonal 
to all three of them}

Similarly to the analysis of amplitudes in the usual Regge regime 
we assume that all $k_t^2\sim m_\perp^2$ where $m_\perp^2$ is of of order of (inverse) transverse separations between Wilson frames. 
Also, we assume that all $s_{ij}$ are of the same order of magnitude $s\gg m_\perp^2$.

The key observation is that as long as there is a sufficient rapidity space  for the evolution of each of Wilson frames these
evolutions are the same as for the two-point correlator of Wilson lines.
To demonstrate this, consider the evolution of $n_1$-parallel Wilson frame schematically depicted by the 
upper gluon ladder in Fig. \ref{fig:3bfklevols}b.  
It is convenient to relate ``triple Sudakov'' variables 
(\ref{tridakov}) to usual Sudakov variables
\begin{equation}
k=\talfa n_1+\tibeta \tiln_1+k_\perp
\label{sudakov}
\end{equation}
where we chose the second light-like vector as
\begin{equation}
\tiln_1~={s_{13}\over 2s_{23}}n_2+{s_{12}\over 2s_{23}}n_3-{n_1\over 4}
\label{sudakov1}
\end{equation}
so that $-2n_1\cdot \tiln_1\equiv\tils={s_{12}s_{13}\over s_{23}}$. 
We can rewrite Eq. (\ref{sudakov}) as follows
\begin{eqnarray}
k~=~\talfa n_1+\tibeta\tiln_2
+\tilk\tile+k_t e_t
\end{eqnarray}
where
\begin{equation}
\tile~\equiv~\Big(n_2\sqrt{s_{13}\over s_{12}s_{23}}-n_3\sqrt{s_{12}\over s_{13}s_{23}}\Big),~~~~~~~~~~~\tile^2=1
\label{defe}
\end{equation}
The relation between variables (\ref{tridakov}) and (\ref{sudakov}) is 
\begin{eqnarray}
\talfa~=~\alpha+{\tibeta\over 4},~~~~\tibeta~=~\beta{s_{23}\over s_{13}}+\gamma{s_{23}\over s_{12}},~~~~
\tilk=\half\sqrt{s_{23}\over s_{12}s_{13}}(\beta s_{12}-\gamma s_{13})
\end{eqnarray}
As we will demonstrate below, characteristic $\tilk$ are of order of $m_\perp$ (see Eq. (\ref{transintegral4}))  so we can define
\begin{equation}
k_\perp^2~\equiv~k_t^2+\tilk^2~\sim~m_\perp^2
\end{equation}
In terms of these variables
\begin{equation}
k^2~=~k_\perp^2-\talfa\tibeta \tils
\end{equation}
and the evolution of the $n_1$-parallel  Wilson lines looks like the evolution considered in  Sect. \ref{sect:2frames}
for the two-point correlator of Wilson frames.
 Thus, we can recycle the result (\ref{evolproj})
\begin{eqnarray}
&&\hspace{-1mm}
\calf_{n_1}\big(l_1;x_{1t}+{w_{1t}\over 2},x_{1t}-{w_{1t}\over 2}\big)~
\label{evolproj1}\\
&&\hspace{-1mm}
=~{iN_c^2\over\pi^3l_1}
\int\! d\nu_1~\nu_1
{ 2^{-4i\nu_1}\Gamma\big({3\over 2}+i\nu_1\big)\Gamma(1-i\nu_1)\over \Gamma\big({3\over 2}-i\nu_1\big)\Gamma(1+i\nu_1)}
(w_{1t}^2)^{-\half+i\nu_1}e^{(Y_1-Y'_1)\aleph(\nu_1)}\calu^{Y'_1}_{\rm conf}(x_{1t},-\nu_1)
\nonumber
\end{eqnarray}
where in the LLA $Y_1=\half\ln {l_1^2\tils\over w_{1t}^2}\simeq \ln l_1+\half\ln\tils m_\perp^2$ and
$Y'_1$ is the rapidity ($\half\ln{\talfa\over\tibeta}$) at which we stop the evolution.  

 To strengthen these coordinate-space arguments  in favor of BFKL evolution in the triple Regge limit,
it is shown in  the Appendix \ref{sect:bfkl} that the standard momentum-space calculation of one-loop
diagrams in the triple Regge limit reproduces the first rung of the BFKL ladder for color dipoles.

It should be noted that the arguments in favor of BFKL evolution in the triple Regge limit presented above are 
somewhat general, so  in  the Appendix \ref{sect:bfkl}
I confirm them by a standard momentum-space calculation of one-loop
diagrams in the triple Regge limit which reproduces the first rung of the BFKL ladder for color dipoles.

The explicit form 
of the conformal dipole (\ref{uconf}) in the coordinates $z_t$ and $\tilz$ reads
\footnote{As mentioned above, in the LLA ${\cal U}_{\rm conf}(z_1,z_2)$ can be replaced by ${\cal U}(z_1,z_2).$}
\begin{eqnarray}
&&\hspace{-1mm}
 {\cal U}_{\rm conf}^a(\nu_1,x_{1t})~
\equiv~{1\over \pi^2}\!\int\! {dz_{1t}d\tilz_1dz_{2t}d\tilz_2\over (z_{12t}^2+\tilz_{12}^2)^2}~
\label{u1}\\
&&\hspace{-1mm}
\times~\Big({z_{12t}^2+\tilz_{12}^2\over [(x_1-z_1)_t^2+\tilz_1^2][(x_1-z_2)_t^2+\tilz_2^2]}\Big)^{\half -i\nu_1}~
 {\cal U}^a(z_{1t}e_t+\tilz_1\tile,z_{2t}e_t+\tilz_2\tile)
\nonumber
\end{eqnarray}

Repeating the same procedure for the  frame parallel to $n_2$  we get
\begin{eqnarray}
&&\hspace{-1mm}
\calf_{n_2}\big(l_2;x_{2t}+{w_{2t}\over 2},x_{2t}-{w_{2t}\over 2}\big)~
\label{evolproj2}\\
&&\hspace{-1mm}
=~{iN_c^2\over\pi^3l_2}
\int\! d\nu_2~\nu_2
{ 2^{-4i\nu_2}\Gamma\big({3\over 2}+i\nu_2\big)\Gamma(1-i\nu_2)\over \Gamma\big({3\over 2}-i\nu_2\big)\Gamma(1+i\nu_2)}
(w_{2t}^2)^{-\half+i\nu_2}e^{(Y_2-Y'_2)\aleph(\nu_2)}\calv^{Y'_2}_{\rm conf}(x_{2t},-\nu_2)
\nonumber
\end{eqnarray}
Here $Y_2\simeq \ln l_2+\half\ln\bres m_\perp^2$,  
$Y'_2$ is the rapidity at which we stop the evolution and
\begin{eqnarray}
&&\hspace{-1mm}
 {\cal V}_{\rm conf}^a(\nu_2,x_{2t})~
\equiv~{1\over \pi^2}\!\int\! {dz_{3t}d\brez_3dz_{4t}d\brez_4\over (z_{34t}^2+\brez_{34}^2)^2}~
\label{v2}\\
&&\hspace{-1mm}
\times~\Big({z_{34t}^2+\brez_{34}^2\over [(x_2-z_3)_t^2+\brez_3^2][(x_2-z_4)_t^2+\brez_4^2]}\Big)^{\half -i\nu_2}~
 {\cal V}^a(z_{3t}e_t+\brez_3,z_{4t}e_t+\brez_4)
\nonumber
\end{eqnarray}
where ${\cal V}$ is the conformal dipole with Wilson lines parallel to $n_2$ and
\begin{equation}
\bre~\equiv~\Big(n_3\sqrt{s_{12}\over s_{13}s_{23}}-n_1\sqrt{s_{23}\over s_{12}s_{13}}\Big),~~~~~\bres\equiv{s_{12}s_{23}\over s_{13}}
\label{defbre}
\end{equation}
Note that the transverse plane $e_t, \bre$ for the evolution of second frame is different from the transverse plane
$e_t, \tile$ for the first frame.

Similarly one can get the result for the evolution of the third frame in the form
\begin{eqnarray}
&&\hspace{-1mm}
\calf_{n_3}\big(l_3;x_{3t}+{w_{3t}\over 2},x_{3t}-{w_{3t}\over 2}\big)~
\label{evolproj3}\\
&&\hspace{-1mm}
=~{iN_c^2\over\pi^3l_3}
\int\! d\nu_3~\nu_3
{ 2^{-4i\nu_3}\Gamma\big({3\over 2}+i\nu_3\big)\Gamma(1-i\nu_3)\over \Gamma\big({3\over 2}-i\nu_3\big)\Gamma(1+i\nu_3)}
(w_{3t}^2)^{-\half+i\nu_3}e^{(Y_3-Y'_3)\aleph(\nu_3)}\calw^{Y'_3}_{\rm conf}(x_{3t},-\nu_3)
\nonumber
\end{eqnarray}
Here $Y_3\simeq\ln l_3+\half\ln\ches m_\perp^2$,  
$Y'_3$ is the rapidity at which we stop the evolution and
\begin{eqnarray}
&&\hspace{-1mm}
 {\cal W}_{\rm conf}^a(\nu_3,x_{3t})~
\equiv~{1\over \pi^2}\!\int\! {dz_{5t}d\chez_5dz_{6t}d\chez_6\over (z_{56t}^2+\chez_{56}^2)^2}~
\label{u2}\\
&&\hspace{-1mm}
\times~\Big({z_{56t}^2+\chez_{56}^2\over [(x_3-z_5)_t^2+\chez_5^2][(x_4-z_6)_t^2+\chez_6^2]}\Big)^{\half -i\nu_3}~
 {\cal W}^a(z_{5t}e_t+\chez_5,z_{6t}e_t+\chez_6)
\nonumber
\end{eqnarray}

where ${\cal W}$ is the conformal dipole with Wilson lines parallel to $n_3$ and
\begin{equation}
\che~\equiv~\Big(n_1\sqrt{s_{23}\over s_{12}s_{13}}-n_2\sqrt{s_{13}\over s_{12}s_{23}}\Big),~~~~~\ches\equiv{s_{12}s_{23}\over s_{13}}
\label{defche}
\end{equation}

\begin{figure}[htb]
\includegraphics[width=141mm]{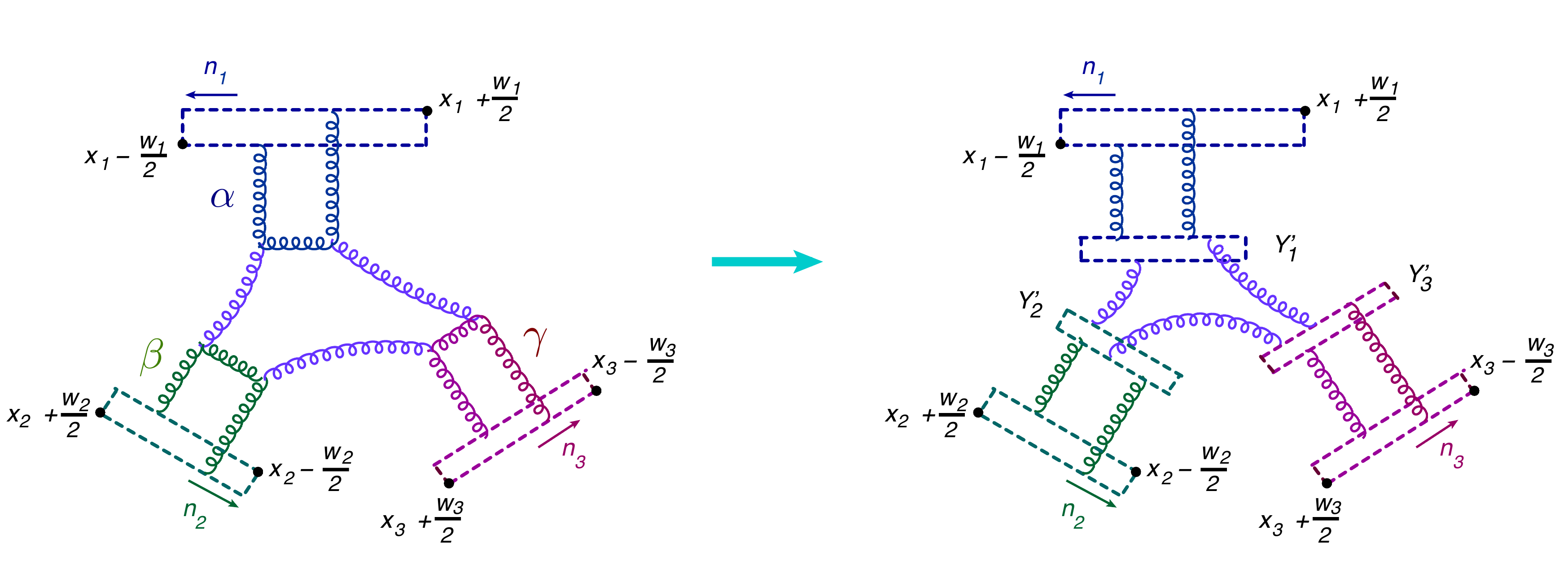}
\caption{Three BFKL loops}
\label{fig:3bfklzlo}
\end{figure}

After three evolutions (\ref{evolproj1}), (\ref{evolproj2}), and (\ref{evolproj3}) we get the correlator of three dipoles
\begin{equation}
\hspace{-1mm}
\langle {\cal U}^a(z_{1t}e_t+\tilz_1\tile,z_{2t}e_t+\tilz_2\tile) {\cal V}^a(z_{3t}e_t+\brez_3\bre,z_{4t}e_t+\brez_4\bre)
\hat{\cal W}^a(z_{5t}e_t+\chez_5\che,z_{6t}e_t+\chez_6\che)\rangle
\end{equation}
with Wilson lines parallel to $n_1, n_2,n_3$ and rapidity cutoffs $Y'_1,Y'_2,Y'_3$.  Moreover, one can think about color dipoles 
$\calu^{Y'_1}$, $\calv^{Y'_2}$, and $\calw^{Y'_3}$ as long Wilson frames with lengths $l'_1=e^{Y'_i}\sqrt{m_\perp^2\tils}$ etc., see Fig. 
\ref{fig:3bfklevols}b. We start the evolution with very long frames and evolve with lengths of these frames.
\begin{figure}[htb]
\includegraphics[width=151mm]{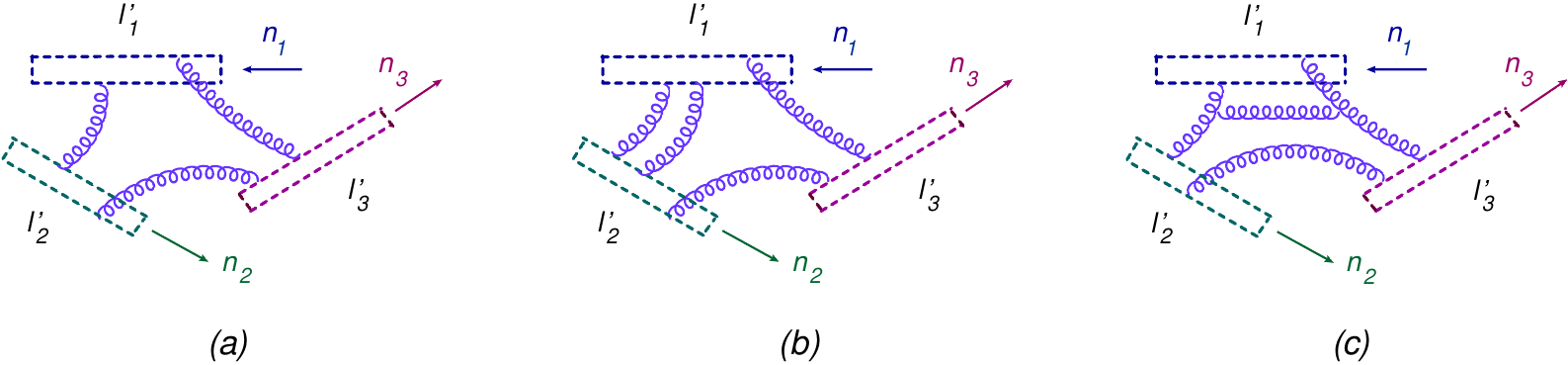}
\caption{Three Wilson frames in the leading order (``a'') and at the one-loop level (``b''+``c''+permutations).}
\label{fig:3framesLOzapas}
\end{figure}
We should stop the evolution if an extra loop in diagrams in Fig. \ref{fig:3framesLOzapas}b and Fig. \ref{fig:3framesLOzapas}c does not
bring an additional BFKL logarithm in comparison to the tree diagram in Fig. \ref{fig:3framesLOzapas}a.
This happens when the relative energy of each pair of dipoles becomes compatible with $m_\perp$, or, in the coordinate space language, 
 when the characteristic longitudinal distances are of the same order as 
the transverse ones so the BFKL approximations break down.  For typical diagrams like in Fig. \ref{fig:3framesLOzapas}b or c
the characteristic longitudinal separations are $\sim l'_il'_j s_{ij}$ so the condition  is $\sim l'_il'_j\geq {1\over m_\perp^2s_{ij}}$. 
Thus, the three BFKL evolutions in diagrams in Fig. \ref{fig:3bfklevols}b  terminate at the rapidities 
\begin{eqnarray}
&&\hspace{-1mm}
l'_1l'_2\geq {1\over m_\perp^2s_{12}}~~\Leftrightarrow~~\ln l'_1+\ln l'_2\geq -\ln {m_\perp^2s_{12}}~~\Leftrightarrow~~Y'_1+Y'_2\geq 0
\nonumber\\
&&\hspace{-1mm}
l'_1l'_3\geq {1\over m_\perp^2s_{13}}~~\Leftrightarrow~~\ln l'_1+\ln l'_3\geq -\ln {m_\perp^2s_{13}}~~\Leftrightarrow~~Y'_1+Y'_3\geq 0
\nonumber\\
&&\hspace{-1mm}
l'_2l'_3\geq {1\over m_\perp^2s_{23}}~~\Leftrightarrow~~\ln l'_2+\ln l'_3\geq -\ln {m_\perp^2s_{23}}~~\Leftrightarrow~~Y'_2+Y'_3\geq 0
\label{rapcuts}
\end{eqnarray}
where $Y'_1=\ln l'_1+\half\ln\tils m_\perp^2$,   $Y'_2=\ln l'_2+\half\ln\bres m_\perp^2$, and  $Y'_3=\ln l'_3+\half\ln\ches m_\perp^2$.
We see that the rapidity at which we stop the evolution of the $n_1$ dipole depends on the where we have 
terminated the evolutions of the second and third dipole which means that we need to integrate over
 all possible choices of ``rapidity stops'' $Y'_i$: 
\begin{eqnarray}
&&\hspace{-1mm}
\langle \calu^{Y_1}_{\rm conf}(x_{1t},-\nu_1)
\calv^{Y_2}_{\rm conf}(x_{2t},-\nu_2)\calw^{Y_3}_{\rm conf}(x_{3t},-\nu_3)\rangle^{\rm Fig. \ref{fig:3bfklzlo}}
\label{result1}\\
&&\hspace{-1mm}
=~{1\over 4}\aleph(\nu_1)\aleph(\nu_2)\aleph(\nu_3)\!\int_{-\infty}^{Y_1}\! dY'_1 \!\int_{-\infty}^{Y_2}\! dY'_2 \!\int_{-\infty}^{Y_3}\! dY'_3~\theta\big(Y'_1+Y'_2\big)
\theta\big(Y'_1+Y'_3\big)\theta\big(Y'_2+Y'_3\big)
\nonumber\\
&&\hspace{-1mm}
\times~e^{(Y_1-Y'_1)\aleph(\nu_3)}e^{(Y_2-Y'_2)\aleph(\nu_3)}e^{(Y_3-Y'_3)\aleph(\nu_3)}\langle \calu_{\rm conf}(x_{1t},-\nu_1)
\calv_{\rm conf}(x_{2t},-\nu_2)\calw_{\rm conf}(x_{3t},-\nu_3)\rangle^{\rm tree}
\nonumber
\end{eqnarray}
The weight of the integrations can be 
figured out from the evolution equations for conformal dipoles up to an overall constant  which will be determined later to be ${1\over 4}$. 
The factors $\aleph(\nu_1)\aleph(\nu_2)\aleph(\nu_3)$ can be understood by considering the lowest-order diagram with three BFKL evolutions 
shown in Fig. \ref{fig:3bfklzlo}. 
The three
integrations over $\alpha$, $\beta$, and $\gamma$ in Fig. \ref{fig:3bfklzlo} for conformal dipoles bring $\aleph(\nu_i)\!\int\! dY_i$ for each of them
so we get
\begin{eqnarray}
&&\hspace{-1mm}
\langle \calu^{Y_1}_{\rm conf}(x_{1t},-\nu_1)
\calv^{Y_2}_{\rm conf}(x_{2t},-\nu_2)\calw^{Y_3}_{\rm conf}(x_{3t},-\nu_3)\rangle
\label{resultlo}\\
&&\hspace{-1mm}
\sim~\aleph(\nu_1)\aleph(\nu_2)\aleph(\nu_3)\!\int_{-\infty}^{Y_1}\! dY'_1 \!\int_{-\infty}^{Y_2}\! dY'_2 \!\int_{-\infty}^{Y_3}\! dY'_3~\theta\big(Y'_1+Y'_2\big)
\theta\big(Y'_1+Y'_3\big)\theta\big(Y'_2+Y'_3\big)
\nonumber\\
&&\hspace{-1mm}
\times~\langle \calu_{\rm conf}(x_{1t},-\nu_1)
\calv_{\rm conf}(x_{2t},-\nu_2)\calw_{\rm conf}(x_{3t},-\nu_3)\rangle^{\rm tree}
\nonumber
\end{eqnarray}

Combining the equations (\ref{evolproj1}), (\ref{evolproj2}), (\ref{evolproj3}) and Eq. (\ref{result1}) we get
\begin{eqnarray}
&&\hspace{-1mm}
\big\langle \calf^{j_1}_{n_1}\big(x_{1t}+{w_{1t}\over 2},x_{1t}-{w_{1t}\over 2}\big)
\calf^{j_2}_{n_2}\big(x_{2t}+{w_{2t}\over 2},x_{2t}-{w_{2t}\over 2}\big)
\calf^{j_3}_{n_3}\big(x_{3t}+{w_{3t}\over 2},x_{3t}-{w_{3t}\over 2}\big)
\big\rangle
\nonumber\\
&&\hspace{-1mm}
=~-i{N_c^6\over\pi^9}\!\int_0^\infty\!
\prod_{n=1}^3\! {dl_n\over l_n}~l_n^{-\omega_n}
\int\!
\prod_{k=1}^3d\nu_k~ {\nu_k2^{-4i\nu_k}\Gamma\big({3\over 2}+i\nu_k\big)\Gamma(1-i\nu_k)
\over \Gamma\big({3\over 2}-i\nu_k\big)\Gamma(1+i\nu_k)}(w_{kt}^2)^{-\half+i\nu_k}
\nonumber\\
&&\hspace{-1mm}
\times~
{1\over 4}\aleph(\nu_1)\aleph(\nu_2)\aleph(\nu_3)
\!\int_{-\infty}^{Y_1}\! dY'_1 \!\int_{-\infty}^{Y_2}\! dY'_2 \!\int_{-\infty}^{Y_3}\! dY'_3~\theta\big(Y'_1+Y'_2\big)
\theta\big(Y'_1+Y'_3\big)\theta\big(Y'_2+Y'_3\big)
\nonumber\\
&&\hspace{-1mm}
\times~e^{(Y_1-Y'_1)\aleph(\nu_1)}e^{(Y_2-Y'_2)\aleph(\nu_3)}e^{(Y_3-Y'_3)\aleph(\nu_3)}
\langle \calu_{\rm conf}(x_{1t},-\nu_1)
\calv_{\rm conf}(x_{2t},-\nu_2)\calw_{\rm conf}(x_{3t},-\nu_3)\rangle^{\rm tree}
\nonumber\\
\label{result1a}
\end{eqnarray}
This result is the integral over $\nu_1,\nu_2$, and $\nu_3$ of the product of longitudinal and transverse integrals which we will consider in turn.

\subsection{Longitudinal integrals}
Let us  integrate the correlator (\ref{result1a}) over  over $l_i$ according to the definition ( \ref{wframe}) of the ``frame with spin j''.
Since 
$$
\int_0^\infty l_1^{-\omega_1-1}~=~\int_{-\infty}^\infty\!dY_1~e^{-\omega_1Y_1}(\tils m_\perp^2)^{\omega_1\over 2}
$$
(and similarly for $Y_2$ and $Y_3$) we get
\begin{eqnarray}
&&\hspace{-1mm}
\int_{-\infty}^\infty\!dY_1dY_2dY_3~e^{-\omega_1 Y_1-\omega_2 Y_2-\omega_3Y_3}
(\tils m_\perp^2)^{\omega_1\over 2}(\bres m_\perp^2)^{\omega_2\over 2}(\ches m_\perp^2)^{\omega_3\over 2}
\label{longint3}\\
&&\hspace{-1mm}
\times~\!\int_{-\infty}^{Y_1}\! dY'_1 \!\int_{-\infty}^{Y_2}\! dY'_2 \!\int_{-\infty}^{Y_3}\! dY'_3~\theta\big(Y'_1+Y'_2\big)
\theta\big(Y'_1+Y'_3\big)\theta\big(Y'_2+Y'_3\big)
\nonumber\\
&&\hspace{-1mm}
\times~e^{(Y_1-Y'_1)\aleph(\nu_1)}e^{(Y_2-Y'_2)\aleph(\nu_3)}e^{(Y_3-Y'_3)\aleph(\nu_3)}
\nonumber\\
&&\hspace{-1mm}  
=~{4s_{12}^{\omega_1+\omega_2-\omega_3\over 2}s_{13}^{\omega_1+\omega_3-\omega_2\over 2}
s_{23}^{\omega_2+\omega_3-\omega_1\over 2}m_\perp^{-\omega_1-\omega_2-\omega_3}
\over 
(\omega_1+\omega_2-\omega_3)(\omega_1+\omega_3-\omega_2)(\omega_2+\omega_3-\omega_1)
[\omega_1-\aleph(\nu_1)][\omega_2-\aleph(\nu_2)][\omega_3-\aleph(\nu_3)]}
\nonumber\\
&&\hspace{-1mm}
\rightarrow~{4\big({s_{12}\over x_{12t}^2}\big)^{\omega_1+\omega_2-\omega_3\over 2}\big({s_{13}\over x_{13t}^2}\big)^{\omega_1+\omega_3-\omega_2\over 2}
\big({s_{23}\over x_{23t}^2}\big)^{\omega_2+\omega_3-\omega_1\over 2}
\over 
(\omega_1+\omega_2-\omega_3)(\omega_1+\omega_3-\omega_2)(\omega_2+\omega_3-\omega_1)
[\omega_1-\aleph(\nu_1)][\omega_2-\aleph(\nu_2)][\omega_3-\aleph(\nu_3)]}
\nonumber
\end{eqnarray}
We have replaced the transverse scale $m_\perp^{-2}$  by $ x_{ijt}^2$ in accordance with general formula (\ref{defclr}) and the result (\ref{baresult}) of explicit
first-order calculation performed in Appendix \ref{sect:firstorder}. 
\footnote{This replacement is within the LLA accuracy and, moreover,   I think that at the NLO level one will get  the third line in Eq. (\ref{longint3})
similarly to the case of the two-frame correlator considered in Ref.   \cite{Balitsky:2013npa}  where the calculation 
at the NLO BFKL level reproduces the 
correct arguments required by general formula (\ref{2pointCF}).}

Also,  as discussed in Ref. \cite{Balitsky:2015oux},   the singularities $(\omega_i-\omega_j-\omega_k)^{-1}$ are of general 
nature since they arise from the fact that the correlator of three Wilson frames (\ref{corr3fs}) acquires boost invariance as $n_j\rightarrow n_k$. 
This property is discussed in Appendix \ref{sect:BKase}.

\subsection{Transverse integral}
Using Eq. (\ref{longint3}) of previous Section, one can rewrite the result (\ref{result1a}) as
\begin{eqnarray}
&&\hspace{-1mm}
\big\langle \calf^{j_1}_{n_1}\big(x_{1t}+{w_{1t}\over 2},x_{1t}-{w_{1t}\over 2}\big)
\calf^{j_2}_{n_2}\big(x_{2t}+{w_{2t}\over 2},x_{2t}-{w_{2t}\over 2}\big)
\calf^{j_3}_{n_3}\big(x_{3t}+{w_{3t}\over 2},x_{3t}-{w_{3t}\over 2}\big)
\big\rangle
\nonumber\\
&&\hspace{-1mm}
=~-i{N_c^6\over\pi^9}
\int\! d\nu_1 d\nu_2 d\nu_3~
\prod_{k=1}^3{ \nu_k2^{-4i\nu_k}\Gamma\big({3\over 2}+i\nu_k\big)\Gamma(1-i\nu_k)
\over \Gamma\big({3\over 2}-i\nu_k\big)\Gamma(1+i\nu_k)}(w_{kt}^2)^{-\half+i\nu_k}
\nonumber\\
&&\hspace{-1mm}
\times~
{\aleph(\nu_1)\aleph(\nu_2)\aleph(\nu_3)
\big({s_{12}\over x_{12t}^2}\big)^{\omega_1+\omega_2-\omega_3\over 2}
\big({s_{13}\over x_{13t}^2}\big)^{\omega_1+\omega_3-\omega_2\over 2}
\big({s_{23}\over x_{23t}^2}\big)^{\omega_2+\omega_3-\omega_1\over 2}
\over 
(\omega_1+\omega_2-\omega_3)(\omega_1+\omega_3-\omega_2)(\omega_2+\omega_3-\omega_1)
[\omega_1-\aleph(\nu_1)][\omega_2-\aleph(\nu_2)][\omega_3-\aleph(\nu_3)]}
\nonumber\\
&&\hspace{-1mm}
\times~\langle \calu^{Y'_1}_{\rm conf}(x_{1t},-\nu_1)\calv^{Y'_2}_{\rm conf}(x_{2t},-\nu_2)\calw^{Y'_3}_{\rm conf}(x_{3t},-\nu_3)\rangle^{\rm tree}
\label{result2}
\end{eqnarray}
where the correlator of three conformal dipoles in the last line should be taken in  the tree approximation.

To calculate this correlator, we rewrite conformal dipoles in terms of usual ones
\begin{eqnarray}
&&\hspace{-1mm}
\pi^6\langle \calu^{Y'_1}_{\rm conf}(x_{1t},-\nu_1)
\calv^{Y'_2}_{\rm conf}(x_{2t},-\nu_2)\calw^{Y'_3}_{\rm conf}(x_{3t},-\nu_3)\rangle^{\rm tree}
~=~\int\! {dz_{1t}d\tilz_1dz_{2t}d\tilz_2\over (z_{12t}^2+\tilz_{12}^2)^2}
\label{transintegral1}\\
&&\hspace{-1mm}
\times~
\int\! {dz_{3t}d\brez_3dz_{4t}d\brez_4\over (z_{34t}^2+\brez_{34}^2)^2}
 {dz_{5t}d\chez_5dz_{6t}d\chez_6\over (z_{56t}^2+\chez_{56}^2)^2}
 \Big({z_{12t}^2+\tilz_{12}^2\over [(x_1-z_1)_t^2+\tilz_1^2][(x_1-z_2)_t^2+\tilz_2^2]}\Big)^{\half +i\nu_1}
\nonumber\\
&&\hspace{-1mm} 
\times~ \Big({z_{34t}^2+\brez_{34}^2\over [(x_2-z_3)_t^2+\brez_3^2][(x_2-z_4)_t^2+\brez_4^2]}\Big)^{\half +i\nu_2}
\Big({z_{56t}^2+\chez_{56}^2\over [(x_3-z_5)_t^2+\chez_5^2][(x_3-z_6)_t^2+\chez_6^2]}\Big)^{\half +i\nu_3}~
\nonumber\\
&&\hspace{-1mm} 
\times~ 
\langle {\cal U}(z_{1t}e_t+\tilz_1\tile,z_{2t}e_t+\tilz_2\tile) {\cal V}(z_{3t}e_t+\brez_3\bre,z_{4t}e_t+\brez_4\bre)
\hat{\cal W}(z_{5t}e_t+\chez_5\che,z_{6t}e_t+\chez_6\che)\rangle
\nonumber
\end{eqnarray}
%
\begin{figure}[htb]
\begin{center}
\includegraphics[width=55mm]{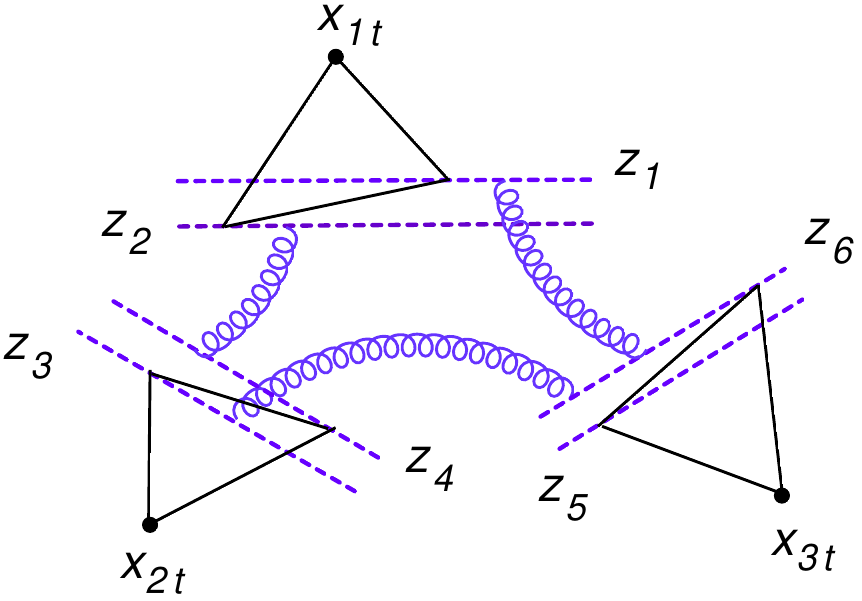}
\caption{The structure of correlator (\ref{transintegral1}). Straight lines denote ``propagators'' $(z_{ij}^2)^\lambda$ or $((x_i-z_j)^2)^{\lambda}$. }
\label{fig:3confdipoles}
\end{center}
\end{figure}
This integral is illustrated on Fig. \ref{fig:3confdipoles}.

\begin{figure}[htb]
\includegraphics[width=131mm]{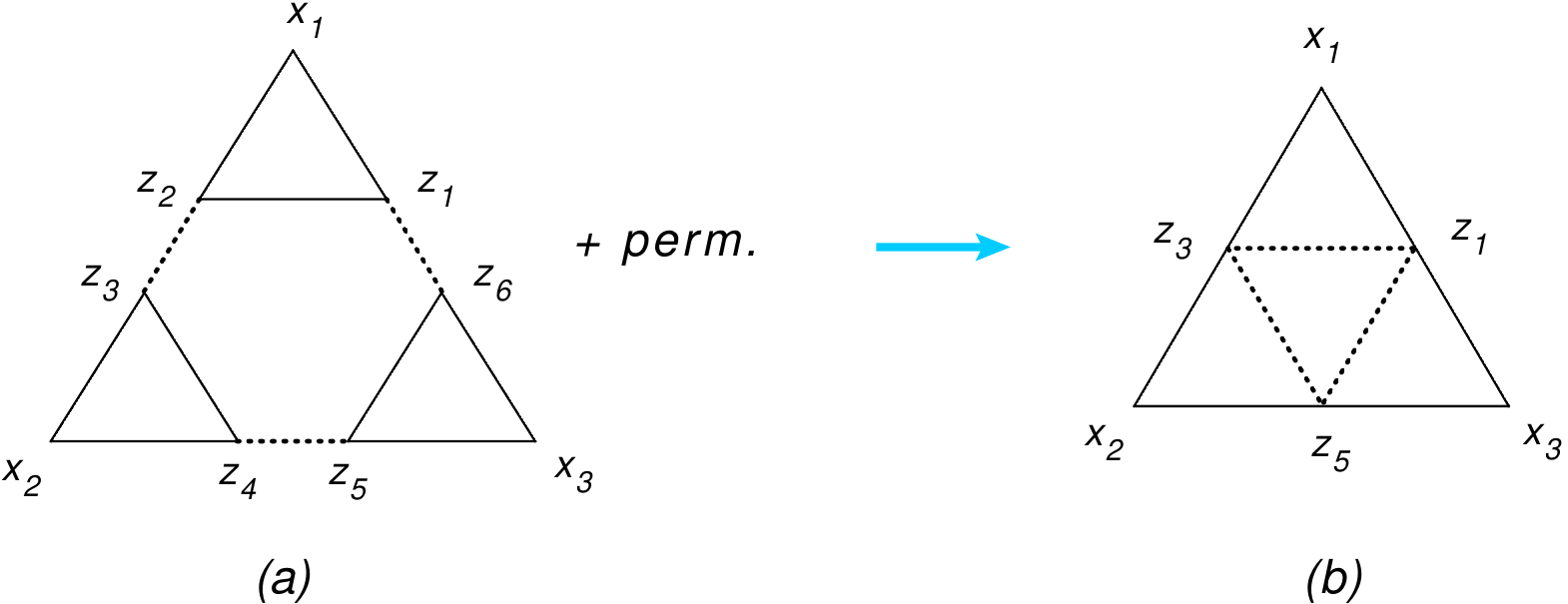}
\caption{Transverse integral. Solid lines denote usual ``propagators'' of the type $(z_i-z_j)^{2\lambda}$ and dotted lines expressions of the type
$[(z_i-z_j)_x^2+(z_i+z_j)_y^2]^\lambda$.}
\label{fig:trint}
\end{figure}

Using the leading-order correlator
\begin{equation}
\hspace{-1mm}
\langle[\infty n_1+x,-\infty n_1+x][\infty n_2+y,-\infty n_2+y]\rangle~=~i\alpha_st^a\otimes t^a\ln \Big[(x-y)^2-2{(x-y)_{n_1}(x-y)_{n_2}\over n_1\cdot n_2}\Big]
\end{equation}
one easily obtains
\begin{eqnarray}
&&\hspace{-1mm}
\langle {U}(z_{1t}e_t+\tilz_1\tile) {V}(z_{3t}e_t+\brez_3\bre)\rangle~=~i\alpha_st^a\otimes t^a\ln [z_{13t}^2+(\tilz_1+\brez_3)^2],
\nonumber\\
&&\hspace{-1mm} 
\langle {U}(z_{1t}e_t+\tilz_1\tile) {W}(z_{5t}e_t+\chez_5\che)\rangle~=~i\alpha_st^a\otimes t^a\ln [z_{15t}^2+(\tilz_1+\chez_5)^2],
\nonumber\\
&&\hspace{-1mm} 
\langle {V}(z_{3t}e_t+\brez_3\bre) {W}(z_{5t}e_t+\chez_5\che)\rangle~=~i\alpha_st^a\otimes t^a\ln [z_{35t}^2+(\brez_3+\chez_5)^2]
\label{treecorrs}
\end{eqnarray}
and therefore the tree-level correlator of three color dipoles reads
\begin{eqnarray}
&&\hspace{-1mm}
\langle {\cal U}^a(z_{1t}e_t+\tilz_1\tile,z_{2t}e_t+\tilz_2\tile) {\cal V}^a(z_{3t}e_t+\brez_3\bre,z_{4t}e_t+\brez_4\bre)
\hat{\cal W}^a(z_{5t}e_t+\chez_5\che,z_{6t}e_t+\chez_6\che)\rangle
\nonumber\\
&&\hspace{-1mm}
=~-i\alpha_s^3{N_c^2-1\over N_c^3}\ln{[z_{13t}^2+(\tilz_1+\brez_3)^2][z_{24t}^2+(\tilz_2+\brez_4)^2]\over [z_{14t}^2+(\tilz_1+\brez_4)^2][z_{23t}^2+(\tilz_2+\brez_3)^2]}
\nonumber\\
&&\hspace{-1mm}
\times~
\ln{[z_{15t}^2+(\tilz_1+\chez_5)^2][z_{26t}^2+(\tilz_2+\chez_6)^2]\over [z_{16t}^2+(\tilz_1+\chez_6)^2][z_{25t}^2+(\tilz_2+\chez_5)^2]}
\ln{[z_{35t}^2+(\brez_3+\chez_5)^2][z_{46t}^2+(\brez_4+\chez_6)^2]\over [z_{36t}^2+(\brez_3+\chez_6)^2][z_{45t}^2+(\brez_4+\chez_5)^2]}
\end{eqnarray}
One obtains
\begin{eqnarray}
&&\hspace{-1mm}
\langle \calu_{\rm conf}(x_{1t},-\nu_1)
\calv_{\rm conf}(x_{2t},-\nu_2)\calw_{\rm conf}(x_{3t},-\nu_3)\rangle^{\rm tree}~
\nonumber\\
&&\hspace{11mm}=~
-i\alpha_s^3{N_c^2-1\over N_c^3\pi^6}I(x_{1t},x_{2t},x_{3t};\nu_1,\nu_2,\nu_3)
\nonumber\\
\label{transintegral2}
\end{eqnarray}
where
\begin{eqnarray}
&&\hspace{-1mm}
I(x_{1t},x_{2t},x_{3t};\nu_1,\nu_2,\nu_3)
~
=~\!\int\! {dz_{1t}d\tilz_1dz_{2t}d\tilz_2\over (z_{12t}^2+\tilz_{12}^2)^2}
{dz_{3t}d\brez_3dz_{4t}d\brez_4\over (z_{34t}^2+\brez_{34}^2)^2}
 {dz_{5t}d\chez_5dz_{6t}d\chez_6\over (z_{56t}^2+\chez_{56}^2)^2}
\nonumber\\
&&\hspace{-1mm} 
\times~ \Big({z_{12t}^2+\tilz_{12}^2\over [(x_1-z_1)_t^2+\tilz_1^2][(x_1-z_2)_t^2+\tilz_2^2]}\Big)^{\half +i\nu_1}
 \Big({z_{34t}^2+\brez_{34}^2\over [(x_2-z_3)_t^2+\brez_3^2][(x_2-z_4)_t^2+\brez_4^2]}\Big)^{\half +i\nu_2}
\nonumber\\
&&\hspace{-1mm} 
\times~ \Big({z_{56t}^2+\chez_{56}^2\over [(x_3-z_5)_t^2+\chez_5^2][(x_3-z_6)_t^2+\chez_6^2]}\Big)^{\half +i\nu_3}
\ln{[z_{13t}^2+(\tilz_1+\brez_3)^2][z_{24t}^2+(\tilz_2+\brez_4)^2]\over [z_{14t}^2+(\tilz_1+\brez_4)^2][z_{23t}^2+(\tilz_2+\brez_3)^2]}
\nonumber\\
&&\hspace{-1mm}
\times~
\ln{[z_{15t}^2+(\tilz_1+\chez_5)^2][z_{26t}^2+(\tilz_2+\chez_6)^2]\over [z_{16t}^2+(\tilz_1+\chez_6)^2][z_{25t}^2+(\tilz_2+\chez_5)^2]}
\ln{[z_{35t}^2+(\brez_3+\chez_5)^2][z_{46t}^2+(\brez_4+\chez_6)^2]\over [z_{36t}^2+(\brez_3+\chez_6)^2][z_{45t}^2+(\brez_4+\chez_5)^2]}
\label{transintegral3}
\end{eqnarray}
The structure of the integrations in this equation is the following: each conformal dipole evolves in its own ``transverse plane'' and
the obtained dipoles interact by logarithmical correlators (\ref{treecorrs}). Fortunately, 
 the integral (\ref{transintegral3}) coincides with an usual two-dimensional integral in the (formal) $(x,y)$ plane
\begin{eqnarray}
&&\hspace{-1mm}
I(x_1,x_2,x_3;\nu_1,\nu_2,\nu_3)
~
=~\!\int\! {d^2z_1d^2z_2\over z_{12}^4}{d^2z_3d^2z_4\over z_{34}^4}{d^2z_5d^2z_6\over z_{56}^4}
\nonumber\\
&&\hspace{-1mm} 
\times~ \Big({z_{12}^2\over (x_1-z_1)^2(x_1-z_2)^2}\Big)^{\half +i\nu_1}
 \Big({z_{34}^2\over (x_2-z_3)^2(x_2-z_4)^2]}\Big)^{\half +i\nu_2}
  \Big({z_{56}^2\over (x_3-z_5)^2(x_3-z_6)^2]}\Big)^{\half +i\nu_2}
\nonumber\\
&&\hspace{-1mm} 
\times~
\ln{[z_{13x}^2+(z_{1y}+z_{3y})^2][z_{24x}^2+(z_{2y}+z_{4y})^2]\over [z_{14x}^2+(z_{1y}+z_{4y})^2][z_{23x}^2+(z_{2y}+z_{3y})^2]}
\ln{[z_{15x}^2+(z_{1y}+z_{5y})^2][z_{26x}^2+(z_{2y}+z_{6y})^2]\over [z_{16x}^2+(z_{1y}+z_{6y})^2][z_{25x}^2+(z_{2y}+z_{5y})^2]}
\nonumber\\
&&\hspace{-1mm}
\times~
\ln{[z_{35x}^2+(z_{3y}+z_{5y})^2][z_{46x}^2+(z_{4y}+z_{6y})^2]\over [z_{36x}^2+(z_{3y}+z_{6y})^2][z_{45x}^2+(z_{4y}+z_{5y})^2]}
\label{transintegral4}
\end{eqnarray}
with $x_{1y}=x_{2y}=x_{3y}=0$, see Fig.\ref{fig:trint}a.

Next, we use formula
$$
\int\! {d^2 z_2\over z_{12}^4}\Big({z_{12}^2\over (x_1-z_1)^2(x_1-z_2)^2}\Big)^{\half+i\nu_1}\ln z_{23}^2
~=~{\pi [(x_1-z_1)^2]^{-{3\over 2}-i\nu_1}\over \big(\half-i\nu_1\big)^2}
\Big[\Big({(x_1-z_3)^2\over z_{13}^2}\Big)^{\half-i\nu_1}-1\Big]
$$
and rewrite it by changing sign of $z_{3y}$ as
\begin{eqnarray}
&&\hspace{-1mm}
\int\! {d^2 z_2\over z_{12}^4}\Big({z_{12}^2\over (x_1-z_1)^2(x_1-z_2)^2}\Big)^{\half+i\nu}\ln [z_{23x}^2+(z_{2y}+z_{3y})^2]
\nonumber\\
&&\hspace{-1mm}
=~{\pi [(x_1-z_1)^2]^{-{3\over 2}-i\nu_1}\over \big(\half-i\nu_1\big)^2}
\Big[\Big({(x_1-z_3)^2\over z_{13x}^2+(z_{1y}+z_{3y})^2}\Big)^{\half-i\nu_1}-1\Big]
\end{eqnarray}
Using this integral and similar integrals for integration over $z_4$ and $z_6$ one gets after some algebra (see Fig.\ref{fig:trint}b)
\begin{eqnarray}
&&\hspace{-3mm}
I(x_1,x_2,x_3;\nu_1,\nu_2,\nu_3)
\\
&&\hspace{-3mm}
=~-{8\pi^3\over\big(\half-i\nu_1\big)^2\big(\half-i\nu_2\big)^2\big(\half-i\nu_3\big)^2}\!\int\! {d^2z_1\over (x_1-z_1)^4}{d^2z_3\over (x_2-z_3)^4}{d^2z_5\over (x_3-z_5)^4}
\nonumber\\
&&\hspace{-3mm}
\times~ \Big[{z_{13x}^2+(z_{1y}+z_{3y})^2\over (x_1-z_1)^2(x_1-z_3)^2}\Big]^{-\half+i\nu_1}
\Big[{z_{35x}^2+(z_{3y}+z_{5y})^2\over (x_2-z_3)^2(x_2-z_5)^2}\Big]^{-\half+i\nu_2}
\Big[{z_{15x}^2+(z_{1y}+z_{5y})^2\over (x_3-z_1)^2(x_3-z_5)^2}\Big]^{-\half+i\nu_3}
\nonumber
\end{eqnarray}
To calculate this integral, we can take $x_1=0$ and perform the inversion $x_i\rightarrow {x_i\over x^2}$ to obtain
\begin{eqnarray}
&&\hspace{-1mm}
I^{\rm inv}(x_1,x_2,x_3;\nu_1,\nu_2,\nu_3)
\nonumber\\
&&\hspace{-1mm}
=~-{8\pi^3(x_2^2)^{1+2i\nu_2}(x_3^2)^{1+2i\nu_3}\over\big(\half-i\nu_1\big)^2\big(\half-i\nu_2\big)^2\big(\half-i\nu_3\big)^2}
\!\int\! d^2z_1{d^2z_3\over (x_2-z_3)^4}{d^2z_5\over (x_3-z_5)^4}
\nonumber\\
&&\hspace{-1mm}
\times~[z_{13x}^2+(z_{1y}+z_{3y})^2]^{-\half+i\nu_1}
\Big[{z_{35x}^2+(z_{3y}+z_{5y})^2\over (x_2-z_3)^2(x_2-z_5)^2}\Big]^{-\half+i\nu_2}
\Big[{z_{15x}^2+(z_{1y}+z_{5y})^2\over (x_3-z_1)^2(x_3-z_5)^2}\Big]^{-\half+i\nu_3}
\nonumber\\
&&\hspace{-1mm}
=~-{8\pi^6(x_2^2)^{1+2i\nu_2}(x_3^2)^{1+2i\nu_3}(x_{23}^2)^{-\half+i(\nu_1-\nu_2-\nu_3)}\over\big(\half-i\nu_1\big)^2\big(\half-i\nu_2\big)^2\big(\half-i\nu_3\big)^2}
\Lambda(\nu_1,\nu_2,\nu_3)
\label{invotvet}
\end{eqnarray}
where
\begin{eqnarray}
&&\hspace{-1mm}
\Lambda(\nu_1,\nu_2,\nu_3)~\equiv~(x_{23}^2)^{\half-i(\nu_1-\nu_2-\nu_3)}
{1\over\pi^3}\!\int\! d^2z_1{d^2z_3\over (x_{23}-z_3)^4}{d^2z_5\over z_5^4}
\nonumber\\
&&\hspace{-1mm}
\times~[z_{13x}^2+(z_{1y}+z_{3y})^2]^{-\half+i\nu_1}
\Big[{z_{35}^2\over (x_{23}-z_3)^2(x_{23}-z_5)^2}\Big]^{-\half+i\nu_2}
\Big[{z_{15}^2\over z_1^2z_5^2}\Big]^{-\half+i\nu_3}
\nonumber\\
&&\hspace{-1mm}
=~
{1\over\pi^3}\!\int\! d^2z_1{d^2z_3\over (1-z_3)^4}{d^2z_5\over z_5^4}
\nonumber\\
&&\hspace{-1mm}
\times~[z_{13x}^2+(z_{1y}+z_{3y})^2]^{-\half+i\nu_1}
\Big[{z_{35}^2\over (1-z_3)^2(1-z_5)^2}\Big]^{-\half+i\nu_2}
\Big[{z_{15}^2\over z_1^2z_5^2}\Big]^{-\half+i\nu_3}
\label{lambda}
\end{eqnarray}
where ``1'' in the denominators stands for the vector (1,0).
 This integral resembles the integral for function $\Omega$ defining three-pomeron vertex \cite{Korchemsky:1997fy}, only with 
``modified propagator'' $[z_{13x}^2+(z_{1y}+z_{3y})^2]^{-\half+i\nu_1}$ instead of usual 
$[z_{13x}^2+z_{13y}^2]^{-\half+i\nu_1}$ in Ref. \cite{Korchemsky:1997fy}.
The function $\Lambda(\nu_1,\nu_2,\nu_3)$ can be represented as four-fold Mellin-Barnes integral, 
see Eq. (\ref{6.40}) below and Eq. (\ref{bLaresult}) in Appendix \ref{sect:lambda}.

Performing inversion of Eq. (\ref{invotvet}) we get
\begin{eqnarray}
&&\hspace{-1mm}
I(x_{1t},x_{2t},x_{3t};\nu_1,\nu_2,\nu_3)~
=~-8\pi^6{\Lambda(\nu_1,\nu_2,\nu_3)
\over \big(\half-i\nu_1\big)^2\big(\half-i\nu_2\big)^2\big(\half-i\nu_3\big)^2}
\nonumber\\
&&\hspace{-1mm}
\times~
(x_{12t}^2)^{-\half-i(\nu_1+\nu_2-\nu_3)}(x_{13t}^2)^{-\half-i(\nu_1+\nu_3-\nu_2)}(x_{23t}^2)^{-\half-i(\nu_2+\nu_3-\nu_1)}
\label{transotvet}
\end{eqnarray}
so that
\begin{eqnarray}
&&\hspace{-1mm}
\langle \calu_{\rm conf}(x_{1t},-\nu_1)
\calv_{\rm conf}(x_{2t},-\nu_2)\calw_{\rm conf}(x_{3t},-\nu_3)\rangle^{\rm tree}~
\label{transresult}\\
&&\hspace{-1mm}
=~
8i\alpha_s^3{N_c^2-1\over N_c^3}\Lambda(\nu_1,\nu_2,\nu_3){(x_{12t}^2)^{-\half-i(\nu_1+\nu_2-\nu_3)}(x_{13t}^2)^{-\half-i(\nu_1+\nu_3-\nu_2)}(x_{23t}^2)^{-\half-i(\nu_2+\nu_3-\nu_1)}
\over \big(\half-i\nu_1\big)^2\big(\half-i\nu_2\big)^2\big(\half-i\nu_3\big)^2}
\nonumber\end{eqnarray}

\subsection{The result}
Substituting the transverse integral (\ref{transresult})  into Eq. (\ref{result2}) we get
\begin{eqnarray}
&&\hspace{-1mm}
\big\langle \calf^{j_1}_{n_1}\big(x_{1t}+{w_{1t}\over 2},x_{1t}-{w_{1t}\over 2}\big)
\calf^{j_2}_{n_2}\big(x_{2t}+{w_{2t}\over 2},x_{2t}-{w_{2t}\over 2}\big)
\calf^{j_3}_{n_3}\big(x_{3t}+{w_{3t}\over 2},x_{3t}-{w_{3t}\over 2}\big)
\big\rangle
\nonumber\\
&&\hspace{-1mm}
=~{64g^6\over \pi^6}(N_c^2-1)
\int_{-\infty}^\infty\! d\nu_1 d\nu_2 d\nu_3~
\prod_{k=1}^3{ \nu_k2^{-4i\nu_k}\Gamma\big({3\over 2}+i\nu_k\big)\Gamma(1-i\nu_k)
\over \Gamma\big({3\over 2}-i\nu_k\big)\Gamma(1+i\nu_k)}(w_{kt}^2)^{-\half+i\nu_k}
\nonumber\\
&&\hspace{-1mm}
\times~
{8\aleph(\nu_1)\aleph(\nu_2)\aleph(\nu_3)
\big({s_{12}\over x_{12t}^2}\big)^{\omega_1+\omega_2-\omega_3\over 2}\big({s_{13}\over x_{13t}^2}\big)^{\omega_1+\omega_3-\omega_2\over 2}
\big({s_{23}\over x_{23t}^2}\big)^{\omega_2+\omega_3-\omega_1\over 2}
\over 
(\omega_1+\omega_2-\omega_3)(\omega_1+\omega_3-\omega_2)(\omega_2+\omega_3-\omega_1)
[\omega_1-\aleph(\nu_1)][\omega_2-\aleph(\nu_2)][\omega_3-\aleph(\nu_3)]}
\nonumber\\
&&\hspace{-1mm}
\times~{(x_{12t}^2)^{-\half-i(\nu_1+\nu_2-\nu_3)}(x_{13t}^2)^{-\half-i(\nu_1+\nu_3-\nu_2)}(x_{23t}^2)^{-\half-i(\nu_2+\nu_3-\nu_1)}
\over \big(\half-i\nu_1\big)^2\big(\half-i\nu_2\big)^2\big(\half-i\nu_3\big)^2}\Lambda(\nu_1,\nu_2,\nu_3)
\label{result3}
\end{eqnarray}
To estimate this integral at small $\omega$'s it is convenient to rewrite it in variables $\gamma_i=2i\nu_i-1$. Defining
$\taleph(\gamma_i)\equiv\aleph(\nu_i)$ and $\tiLa(\gamma_i)\equiv\Lambda(\nu_i)$ we get
\begin{eqnarray}
&&\hspace{-1mm}
\big\langle \calf^{j_1}_{n_1}\big(x_{1t}+{w_{1t}\over 2},x_{1t}-{w_{1t}\over 2}\big)
\calf^{j_2}_{n_2}\big(x_{2t}+{w_{2t}\over 2},x_{2t}-{w_{2t}\over 2}\big)
\calf^{j_3}_{n_3}\big(x_{3t}+{w_{3t}\over 2},x_{3t}-{w_{3t}\over 2}\big)
\big\rangle
\nonumber\\
&&\hspace{-1mm}
=~-{8g^6(N_c^2-1)\over \pi^6x_{12t}^2x_{13t}^2x_{23t}^2}
{\big({s_{12}\over x_{12t}^2}\big)^{\omega_1+\omega_2-\omega_3\over 2}
\big({s_{13}\over x_{13t}^2}\big)^{\omega_1+\omega_3-\omega_2\over 2}
\big({s_{23}\over x_{23t}^2}\big)^{\omega_2+\omega_3-\omega_1\over 2}
\over 
(\omega_1+\omega_2-\omega_3)(\omega_1+\omega_3-\omega_2)(\omega_2+\omega_3-\omega_1)}
\nonumber\\
&&\hspace{-1mm}
\times~\int_{-1-i\infty}^{-1+i\infty}\! d\gamma_1 d\gamma_2 d\gamma_3~
\prod_{k=1}^3{(1+ \gamma_k)2^{-2\gamma_k}\Gamma\big(2+{\gamma_k\over 2}\big)\Gamma\big(\half-{\gamma_k\over 2}\big)
\over \Gamma\big(1-{\gamma_k\over 2}\big)\Gamma\big({3\over 2}+{\gamma_k\over 2}\big)\gamma_k^2}
\nonumber\\
&&\hspace{-1mm}
\times~
{\taleph(\gamma_1)\taleph(\gamma_2)\taleph(\gamma_3)\tiLa(\gamma_1,\gamma_2,\gamma_3)
\over 
[\omega_1-\taleph(\gamma_1)][\omega_2-\taleph(\gamma_2)][\omega_3-\taleph(\gamma_3)]}
\Big({w_{1t}^2x_{23t}^2\over x_{12t}^2x_{13t}^2}\Big)^{\gamma_1\over 2}\Big({w_{2t}^2x_{13t}^2\over x_{12t}^2x_{23t}^2}\Big)^{\gamma_2\over 2}
\Big({w_{3t}^2x_{12t}^2\over x_{13t}^2x_{23t}^2}\Big)^{\gamma_3\over 2}
\label{result4}
\end{eqnarray}
where  contours over real $\nu_i$ transform to the contours parallel to imaginary axis since $\gamma_i=2i\nu_i-1$.
The function $\tiLa(\gamma_i)$ (defined by Eq. (\ref{lambda})) is represented in Appendix  \ref{sect:lambda} 
as 
\begin{eqnarray}
&&\hspace{-1mm}
\bLa(\epsilon_1,\epsilon_2,\epsilon_3)~
=~-{\sin^2\pi\epsilon_2\sin^2\pi\epsilon_3\over \pi^3\sin\pi(\epsilon_2+\epsilon_3)\Gamma^2(\epsilon_1)}I(\epsilon_1,\epsilon_2,\epsilon_3)
\label{6.40}
\end{eqnarray}
where $\bLa(\epsilon_i)\equiv\tiLa(\gamma_i)$, $\epsilon_i\equiv -{\gamma_i\over 2}$ and 
\begin{eqnarray}
&&\hspace{-1mm}
I(\epsilon_1,\epsilon_2,\epsilon_3)~
\label{bigI}\\
&&\hspace{-1mm}
=~\!\int_C \! {ds_1ds_2ds_3ds_4\over (2\pi i)^4}~
\Gamma(\epsilon_1-s_1-s_2)\Gamma(s_1)\Gamma(s_2)\Gamma(\epsilon_1-s_3-s_4)\Gamma(s_3)\Gamma(s_4)
\nonumber\\
&&\hspace{-1mm}
\times~{\Gamma(\epsilon_2-1-s_1)\Gamma(s_1+1)
\over\Gamma(\epsilon_2)}~
{\Gamma(1+\epsilon_3-s_3)\Gamma(-1+s_3)
\over\Gamma(\epsilon_3)}~
\nonumber\\
&&\hspace{-1mm}
\times~
{\Gamma(\epsilon_2-1-s_4)\Gamma(1-\epsilon_1-\epsilon_2+s_3+s_4)\over\Gamma(-\epsilon_1+s_3)}
\nonumber\\
&&\hspace{-1mm}
\times~{\Gamma(1+\epsilon_3-s_2)\Gamma(1-\epsilon_1-\epsilon_3+s_1+s_2)\over\Gamma(2-\epsilon_1+s_1)}
{\Gamma(\epsilon_2-s_1)\Gamma(\epsilon_3-s_3)\over \Gamma(\epsilon_2+\epsilon_3-s_1-s_3)}
{\Gamma(\epsilon_2-s_2)\Gamma(\epsilon_3-s_4)\over \Gamma(\epsilon_2+\epsilon_3-s_2-s_4)}
\nonumber
\end{eqnarray}
is a four-fold Mellin-Barnes integral
with the contour C specified in the Appendix.

In the limit $w_{it}^2\rightarrow 0$ the contours of integration over $\gamma_i$  can be moved to the right so  the integrals 
are determined by the residues lying on the real axis to the right of point $\gamma_i=-1$.   
As demonstrated in Appendix \ref{sect:lambda}, the function $\Lambda$ is regular at small $\gamma_i$ 
\begin{equation}
\tiLa(\gamma_i)~=~1+O(\gamma_i)
\label{tila}
\end{equation}
so the analytic structure of the integral (\ref{result4}) is determined by the poles in $\Gamma$-functions   and in $[\omega-\aleph(\gamma)]^{-1}$ denominators.
The leftmost of such poles are located at $\gamma_i=0$ and at $\gamma_i^\ast(\omega_i,g^2)$ = roots of equation
 (\ref{gammast}) which simplifies to 
\begin{equation}
\omega_i~=~\taleph(\gamma_i,g^2)~=~4g^2\big[2\psi(1)-\psi\big(-{\gamma_i\over 2}\big)-\psi\big(1+{\gamma_i\over 2}\big)\big]
\label{simproot}
\end{equation}
 in the leading log approximation, see the discussion in Sect. \ref{general}.
 
First, we consider poles at $\gamma_i^\ast(\omega_i,g^2)$. Taking residues in these poles we obtain 
\begin{eqnarray}
&&\hspace{-1mm}
\big\langle \calf^{j_1}_{n_1}\big(x_{1t}+{w_{1t}\over 2},x_{1t}-{w_{1t}\over 2}\big)
\calf^{j_2}_{n_2}\big(x_{2t}+{w_{2t}\over 2},x_{2t}-{w_{2t}\over 2}\big)
\calf^{j_3}_{n_3}\big(x_{3t}+{w_{3t}\over 2},x_{3t}-{w_{3t}\over 2}\big)
\big\rangle
\nonumber\\
&&\hspace{-1mm}
=~i{64g^6(N_c^2-1)\over \pi^3x_{12t}^2x_{13t}^2x_{23t}^2}
{\omega_1\omega_2\omega_3\big({s_{12}\over x_{12t}^2}\big)^{\omega_1+\omega_2-\omega_3\over 2}
\big({s_{13}\over x_{13t}^2}\big)^{\omega_1+\omega_3-\omega_2\over 2}
\big({s_{23}\over x_{23t}^2}\big)^{\omega_2+\omega_3-\omega_1\over 2}
\over 
(\omega_1+\omega_2-\omega_3)(\omega_1+\omega_3-\omega_2)(\omega_2+\omega_3-\omega_1)}
\tiLa(\gamma^\ast_1,\gamma^\ast_2,\gamma^\ast_3)
\nonumber\\
&&\hspace{-1mm}
\times~
\prod_{k=1}^3{(1+ \gamma_k^\ast)2^{-2\gamma^\ast_k}\Gamma\big(2+{\gamma_k^\ast\over 2}\big)\Gamma\big(\half-{\gamma_k^\ast\over 2}\big)
\over \Gamma\big(1-{\gamma_k^\ast\over 2}\big)
\Gamma\big({3\over 2}+{\gamma_k^\ast\over 2}\big){\gamma_k^\ast}^2\talef'(\gamma_k^\ast)}
\Big({w_{1t}^2x_{23t}^2\over x_{12t}^2x_{13t}^2}\Big)^{\gamma^\ast_1\over 2}\
\Big({w_{2t}^2x_{13t}^2\over x_{12t}^2x_{23t}^2}\Big)^{\gamma^\ast_2\over 2}
\Big({w_{3t}^2x_{12t}^2\over x_{13t}^2x_{23t}^2}\Big)^{\gamma^\ast_3\over 2}
\label{result}
\end{eqnarray}

Let us present this result in the ${g^2\over\omega}\ll 1$ limit. In this limit $\taleph(\gamma)\simeq -{8g^2\over\gamma}$ 
(see Eq. (\ref{gammast})) so
\begin{eqnarray}
&&\hspace{-1mm}
\big\langle \calf^{j_1}_{n_1}\big(x_{1t}+{w_{1t}\over 2},x_{1t}-{w_{1t}\over 2}\big)
\calf^{j_2}_{n_2}\big(x_{2t}+{w_{2t}\over 2},x_{2t}-{w_{2t}\over 2}\big)
\calf^{j_3}_{n_3}\big(x_{3t}+{w_{3t}\over 2},x_{3t}-{w_{3t}\over 2}\big)
\big\rangle
\nonumber\\
&&\hspace{-1mm}
=~{i(N_c^2-1)\over \pi^3x_{12t}^2x_{13t}^2x_{23t}^2}
{\omega_1\omega_2\omega_3\big({s_{12}\over x_{12t}^2}\big)^{\omega_1+\omega_2-\omega_3\over 2}
\big({s_{13}\over x_{13t}^2}\big)^{\omega_1+\omega_3-\omega_2\over 2}
\big({s_{23}\over x_{23t}^2}\big)^{\omega_2+\omega_3-\omega_1\over 2}
\over 
(\omega_1+\omega_2-\omega_3)(\omega_1+\omega_3-\omega_2)(\omega_2+\omega_3-\omega_1)}
\nonumber\\
&&\hspace{-1mm}
\times~
\Big({w_{1t}^2x_{23t}^2\over x_{12t}^2x_{13t}^2}\Big)^{\gamma^\ast_1\over 2}\Big({w_{2t}^2x_{13t}^2\over x_{12t}^2x_{23t}^2}\Big)^{\gamma^\ast_2\over 2}
\Big({w_{3t}^2x_{12t}^2\over x_{13t}^2x_{23t}^2}\Big)^{\gamma^\ast_3\over 2}\Big[1~+~O\big({g^2\over\omega}\big)\Big]
\label{resulta}
\end{eqnarray}
where $\gamma^\ast_i=-{8g^2\over \pi\omega_i}$.
This result should agree with the first perturbative diagrams calculated in Appendix \ref{sect:firstorder}. 
Indeed, there are poles in the integral (\ref{result4}) at $\gamma_i=0$ which give
\begin{eqnarray}
&&\hspace{-1mm}
-i{(N_c^2-1)\over \pi^3x_{12t}^2x_{13t}^2x_{23t}^2}
{\omega_1\omega_2\omega_3\big({s_{12}\over x_{12t}^2}\big)^{\omega_1+\omega_2-\omega_3\over 2}
\big({s_{13}\over x_{13t}^2}\big)^{\omega_1+\omega_3-\omega_2\over 2}
\big({s_{23}\over x_{23t}^2}\big)^{\omega_2+\omega_3-\omega_1\over 2}
\over 
(\omega_1+\omega_2-\omega_3)(\omega_1+\omega_3-\omega_2)(\omega_2+\omega_3-\omega_1)}
\label{polatzero}
\end{eqnarray}
(recall that $\tiLa(0,0,0)=1$).
Similarly to the case of correlator of two light-rays, this term should cancel with the lowest-order diagrams
shown in Fig. \ref{fig:3framesLOzapas}a. At the tree level the limit $w_{it}\rightarrow 0$ is trivial so one gets the diagrams in Fig. \ref{triplo}
which yield Eq. (\ref{treesult}). We see that the result (\ref{polatzero}) cancels with that of Eq. (\ref{treesult}) 
which justifies our choice of constant ${1\over 4}$ in Eq. (\ref{result1}). 

\vspace{-0mm}
\begin{figure}[htb]
\vspace{-0mm}
\hspace{-0mm}
\includegraphics[width=150mm]{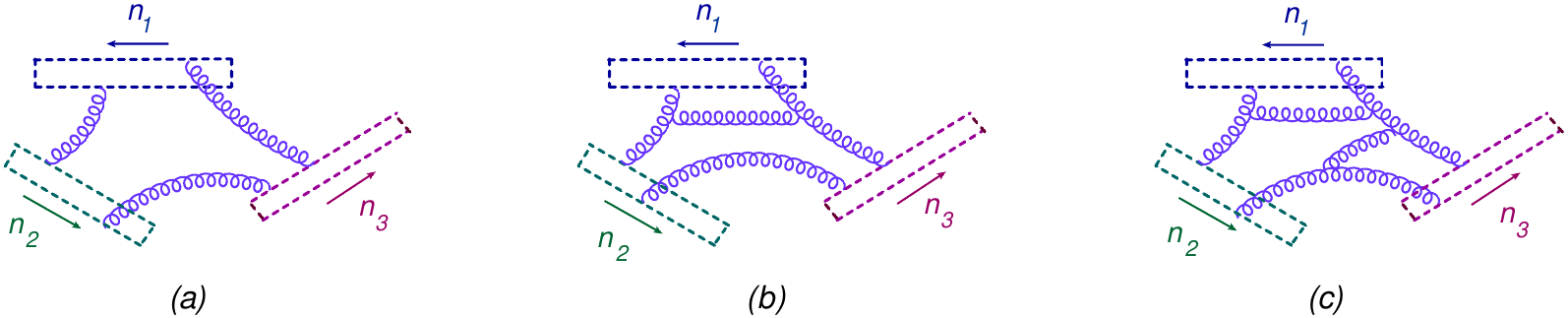}
\vspace{-0mm}
\caption{Typical diagrams for correlator of three Wilson frames in the leading order (a), at the one-loop level (b)  and at the two-loop level (c) }
\label{fig:Figura}
\end{figure}

There are also ``mixed'' poles at $\gamma_i=0,\gamma^\ast_j,\gamma^\ast_k$ and  $\gamma_i=0,\gamma_j=0,\gamma^\ast_k$.
They should cancel with the contribution of diagrams of the type shown in Fig. \ref{fig:Figura} b and c (recall that our three-dipole
 approximations are correct starting from the $g^6$ diagrams of the Fig. \ref{fig:3bfklzlo}a.)

Thus, the result for all diagrams  (Fig.  \ref{fig:3bfklevols}   + Fig. \ref{fig:Figura}) is given by Eq. (\ref{result}) which translates to the structure 
constant $C(\omega_i,g^2)$ given by Eq. (\ref{Cgeneral}) 
\begin{eqnarray}
&&\hspace{-1mm}
F(\gamma^\ast_1,\gamma^\ast_2,\gamma^\ast_3)~=~64g^6
\tiLa(\gamma^\ast_1,\gamma^\ast_2,\gamma^\ast_3)
\prod_{k=1}^3{(1+ \gamma_k^\ast)2^{-2\gamma^\ast_k}\Gamma\big(2+{\gamma_k^\ast\over 2}\big)\Gamma\big(\half-{\gamma_k^\ast\over 2}\big)
\over \Gamma\big(1-{\gamma_k^\ast\over 2}\big)
\Gamma\big({3\over 2}+{\gamma_k^\ast\over 2}\big){\gamma_k^\ast}^2\talef'(\gamma_k^\ast)}
\label{Fotvet}
\end{eqnarray}
where $\gamma^\ast$ is the solution  of Eq. (\ref{simproot}) and $\tiLa(\gamma_i)=\bLa(-2\epsilon_i)$ is
given by Eq. (\ref{bLaresult}).  It is easy to see that at small $\gamma^\ast_i$ ($\Leftrightarrow ~g^2\ll\omega_i\ll 1$)
\begin{equation}
F(\gamma^\ast_1,\gamma^\ast_2,\gamma^\ast_3)~=~1+O(\gamma_i^\ast)~=1+O\big({g^2\over\omega_i}\big)
\end{equation}

\section{Conclusions}
Let us summarize the results of this paper. The correlator of three ``forward''  light-ray operators (\ref{sjlrs}) has the form
\begin{eqnarray}
&&\hspace{-1mm}
\langle S_{1n_1}^{j_1}(z_{1_t})S_{1n_2}^{j_2}(z_{2_t})S_{1n_3}^{j_3}(z_{3_t})\rangle
\label{conc1}\\
&&\hspace{-1mm}
=~C(\Delta_i,j_i,g^2){(-2n_1\cdot n_2)^{j_1+j_2-j_3-1\over 2}\over |z_{12_t}|^{\Delta_1+\Delta_2-\Delta_3-1}}
{(-2n_1\cdot n_3)^{j_1+j_3-j_2-1\over 2}\over |z_{13_t}|^{\Delta_1+\Delta_3-\Delta_2-1}}
{(-2n_2\cdot n_3)^{j_2+j_3-j_1-1\over 2}\over |z_{23_t}|^{\Delta_2+\Delta_3-\Delta_1-1}}
\mu^{-\gamma_1-\gamma_2-\gamma_3}
\nonumber
\end{eqnarray}
with the structure constant
\begin{eqnarray}
&&\hspace{-1mm}
C(\omega_i,g^2)~
=~{iN_c^2\omega_1\omega_2\omega_3F(\omega_i,g^2)
\over \pi^3(\omega_1+\omega_2-\omega_3) (\omega_1+\omega_3-\omega_2) (\omega_2+\omega_3-\omega_1)}
\label{conc2}
\end{eqnarray}
At small $\omega_i$ the operator  $S_1^j$ can be identified with gluon light-ray operator $\calf^j$ given by Eq. (\ref{lrays}).
In the tree approximation,  the correlator of three gluon operators is given by Eq. (\ref{conc1}) with $F=1+O(\omega_i)$ as follows 
from Eq. (\ref{treesultF}). 
In  the BFKL regime ($g^2,\omega_i\ll 1$, ${g^2\over\omega_i}\sim 1$) the function $F$  has the form (\ref{Fotvet})
\begin{eqnarray}
&&\hspace{-1mm}
F(\gamma^\ast_1,\gamma^\ast_2,\gamma^\ast_3)~=~64g^6
\tiLa(\gamma^\ast_1,\gamma^\ast_2,\gamma^\ast_3)
\prod_{k=1}^3{(1+ \gamma_k^\ast)2^{-2\gamma^\ast_k}\Gamma\big(2+{\gamma_k^\ast\over 2}\big)\Gamma\big(\half-{\gamma_k^\ast\over 2}\big)
\over \Gamma\big(1-{\gamma_k^\ast\over 2}\big)
\Gamma\big({3\over 2}+{\gamma_k^\ast\over 2}\big){\gamma_k^\ast}^2\talef'(\gamma_k^\ast)}
\label{concl3}
\end{eqnarray}
where $\gamma^\ast$ is the solution  of Eq. (\ref{simproot}).

\begin{figure}[htb]
\begin{center}
\includegraphics[width=131mm]{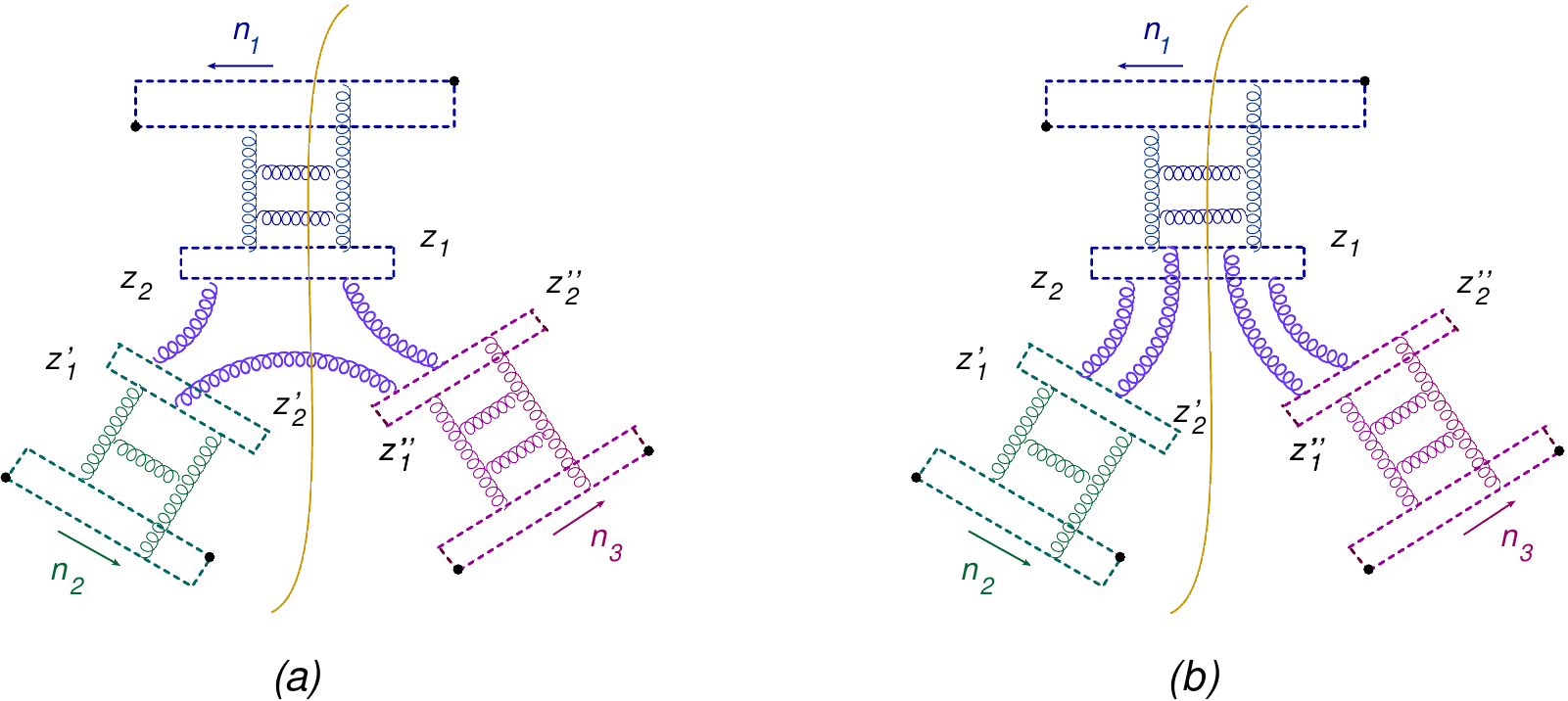}
\caption{Typical ``cut diagrams'' for the correlator of three Wilson frames }
\label{fig:3bfklzloim}
\end{center}
\end{figure}

Let us now discuss main features of the result (\ref{concl3}).  First, note
that since $\gamma^\ast$ is real, in our LLA approximation the constant $F$ is real since all the 
functions in the r.h.s. of Eq. (\ref{Fotvet}) are real.
Indeed, for $\Gamma$-functions it is trivial and  for $\tiLa(\gamma^\ast_1,\gamma^\ast_2,\gamma^\ast_3)$ 
it follows from the explicit expression (\ref{sumj}). 
This is in accordance with the fact that physical s-channel imaginary part of the amplitude (\ref{general1}) vanishes in our approximation. Indeed, it would correspond to ``cut'' diagram
of Fig.  \ref{fig:3bfklzloim}a type 
and cut propagator connecting two infinite Wilson lines in $n_2$ and $n_3$ directions 
vanishes (see the last line in the Eq. (\ref{treecorrels}) in Appendix \ref{sect:bfkl}). 
The imaginary part comes from the next terms of the expansion in powers of $g^2$ and $\omega$. The imaginary part $\sim\omega$
is given by the second term in the square brackets in the r.h.s. of tree-level expression (\ref{treesult}). As to imaginary part $\sim g^2$,  
it comes from the diagrams of the type shown in Fig. \ref{fig:3bfklzloim}b. 
These diagrams were calculated in the $n_2\rightarrow n_3$ 
limit in Refs. \cite{Balitsky:2015tca,Balitsky:2015oux}  and the result is given by Eq. (\ref{3poms12}) or Eq. (\ref{3poms14}) for small $g^2/\omega$. 
Note that the result  (\ref{3poms11}) has the same structure as Eq. (\ref{concl3}).

We saw that the structure constant has poles at $\omega_i=\omega_j+\omega_k$ reflecting boost invariance at $n_j\rightarrow n_k$. An interesting question
is what are singularities in the function $F(\omega_i,g^2)$ apart from obvious singular point $\omega_i=0$, for example like ${1\over\omega_1-\omega_2}$.
It is worthwhile to note that such terms appeared in the intermediate steps in the calculation of $\Lambda$ at small ${g^2\over \omega_i}$, for example 
the term $J_6$ contains ${1\over\epsilon_1-\epsilon_2}\rightarrow {g^2\omega_1\omega_2\over\omega_1-\omega_2}$, see Eq. (\ref{Jayz}). There were also
other singularities which all canceled in the final result (\ref{Lambdaresult}) so it suggests that the  function $F$ 
is finite at $\omega_i\neq 0$.

In conclusion, let us discuss the applicability of our results to QCD correlators of gluon light-ray operators. In the leading log
approximation considered here the formulas for correlators will be the same as in ${\cal N}=4$ case since running of the coupling constant
is beyond the LLA approximation, and since the contribution of scalar and gluino operators is negligible at small $\omega$. 
At the NLO level, in  ${\cal N}=4$ case we expect only corrections to structure constant of the type of Eq. (\ref{concl3}), but in QCD 
the functional form of two- and three-point correlators may change. An example of such change is the modification of the
formula (\ref{cornalba}) for the  $\gamma^\ast\gamma^\ast$ amplitude in QCD calculated at the NLO BFKL level in 
Ref. \cite{Chirilli:2013kca,Chirilli:2014dcb}. It would be interesting to write down such modifications for the correlators of gluon-light-ray operators at the NLO BFKL level in QCD.

The author is grateful to V. Kazakov, G. Korchemsky, and E. Sobko for valuable discussions. This  work is supported by contract  DE-AC05-06OR23177 under which the Jefferson Science Associates, LLC operate the Thomas Jefferson National Accelerator Facility, 
and by the grant DE-FG02-97ER41028.

\section{Appendix}
\subsection{BFKL kernel in the triple Regge limit\label{sect:bfkl}}
\begin{figure}[htb]
\begin{center}
\includegraphics[width=77mm]{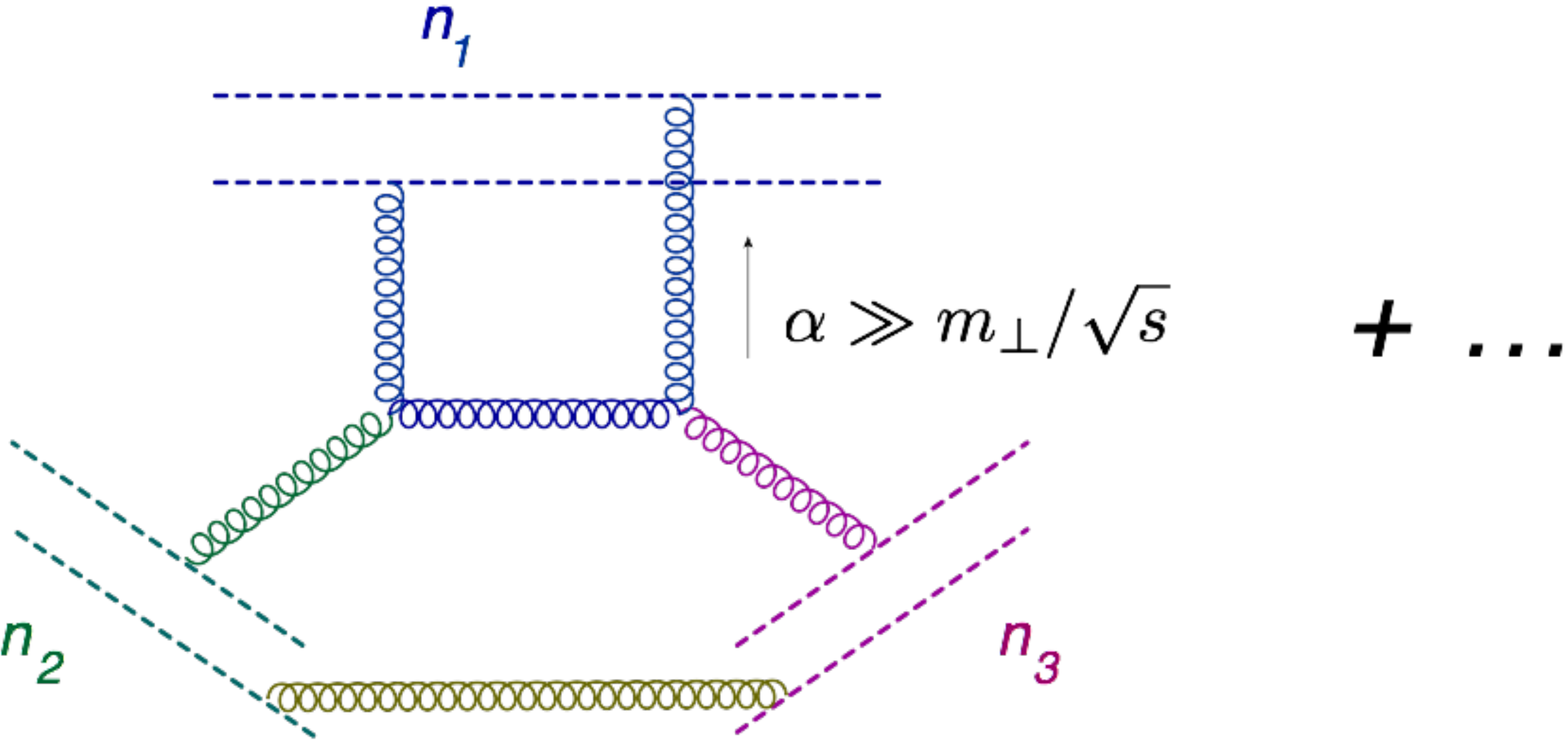}
\end{center}

\vspace{-11mm}
\caption{Leading diagrams with logarithmic integrals over  $\alpha$ \label{fig:bfkl3a}}
\end{figure}

In this Section I will demonstrate how the BFKL kernel comes out of the  conventional momentum-space calculation  in the triple Regge limit. 
Let us again consider the first diagram in Fig. \ref{fig:3bfklzlo}. Our LLA approximation (\ref{rapcuts}) in the momentum space reads\\ 
\begin{equation}
\alpha\beta s_{12}\gg m_\perp^2,~~~~\alpha\gamma s_{13}\gg m_\perp^2,~~~~\beta\gamma s_{23} \gg m_\perp^2
\end{equation}
 If all $s_{ij}$ are of the same order, this translates to $\alpha\gg m_\perp\sqrt{s_{23}\over s_{12}s_{13}}$ etc.
\footnote{
Indeed, if $\alpha\beta s_{12}=\lambda_3 m_\perp^2,\alpha\gamma s_{13}=\lambda_2 m_\perp^2,\beta\gamma s_{13}=\lambda_1m_\perp^2$ we get 
$\alpha=\sqrt {\lambda_2\lambda_3\over\lambda_1}m_\perp\sqrt{s_{23}\over s_{12}s_{13}} $
and therefore $\alpha\gg m_\perp\sqrt{s_{23}\over s_{12}s_{13}}$ if all $\lambda$'s are large and of the same order.}
Suppose now that we already performed integrals over $\beta$ and $\gamma$ which results in logs multiplied
by (conformal) dipoles and we would like to consider the last integral over $\alpha$ coming from the diagrams of Fig. \ref{fig:bfkl3a} type.

\vspace{-1mm}

\begin{figure}[htb]
\begin{center}
\includegraphics[width=151mm]{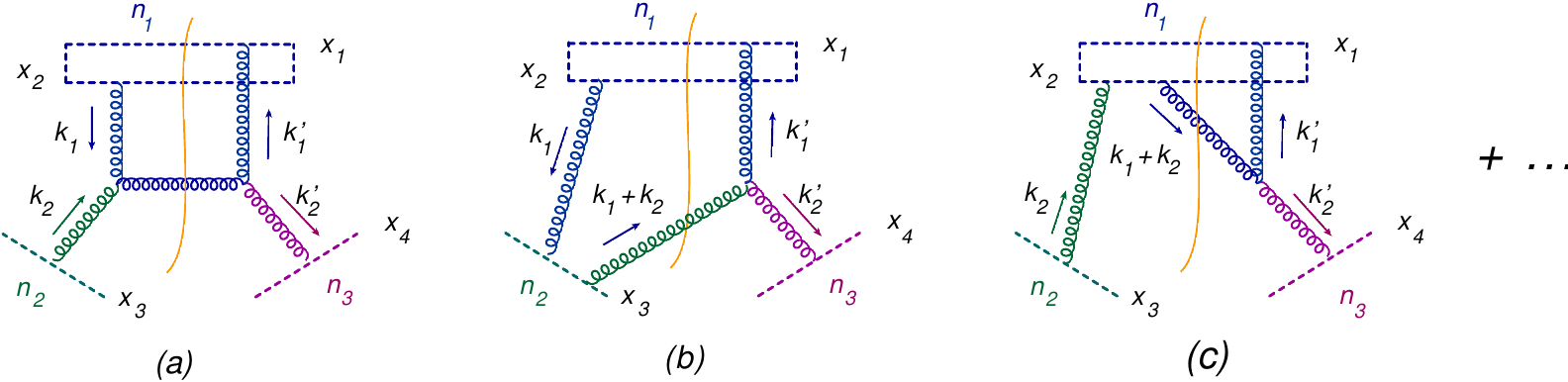}
\end{center}
\caption{First rung of the BFKL evolution in the triple Regge limit \label{fig:bfkl3aa}}
\end{figure}

To avoid cluttering of formulas, we will disregard the bottom gluon connecting Wilson lines parallel to $n_2$ and $n_3$. Indeed, 
the corresponding factor
$$
\langle {V}(z_{3t}e_t+\brez_3\bre) {W}(z_{5t}e_t+\chez_5\che)\rangle~=~i\alpha_st^a\otimes t^a\ln [z_{35t}^2+(\brez_3+\chez_5)^2]
$$
(plus permutations) simply multiplies  contributions of diagrams in Fig. \ref{fig:bfkl3aa}  and has nothing to do with logarithm coming out of $\alpha$ integration.  

To simplify our formulas, let us calculate the ``cut diagram''  shown in Fig. \ref{fig:bfkl3aa}.  It can be represented by a functional integral
over double set of variables: fields to the left and to the right of the cut which coincide at $t=\infty$.  
\footnote{If $n_2=n_3$ and the corresponding dipoles 
are the same, the double functional integral for the cut diagram gives the imaginary  part of the non-cut diagram, see e.g. the discussion in 
Refs. \cite{Balitsky:1988fi,Balitsky:1990ck,Balitsky:1991yz}.}
We get
\begin{eqnarray}
&&\hspace{-1mm}
\langle 
{\rm Tr}\{\tilU^\dagger(x_1) U(x_1)U^\dagger(x_2)\tilU(x_2)\}\tilV^\dagger(x_3)W(x_4)\rangle
\label{1strung}\\
&&\hspace{-1mm}
=~
g_{{}_{\rm YM}}^6N_ct^a\otimes t^a\!\int\!\dhd^4 k_1\dhd^4 k'_1\dhd^4 k_2\dhd^4 k'_2~
\dbar^4(k_2+k_1-k'_2-k'_1)~
\dbar ((k_2+k_1)^2)\theta(k_2+k_1)_0
\nonumber\\
&&\hspace{-1mm}
\times~
\dbar(k_2\cdot n_2)\dbar(k_1\cdot n_1)\dbar(k'_2\cdot n_3)\dbar(k'_1\cdot n_1)~ 
{L^{\lambda}(k_2,k_1)
L_{\lambda}(k'_2,k'_1)\over  k_1^2k_2^2{k'}_1^2{k'}_2^2}
~e^{ik_{2t}x_{3t}+i\brek_2\brex_3}
\nonumber\\
&&\hspace{-1mm}
\times~
\big(e^{ik_{1t}x_{1t}+i\tilk_1\tilx_1}-e^{ik_{1t}x_{2t}+i\tilk_1\tilx_2}\big)
e^{-ik'_{2t}x_{4t}-i\chek'_2\chex_4}
\big(e^{-ik'_{1t}x_{1t}-i\tilk'_1\tilx_1}-e^{-ik'_{1t}x_{2t}-i\tilk'_1\tilx_2}\big)
\nonumber
\end{eqnarray}
where we denoted Wilson lines to the left of the cut by tilde. 
Here we use space-saving notations $\dhd^n k\equiv {d^n k\over (2\pi)^n}$ and $\dbar^{(n)}(k)\equiv (2\pi)^n\delta^{(n)}(k)$.
The Lipatov vertex  of gluon emission can be taken e.g. from Ref. \cite{Babansky:2002my}
\begin{eqnarray}
{2\over s_{12}}L(k_2,k_1)&=&(k_1-k_2)
-\Big[2{k_{1n_2}\over n_{12}}+{k_1^2\over k_{2n_1}}\Big]n_1
+\Big[2{k_{2n_1}\over n_{12}}+{k_2^2\over k_{1n_2}}\Big]n_2
\nonumber\\
{2\over s_{13}}L(k'_2,k'_1)&=&(k'_1-k'_2)
-\Big[2{k'_{1n_3}\over n_{13}}+{{k'}_1^2\over k'_{2n_1}}\Big]n_1
+\Big[2{k'_{2n_1}\over n_{13}} +{{k'}_2^2\over k'_{1n_3}}\Big]n_3
\label{els}
\end{eqnarray}
Rewriting these formulas in terms of triple Sudakov variables (\ref{tridakov}) and taking into account $\delta$-functions in the r.h.s. of Eq. (\ref{1strung}) we obtain
\begin{eqnarray}
&&\hspace{-1mm}
{2\over s_{12}}L(k_1,k_2)~=~(k_1-k_2)_t+
\label{elses}\\
&&\hspace{-1mm}
+~n_1\Big[-\alpha_1-\alpha_2-{k_1^2\over k_{2n_1}}
-2\gamma_1{n_{23}\over n_{12}}\Big]
+n_2\Big[\beta_1+\beta_2+{k_2^2\over k_{1n_2}}+2\gamma_2{n_{13}\over n_{12}}\Big]
+(\gamma_1-\gamma_2)n_3
\nonumber\\
&&\hspace{-1mm}
{2\over s_{13}}L(k'_1,k'_2)~=~(k'_1-k'_2)_t
\nonumber\\
&&\hspace{-1mm}
+~n_1\Big[-\alpha'_1-\alpha'_2-{{k'}_1^2\over k_{2n_1}}-2\beta'_1{n_{23}\over n_{13}}\Big]
+(\beta'_1-\beta'_2)n_2
+n_3\Big[\gamma'_1+\gamma'_2+{{k'}_2^2\over k'_{1n_3}}+2\beta'_2{n_{12}\over n_{13}}\Big]
\nonumber
\end{eqnarray}
In our LLA  approximation
 $\alpha_1\gg m\sqrt{s_{23}\over s_{12}s_{13}}$ and $\beta_1\sim\beta'_1\sim \beta'_2\sim m\sqrt{s_{13}\over s_{12}s_{23}}$, $\alpha_2\sim\alpha'_2\sim m\sqrt{s_{23}\over s_{12}s_{13}}$, $\gamma_1\sim\gamma_2\sim m\sqrt{s_{12}\over s_{13}s_{23}}$. Moreover,  
from $\delta$-function $\delta ((k_2+k_1)^2)\theta(k_2+k_1)_0$ 
we see that $\xi_2=\beta_2-\alpha_2{s_{13}\over s_{23}}={s_{13}\over s_{12}}\big(\gamma'_2-\alpha'_2{s_{12}\over s_{23}}\big) \ll {m\over\sqrt s}$.  
Using these approximations, one obtains after some algebra 
\begin{eqnarray}
&&\hspace{-1mm}
{2\over s_{12}s_{13}}L(k_2,k_1)L(k'_2,k'_1)~=~
-(k_1-k'_1)_t^2-{s_{12}s_{23}\over s_{13}}(\beta_1-\beta'_1)^2
\label{bfklkernel}\\
&&\hspace{-1mm}
+~{\big(k_{1t}^2+{s_{12}s_{23}\over s_{13}}\beta_1^2\big)\big({k'_{2t}}^2+{s_{12}s_{23}\over s_{13}}{\beta'}_2^2\big)
\over (k_1+k_2)_t^2+{s_{12}s_{23}\over s_{13}}(\beta_1+\beta_2)^2}
+{\big(k_{2t}^2+{s_{12}s_{23}\over s_{13}}\beta_2^2\big)\big({k'_{1t}}^2+{s_{12}s_{23}\over s_{13}}{\beta'}_1^2\big)
\over (k_1+k_2)_t^2+{s_{12}s_{23}\over s_{13}}(\beta_1+\beta_2)^2}
~+~O\big({k_t^2\over\alpha_1s}\big)
\nonumber
\end{eqnarray}
which is a ``real part'' of the BFKL kernel.

The amplitude can be rewtitten as 
\begin{eqnarray}
&&\hspace{-1mm}
\langle 
{\rm Tr}\{\tilU^\dagger(x_1) U(x_1)U^\dagger(x_2)\tilU(x_2)\}\tilV^\dagger(x_3)W(x_4)\rangle
\label{fla34}\\
&&\hspace{-1mm}
=~
{g_{{}_{\rm YM}}^6N_c\over 4\pi}t^a\otimes t^a\!\int_0^\infty\!{d\alpha_1\over\alpha_1}\!\int\!\dhd k_{1t}\dhd k'_{1t}\dhd k_{2t}\dhd k'_{2t}~\dbar(k_1+k_2-k'_1-k'_2)_t
\nonumber\\
&&\hspace{-1mm}
\times~\int\!
\dhd\tilk_1\dhd\tilk'_1\dhd\brek_2\dhd\chek'_2
~\dbar\big(\tilk_1-\brek_2-\tilk'_1+\chek'_2\big)
\nonumber\\
&&\hspace{-1mm}
\times~
\big(e^{ik_{1t}x_{1t}+i\tilk_1\tilx_1}-e^{ik_{1t}x_{2t}+i\tilk_1\tilx_2}\big)
\big(e^{-i(k'_1,x_1)_t}-e^{-i(k'_1,x_2)_t}\big)
e^{ik_{2t}x_{3t}+i\brek_2\brex_3}e^{-ik'_{2t}x_{4t}-i\chek'_2\chex_4}
\nonumber\\
&&\hspace{-1mm}
\times~{-(k_1-k'_1)_t^2-(\tilk_1-\tilk'_1)^2+{(k_{1t}^2+\tilk_1^2)({k'}_{2t}^2+\tilde{k'}_2^2)+(k_{2t}^2+\tilk_2^2)({k'}_{1t}^2+\tilde{k'}_1^2)\over (k_1+k'_1)_t^2+(\tilk_1+\tilk'_1)^2}
\over(k_{1t}^2+\tilk_1^2)({k'}_{1t}^2+{\tilde{k'}}_1^2)(k_{2t}^2+\brek_2^2)({k'}_{2t}^2+{\check{k'}_2}^2)}
\nonumber
 \end{eqnarray}
Note that at $n_2=n_3$ this formula reduces to the first rung of the BFKL ladder for dipole-dipole cross section \cite{Babansky:2002my}

In this form it coincides with the first iteration of the evolution equation for color dipoles. 
Let us demonstrate this for the simple term in the BFKL kernel $2k_{1t}k'_{1t}+2\tilk_1\tilk'_1$.
Performing momentum integrals one obtains
\begin{eqnarray}
&&\hspace{-1mm}
{g_{{}_{\rm YM}}^6N_c\over 4\pi}t^a\otimes t^a\!\int_0^\infty\!{d\alpha_1\over\alpha_1}\!\int\!\dhd k_{1t}\dhd k'_{1t}\dhd k_{2t}\dhd k'_{2t}~\dbar(k_1+k_2-k'_1-k'_2)_t
\label{sampleterm}\\
&&\hspace{-1mm}
\times~\int\!
\dhd\tilk_1\dhd\tilk'_1\dhd\brek_2\dhd\chek'_2
~\dbar\big(\tilk_1-\brek_2-\tilk'_1+\chek'_2\big)
{2k_{1t}k'_{1t}+2\tilk_1\tilk'_1
\over(k_{1t}^2+\tilk_1^2)({k'}_{1t}^2+{\tilde{k'}}_1^2)(k_{2t}^2+\brek_2^2)({k'}_{2t}^2+{\check{k'}_2}^2)}
\nonumber\\
&&\hspace{-1mm}
\times~
\big(e^{ik_{1t}x_{1t}+i\tilk_1\tilx_1}-e^{ik_{1t}x_{2t}+i\tilk_1\tilx_2}\big)
\big(e^{-i(k'_1,x_1)_t}-e^{-i(k'_1,x_2)_t}\big)
e^{ik_{2t}x_{3t}+i\brek_2\brex_3}e^{-ik'_{2t}x_{4t}-i\chek'_2\chex_4}
\nonumber\\
&&\hspace{-1mm}
=~{\alpha_s^3N_c\over 2\pi^2}t^a\otimes t^a\!\int_0^\infty\!{d\alpha_1\over\alpha_1}
\nonumber\\
&&\hspace{11mm}
\times~\!\int\! d^2z_\perp{(x_1-x_2)_\perp^2\over (x_1-z)^2(x_2-z)^2}
\ln[(x_{3t}-z_t)^2+(\brex_3+\tilz)^2]\ln[(x_{4t}-z_t)^2+(\chex_4+\tilz)^2]
\nonumber
\end{eqnarray}
where $z_\perp\equiv z_t,\tilz$.

On the other hand, the (linearized) evolution equation for color dipoles (in the double functional integral formalism) reads
 \cite{Balitsky:1997mk,Balitsky:2001gj}
\begin{eqnarray}
&&\hspace{-1mm}
{d\over dY}{\rm Tr}\{\tilU^\dagger(x_1)U(x_1)U^\dagger(x_2)\tilU(x_2)\}~
\nonumber\\
&&\hspace{-1mm}
=~{\alpha_sN_c\over 2\pi}\!\int\! d^2z_\perp ~{(x_1-x_2)_\perp^2\over (x_1-z)_\perp^2(x_2-z)_\perp^2}
\big[{\rm Tr}\{\tilU^\dagger(x_1)U(x_1)U^\dagger(z)\tilU(z)\}
\nonumber\\
&&\hspace{-1mm}
+~{\rm Tr}\{\tilU^\dagger(z)U(z)U^\dagger(x_2)\tilU(x_2)\}
-{\rm Tr}\{\tilU^\dagger(x_1)U(x_1)U^\dagger(x_2)\tilU(x_2)\}\big]
\label{lbkdiff}
\end{eqnarray}
where $Y\equiv\ln\alpha$. 
The term (\ref{sampleterm}) comes from the correlator
\begin{eqnarray}
&&\hspace{-1mm}
{\alpha_sN_c\over 2\pi}\!\int_0^\infty\!{d\alpha\over\alpha}\!\int\! d^2z_\perp ~{-2(x_1-z,x_2-z)_\perp\over (x_1-z)_\perp^2(x_2-z)_\perp^2}
\nonumber\\
&&\hspace{33mm}
\times~
\langle {\rm Tr}\{U^\dagger(z)\tilU(z)+\tilU^\dagger(z)U(z)\}\tilV^\dagger(x_3)W(x_4)\rangle
\label{fla79}
\end{eqnarray}
Using the tree-level correlators of Wilson lines 
\begin{eqnarray}
&&\hspace{-1mm}
\langle \tilde{U}(z_{t}e_t+\tilz\tile) \tilde{V}(x_{3t}e_t+\brex_3\bre)\rangle~=~i\alpha_st^a\otimes t^a\ln [(z-x_3)_t^2+(\tilz+\brex_3)^2],
\nonumber\\
&&\hspace{-1mm} 
\langle {U}(z_{t}e_t+\tilz\tile) {W}(x_{4t}e_t+\chez_4\che)\rangle~=~-i\alpha_st^a\otimes t^a\ln [(z-x_4)_t^2+(\tilz+\chex_4)^2],
\nonumber\\
&&\hspace{-1mm} 
\langle \tilde{U}(z_{t}e_t+\tilz\tile)  {W}(x_{4t}e_t+\chex_4\che)\rangle~=~\langle U(z_{t}e_t+\tilz\tile)  \tilde{V}(x_{3t}e_t+\chex_3\che)\rangle~=~0
\label{treecorrels}
\end{eqnarray}
it is easy to see that Eq.(\ref{fla79}) coincides with the r.h.s. of Eq. (\ref{sampleterm}). Similarly, one can check that other terms in the BFKL kernel
correspond to linear part of the evolution equation for color dipoles (\ref{lbkdiff}), see e.g. the book \cite{Kovchegov:2012mbw}.

\subsection{Correlator of three twist-2 LR operators in the tree approximation \label{sect:firstorder}}

First, we calculate the correlator of two light-ray gluon operators. 
Using bare propagator 
\begin{eqnarray}
&&\hspace{-1mm}   
\pi^2\langle F_{n_1}^{~\mu}(u_1 n_1+x_{1_\perp})F_{n_2}^{~\nu}(u_2 n_2+x_{2_\perp})\rangle
\label{fla142}\\
&&\hspace{-1mm}
=~-{g^{\mu\nu}\delta^{ab}s_{12}x_{12_\perp}^2\over 2(u_1u_2s_{12}
+x_{12_\perp}^2+i\epsilon)^3}-{\delta^{ab}n_2^\mu n_1^\nu\over (u_1u_2s_{12}+x_{12_\perp}^2+i\epsilon)^2}
\nonumber\\
&&\hspace{5mm}
-~{s_{12}\delta^{ab}\over  (u_1u_2s_{12}+x_{12_\perp}^2+i\epsilon)^3}\big[u_2x_{12}^\mu n_2^\nu
-u_1n_1^\mu x_{12}^\nu-x_{12}^\mu x_{12}^\nu
+u_1u_2(n_1^\mu n_2^\nu-n_2^\mu n_1^\nu)
\big]
\nonumber
\end{eqnarray}
after simple integration we get the  tree-level correlator in the form
\footnote{As we discussed above, $\delta(j-j')$ actually means ``analytic continuation'' of $\delta(\nu-\nu')$ for $j=\half +i\nu,~j'=\half +i\nu'$.}
\begin{equation}
\hspace{-1mm}
{1\over N_c^2-1}\langle \calf^{j}_{n_1}(x_{1_\perp})\calf^{j'}_{n_2}(x_{2_\perp})
\big\rangle
~=~-{is^4\over 2\pi}\delta(\omega-\omega'){1-e^{i\pi\omega}\over\sin\pi\omega}
{\omega(1+\omega)\over (x_{12_\perp}^2)^{2+\omega}s^{-\omega}}{\Gamma(2+\omega)\Gamma(4+\omega)\over\Gamma(4+2\omega)}
\label{corr2LR}
\end{equation}
which agrees with Eq. (\ref{ostatok}) in the limit $\omega\rightarrow 0$
\begin{eqnarray}
&&\hspace{-1mm}
{1\over N_c^2-1}\langle \calf^{j}_{n_1}(x_{1_\perp})\calf^{j'}_{n_2}(x_{2_\perp})
\big\rangle~
\simeq~-\delta(j-j'){\omega\over 2\pi x_{12_\perp}^4}{s^\omega\over x_{12_\perp}^{2\omega}}
\label{corr2LRs}
\end{eqnarray}
as discussed in the end of Sect. \ref{sect:2frames}.

\begin{figure}[htb]
\begin{center}
\includegraphics[width=121mm]{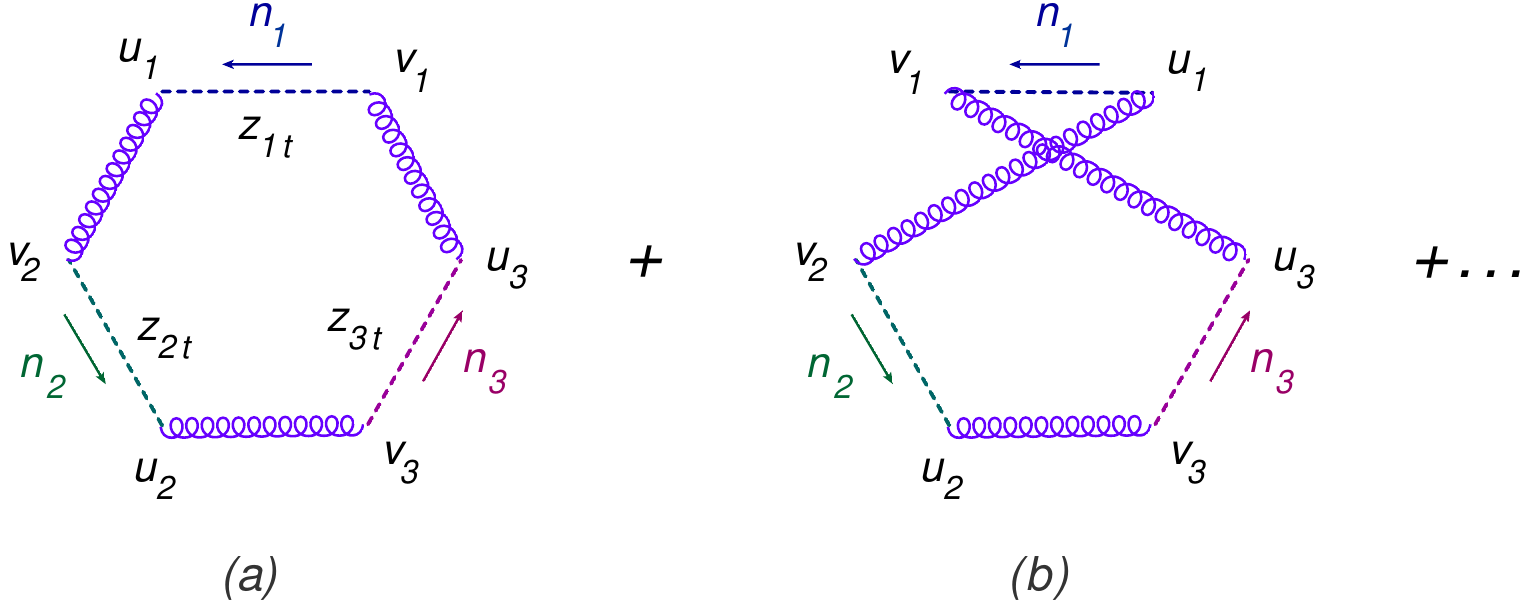}
\end{center}
\caption{Tree-level correlator of three twist-two gluon LR operators \label{triplo}}
\end{figure}

Next, consider diagrams in Fig. \ref{triplo} representing the 3-point correlator of gluon light-ray operators 
in the leading perturbative order.

After some algebra one obtains
the result for the first diagram in Fig. \ref{triplo}a in the form 
\begin{eqnarray}
&&\hspace{-1mm}   
\langle 
F_{n_1}^{a\mu}(u_1 n_1+z_{1t})F_{n_2}^{b~\nu}(v_2 n_2+z_{2t})\rangle
\langle F^a_{n_1\mu}(v_1 n_1+z_{1t})F_{n_3}^{c\lambda}(u_3 n_3+z_{3t})\rangle
\nonumber\\
&&\hspace{33mm}
\times~\langle F^b_{n_2\nu}(u_2 n_2+z_{2t})F^c_{n_3\lambda}(v_3 n_3+z_{3t})\rangle~
\nonumber\\
&&\hspace{-1mm}
=~{N_c^2-1\over 4\pi^6s_{12}s_{23}s_{13}}{z_{12_t}^2z_{13_t}^2z_{23_t}^2\over 
(u_1v_2s_{12}+z_{12_t}^2+i\epsilon)^3(v_1u_3s_{13}+z_{13_t}^2+i\epsilon)^3(u_2v_3s_{23}+z_{23_t}^2+i\epsilon)^3}
\nonumber\\
\label{gluotvet}
\end{eqnarray}
where $s_{12}\equiv -2n_1\cdot n_2$ etc. Adding the diagrams with permutations we obtain
\begin{eqnarray}
&&\hspace{-1mm}   
\int_{-\infty}^\infty\! du_1 dv_1du_2 dv_2du_3 dv_3~\theta(u_1-v_1)\theta(u_2-v_2)\theta(u_3-v_3)~\nonumber\\
&&\hspace{11mm}
\times~(u_1-v_1)^{-\omega_1}
(u_2-v_2)^{-\omega_2}(u_3-v_3)^{-\omega_3}
\langle 
F_{n_1}^{a\mu}(u_1 n_1+z_{1t})F^a_{n_1\mu}(v_1 n_1+z_{1t})
\nonumber\\
&&\hspace{11mm}
\times~
F_{n_2}^{b~\nu}(v_2 n_2+z_{2t})F^b_{n_2\nu}(u_2 n_2+z_{2t})F_{n_3}^{c\lambda}(u_3 n_3+z_{3t})F^c_{n_3\lambda}(v_3 n_3+z_{3t})\rangle~
\nonumber\\
&&\hspace{-1mm}
=~{N_c^2-1\over 4\pi^6s_{12}s_{23}s_{13}}
\int_{-\infty}^\infty\! du_1 dv_1du_2 dv_2du_3 dv_3~\big[\theta(u_1-v_1)(u_1-v_1)^{-\omega_1}+u_1\leftrightarrow v_1\big]
\nonumber\\
&&\hspace{11mm}
\times~\big[\theta(u_2-v_2)(u_2-v_2)^{-\omega_2}+u_2\leftrightarrow v_2\big]
\big[\theta(u_3-v_3)(u_3-v_3)^{-\omega_3}+u_3\leftrightarrow v_3\big]
\nonumber\\
&&\hspace{11mm}{z_{12_t}^2z_{13_t}^2z_{23_t}^2\over 
\times~(u_1v_2s_{12}+z_{12_t}^2+i\epsilon)^3(v_1u_3s_{13}+z_{13_t}^2+i\epsilon)^3(u_2v_3s_{23}+z_{23_t}^2+i\epsilon)^3}
\nonumber\\
\label{gluotvet1}
\end{eqnarray}

The integration over light-ray variables $u_i,v_i$ is done with the help of two formulas:
\begin{eqnarray}
&&\hspace{-1mm} 
\int_{-\infty}^\infty\! dt_1dt_2 dt_3\!\int_0^\infty\! dr_1dr_2 dr_3
{1\over [(r_1+t_1)(r_2+t_2)+a_{12}^2+i\epsilon]}
\label{1stpartotvet}\\
&&\hspace{-1mm}
\times~\Big\{
{r_1^{-\omega_1}r_2^{-\omega_2}r_3^{-\omega_3}\over [t_1(r_3+t_3)+a_{13}^2+i\epsilon](t_2t_3+a_{23}^2+i\epsilon)}
+{r_1^{-\omega_1}r_2^{-\omega_2}r_3^{-\omega_3}\over (t_1t_3+a_{13}^2+i\epsilon)[t_2(r_3+t_3) +a_{23}^2+i\epsilon]}\Big\}
\nonumber\\
&&\hspace{-2mm}
=
-4\sin{\pi\omega_3\over 2}\sin{\pi\over 2}(\omega_1+\omega_2-\omega_3)\cos{\pi\over 2}(\omega_1-\omega_2)\Gamma(1-\omega_1)\Gamma(1-\omega_2)\Gamma(1-\omega_3)
\nonumber\\
&&\hspace{-1mm}
\times~\Gamma^2\big({\omega_1+\omega_2-\omega_3\over 2}\big)\Gamma^2\big({\omega_2+\omega_3-\omega_1\over 2}\big)\Gamma^2\big({\omega_1+\omega_3-\omega_2\over 2}\big)
a_{12}^{\omega_3-\omega_1-\omega_2}a_{13}^{\omega_2-\omega_1-\omega_3}a_{23}^{\omega_1-\omega_2-\omega_3}
\nonumber
\end{eqnarray}
and
\begin{eqnarray}
&&\hspace{-1mm} 
\int_{-\infty}^\infty\! dt_1dt_2 dt_3\!\int_0^\infty\! dr_1dr_2 dr_3
\nonumber\\
&&\hspace{-1mm}
\times~\Big[
{r_1^{-\omega_1}r_2^{-\omega_2}r_3^{-\omega_3}\over [(r_1+t_1)t_2+a_{12}+i\epsilon][t_1(r_3+t_3)+a_{13}+i\epsilon][(r_2+t_2)t_3+a_{23}+i\epsilon]}
\nonumber\\
&&\hspace{11mm}
+~{r_1^{-\omega_1}r_2^{-\omega_2}r_3^{-\omega_3}\over [t_1(r_2+t_2)+a_{12}+i\epsilon][(r_1+t_1)t_3+a_{13}+i\epsilon][t_2(r_3+t_3)+a_{23}+i\epsilon]}\Big]
\nonumber\\
&&\hspace{-1mm}
=~\Gamma(1-\omega_1)\Gamma(1-\omega_2)\Gamma(1-\omega_3)
\Gamma^2\big({\omega_1+\omega_3-\omega_2\over 2}\big)
\Gamma^2\big({\omega_2+\omega_3-\omega_1\over 2}\big)
\Gamma^2\big({\omega_1+\omega_2-\omega_3\over 2}\big)
\nonumber\\
&&\hspace{11mm}
\times~a_{12}^{\omega_3-\omega_1-\omega_2}
a_{13}^{\omega_2-\omega_1-\omega_3}a_{23}^{\omega_1-\omega_2-\omega_3}
\nonumber\\
&&\hspace{11mm}
\times~\big[e^{i\pi(\omega_1+\omega_2-\omega_3)}+e^{i\pi(\omega_2+\omega_3-\omega_1)}
+e^{i\pi(\omega_1+\omega_3-\omega_2)}-e^{i\pi(\omega_1+\omega_2+\omega_3)}-2\big]
\label{2intlotvet}
\end{eqnarray}
Using these formulas with $a^2_{ij}={z_{ij_t}^2\over s_ij}$ it is easy to get
\begin{eqnarray}
&&\hspace{-1mm}
\int_{-\infty}^\infty\! du_1 dv_1\big[(u_1-v_1)^{-\omega_1}\theta(u_1-v_1)+u_1\leftrightarrow v_1\big]
\!\int_{-\infty}^\infty\! du_2 dv_2\big[(u_2-v_2)^{-\omega_2}\theta(u_2-v_2)
\label{mastegral}\\
&&\hspace{-1mm}
+~u_2\leftrightarrow v_2\big]
\!\int_{-\infty}^\infty\! du_3 dv_3\big[(u_3-v_3)^{-\omega_3}\theta(u_3-v_3)+u_3\leftrightarrow v_3\big]
\nonumber\\
&&\hspace{11mm}
\times~{1\over (u_1v_2s_{12}+z_{12}^2+i\epsilon)(v_1u_3s_{13}+z_{13}^2+i\epsilon)(u_2v_3s_{23}+z_{23}^2+i\epsilon)}
\nonumber\\
&&\hspace{-1mm}
=~{1\over s_{12}s_{13}s_{23}}\Phi(\omega_1,\omega_2,\omega_3)
\Gamma(1-\omega_1)\Gamma(1-\omega_2)\Gamma(1-\omega_3)
\Gamma^2\big({\omega_1+\omega_2-\omega_3\over 2}\big)
 \nonumber\\
&&\hspace{11mm}
\times~\Gamma^2\big({\omega_2+\omega_3-\omega_1\over 2}\big)\Gamma^2\big({\omega_1+\omega_3-\omega_2\over 2}\big)
\big({s_{12}\over z_{12}^2}\big)^{\omega_1+\omega_2-\omega_3\over 2}\big({s_{13}\over z_{13}^2}\big)^{\omega_1+\omega_3-\omega_2\over 2}
\big({s_{23}\over z_{23}^2}\big)^{\omega_2+\omega_3-\omega_1\over 2}
\nonumber
\end{eqnarray}
where 
\begin{eqnarray}
&&\hspace{-1mm}
\Phi(\omega_1,\omega_2,\omega_3)~
\nonumber\\
&&\hspace{-1mm}
=~
-4\sin{\pi\omega_1\over 2}\sin{\pi\over 2}(\omega_2+\omega_3-\omega_1)
\cos{\pi\over 2}(\omega_2-\omega_3)
-4\sin{\pi\omega_2\over 2}\sin{\pi\over 2}(\omega_1+\omega_3-\omega_2)
\nonumber\\
&&\hspace{-1mm}
\times~\cos{\pi\over 2}(\omega_1-\omega_3)
-4\sin{\pi\omega_3\over 2}\sin{\pi\over 2}(\omega_1+\omega_2-\omega_3)\cos{\pi\over 2}(\omega_1-\omega_2)
\nonumber\\
&&\hspace{-1mm}
+~
e^{i\pi(\omega_1+\omega_2-\omega_3)}+e^{i\pi(\omega_2+\omega_3-\omega_1)}
+e^{i\pi(\omega_1+\omega_3-\omega_2)}-e^{i\pi(\omega_1+\omega_2+\omega_3)}-2
\end{eqnarray}

Now, differentiating Eq. (\ref{mastegral}) two times with respect to each $x_{ij_t}^2$ one obtains
\begin{eqnarray}
&&\hspace{-1mm}   
\langle \calf_{n_1}(\omega_1,z_{1_t})\calf_{\omega_2,n_2}(z_{2_t})\calf_{\omega_3,n_3}(z_{3_t})\rangle~
\nonumber\\
&&\hspace{-1mm}
=~{N_c^2-1\over 32\pi^6z_{12}^2z_{13}^2z_{23}^2}
\Gamma\big({\omega_1+\omega_2-\omega_3\over 2}\big)\Gamma\big({\omega_2+\omega_3-\omega_1\over 2}\big)
\Gamma\big({\omega_1+\omega_3-\omega_2\over 2}\big)
\nonumber\\
&&\hspace{-1mm}
\times~
\Gamma\big({\omega_1+\omega_2-\omega_3\over 2}+2\big)
\Gamma\big({\omega_2+\omega_3-\omega_1\over 2}+2\big)
\Gamma\big({\omega_1+\omega_3-\omega_2\over 2}+2\big)
\Phi(\omega_1,\omega_2,\omega_3)
\nonumber\\
&&\hspace{-1mm}
\times~\Gamma(1-\omega_1)\Gamma(1-\omega_2)\Gamma(1-\omega_3)
\big({s_{12}\over z_{12}^2}\big)^{\omega_1+\omega_2-\omega_3\over 2}
\big({s_{13}\over z_{13}^2}\big)^{\omega_1+\omega_3-\omega_2\over 2}
\big({s_{23}\over z_{23}^2}\big)^{\omega_2+\omega_3-\omega_1\over 2}
\label{baresult}
\end{eqnarray}
A quick check of this formula can be obtained by Eq. (\ref{operff}) which states that as $\omega_i\rightarrow 1$ the coefficient in front of 
$\Gamma(1-\omega_1)\Gamma(1-\omega_2)\Gamma(1-\omega_3)$ is represented by the three-point correlator of local two-gluon operators
\begin{equation}
\hspace{-1mm} 
\int\! dudvdw\langle F^a_{n_1\xi}F_{n_1}^{a\xi}(un_1+x_{1t})F^b_{n_2\eta}F_{n_2}^{b\eta}(vn_2+x_{2t})F^c_{n_3\zeta}F_{n_3}^{c\zeta}(wn_3+x_{3t})
\end{equation}
Using tree-level correlator
\begin{eqnarray}
&&\hspace{-1mm}   
\langle F_{n_1}^{a\mu}F^a_{n_1\mu}(u_1 n_1+x_{1t})F_{n_2}^{b\nu}F^b_{n_2\nu}(u_2 n_2+x_{2t})
F_{n_3}^{c\lambda}F^c_{n_3\lambda}(u_3 n_3+x_{3t})\rangle^{\rm tree}~
\nonumber\\
&&\hspace{-1mm}
=~{2(N_c^2-1)s_{12}s_{13}s_{23} x_{12t}^2x_{13t}^2x_{23t}^2\over \pi^6(uvs_{12}+x_{12t}^2+i\epsilon)^3
(uws_{13}+x_{13t}^2+i\epsilon)^3(vws_{23}+x_{23t}^2+i\epsilon)^3}
\end{eqnarray}
and the integral
\begin{eqnarray}
&&\hspace{-11mm}
\int\! du dv dw~{1\over [uv+a+i\epsilon][vw+b+i\epsilon][uw+c+i\epsilon]}~=~-{2\pi^3\over\sqrt{abc}}
\end{eqnarray}
one obtains
\begin{eqnarray}
&&\hspace{-1mm} 
\int\! dudvdw\langle F^a_{n_1\xi}F_{n_1}^{a\xi}(un_1+x_{1t})F^b_{n_2\eta}F_{n_2}^{b\eta}(vn_2+x_{2t})F^c_{n_3\zeta}F_{n_3}^{c\zeta}(wn_3+x_{3t})
\nonumber\\
&&\hspace{-1mm}
=~-{27(N_c^2-1)\sqrt{s_{12}s_{13}s_{23}}\over 128\pi^3(x_{12t}^2x_{13t}^2x_{23t}^2)^{3\over 2}}  
\end{eqnarray}
which agrees with Eq. (\ref{baresult}) at $\omega_i\rightarrow 1$ since $\Phi(1,1,1)=-16$.

For the BFKL limit we need the behavior of the tree-level correlator (\ref{baresult}) as $\omega_i\rightarrow 0$.
It is easy to see that at small $\omega_i$
\begin{equation}
\Phi(\omega_1,\omega_2,\omega_3)~=~4i\pi^3\omega_1\omega_2\omega_3-\pi^4\omega_1\omega_2\omega_3(\omega_1+\omega_2+\omega_3)~+~O(\omega^5)
\end{equation}
and therefore 
\begin{eqnarray}
&&\hspace{-0mm}   
\langle \calf_{n_1}(\omega_1,z_{1_t})\calf_{\omega_2,n_2}(z_{2_t})\calf_{\omega_3,n_3}(z_{3_t})\rangle^{\rm tree}~
\nonumber\\
&&\hspace{11mm}
=~{i(N_c^2-1)\over \pi^3z_{12}^2z_{13}^2z_{23}^2}
{\omega_1\omega_2\omega_3\big[1+{i\pi\over4}(\omega_1+\omega_2+\omega_3)~+~O(\omega^2)\big]
\over(\omega_1+\omega_2-\omega_3)(\omega_2+\omega_3-\omega_1)(\omega_1+\omega_3-\omega_2)}
\nonumber\\
&&\hspace{22mm}
\times~
\big({s_{12}\over z_{12}^2}\big)^{\omega_1+\omega_2-\omega_3\over 2}
\big({s_{13}\over z_{13}^2}\big)^{\omega_1+\omega_3-\omega_2\over 2}
\big({s_{23}\over z_{23}^2}\big)^{\omega_2+\omega_3-\omega_1\over 2}
\label{treesult}
\end{eqnarray}
which corresponds to
\begin{eqnarray}
&&\hspace{-1mm}
F(\omega_i,g^2)\Big|_{g^2=0}~=~1+{i\pi\over4}(\omega_1+\omega_2+\omega_3)~+~O(\omega^2)
\label{treesultF}
\end{eqnarray}
in the notations of Eq. (\ref{Cgeneral}) parametrization.

As demonstrated in the next Section, the singularities at $\omega_i=\omega_j+\omega_k$ originate from boost invariance 
of the correlator of three light-ray operators at $n_j=n_k$. Note, however, that such singularity is absent in the correlator
of three local operators $F_n^j(x_t)$ (see Eq. (\ref{lops}) for definition) since $\Phi(j_i,j_k,j_i+j_k)=0$ for integer $j$'s.

\subsection{Boost invariance and singularities of structure constants \label{sect:BKase}}
\begin{figure}[htb]
\hspace{-1mm}
\includegraphics[width=99mm]{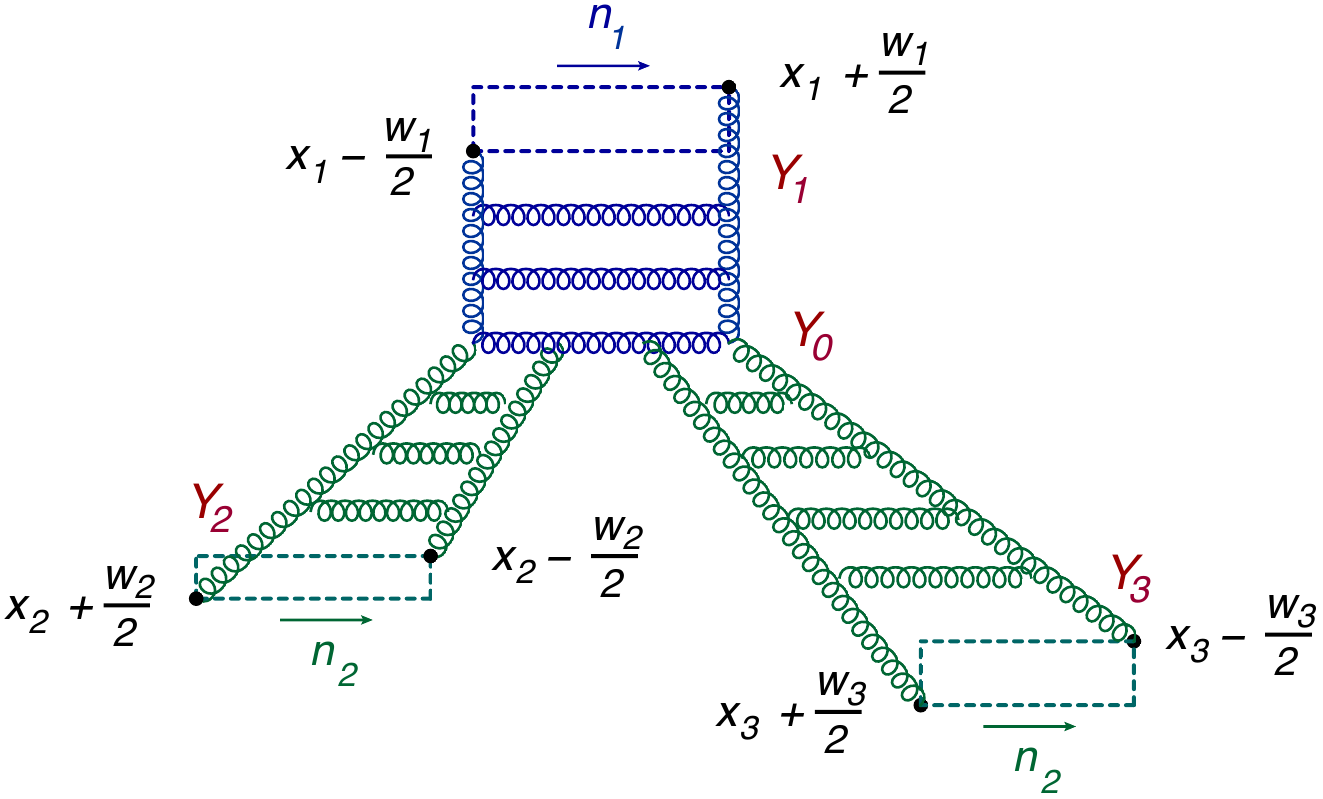}
\caption{Triple BFKL evolution at $n_2=n_3$ with BK vertex at the rapidity $Y_0$.}
\label{fig:3poms}
\end{figure}

As we mentioned above, the singularities at $\omega_i=\omega_j+\omega_k$ are related to boost invariance. 
To demonstrate this, let us follow
 Ref. \cite{Balitsky:2015tca} and consider the correlator of Wilson frame in $n_1$ direction and two Wilson frames in $n_2$ directions,
see Fig. \ref{fig:3poms}
%
\begin{equation}
\big\langle \calf^{j_1}_{n_1}\big(x_{1t}+{w_{1t}\over 2},x_{1t}-{w_{1t}\over 2}\big)
\calf^{j_2}_{n_2}\big(x_{2t}+{w_{2t}\over 2},x_{2t}-{w_{2t}\over 2}\big)
\calf^{j_3}_{n_2}\big(x_{3t}+{w_{3t}\over 2},x_{3t}-{w_{3t}\over 2}\big)\big\rangle
\label{3poms1}
\end{equation}
As we discussed in Sect. \ref{sect:2frames}, this correlator in the Regge limit can be represented by the correlator of three
conformal dipoles, one in $n_1$ direction and two in $n_2$ directions. We get from Eq. (\ref{ffif})
\begin{eqnarray}
&&\hspace{-3mm}
\big\langle \calf^{j_1}_{n_1}\big(x_{1t}+{w_{1t}\over 2},x_{1t}-{w_{1t}\over 2}\big)
\calf^{j_2}_{n_2}\big(x_{2t}+{w_{2t}\over 2},x_{2t}-{w_{2t}\over 2}\big)
\calf^{j_3}_{n_2}\big(x_{3t}+{w_{3t}\over 2},x_{3t}-{w_{3t}\over 2}\big)
\big\rangle
\label{3poms2}\\
&&\hspace{-1mm}
=~-i{N_c^6\over\pi^9}\!\int_0^\infty\! {dl_1dl_2dl_3\over l_1l_2l_3}~l_1^{-\omega_1}l_2^{-\omega_2}l_3^{-\omega_3}
\!\int\! d\nu_1 d\nu_2 d\nu_3(w_{1t}^2)^{-\half+i\nu_1}(w_{2t}^2)^{-\half+i\nu_2}(w_{3t}^2)^{-\half+i\nu_3}
\nonumber\\
&&\hspace{3mm}
\times~\langle \calu^{Y_1}_{\rm conf}(x_{1t},-\nu_1)\calv^{Y_2}_{\rm conf}(x_{2t},-\nu_2)\calv^{Y_3}_{\rm conf}(x_{3t},-\nu_3)\rangle
\prod_{i=1,2,3}\nu_i
{ 2^{-4i\nu_i}\Gamma\big({3\over 2}+i\nu_i\big)\Gamma(1-i\nu_i)\over \Gamma\big({3\over 2}-i\nu_i\big)\Gamma(1+i\nu_i)}
\nonumber
\end{eqnarray}
where $Y_i\simeq\ln l_i+\half\ln s m_\perp^2$  ($s\equiv s_{12}$).

As discussed in Refs.  \cite{Balitsky:2015tca} and \cite{Balitsky:2015oux}, 
the BK equation for color dipoles leads to the following structure of the correlator of a conformal dipole in $n_1$ direction and two dipoles in $n_2$ direction:
\begin{equation}
\int\! dY_0~\theta(Y_0+Y_2)\theta(Y_0+Y_3)~\big({\rm evolution}~\calu^{Y_1}\rightarrow \calu^{Y_0}\big)\otimes({\rm BK~vertex~at~}Y_0)\otimes
\bigg(\begin{array}{c}\langle \calu^{Y_0}\calv^{Y_2}\rangle\\
\langle \calu^{Y_0}\calv^{Y_3}\rangle
\end{array}\bigg)
\label{3poms3}
\end{equation}
where the integral over $Y_0$ comes from the fact that the splitting of one dipole into two (described by the BK vertex) can occur at any 
rapidity  between the $n_1$ dipole and the most energetic of $n_2$ dipoles. Specifically,  $\theta(Y_0+Y_i)$ reflects the fact that there 
should be sufficient energy between dipoles $\calu^{Y_0}$ and $\calv^{Y_i}$ to 
apply high-energy approximation, see the footnote \ref{futnout} at page 14. 

Rewrtiting Eq. (18) from Ref. \cite{Balitsky:2015tca} in terms of conformal dipoles, one gets
\begin{eqnarray}
&&\hspace{-1mm}
\langle \calu^{Y_1}_{\rm conf}(x_{1t},-\nu_1)\calv^{Y_2}_{\rm conf}(x_{2t},-\nu_2)\calv^{Y_3}_{\rm conf}(x_{3t},-\nu_3)\rangle
~=~-{\alpha_s^5N_c\over 4\pi^4}{1\over\big({1\over 4}+\nu_2^2\big)^2\big({1\over 4}+\nu_3^2\big)^2}
\nonumber\\
&&\hspace{-1mm}
\times~
\!\int_{-\infty}^{Y_1}\! dY_0\!\int\! dY_2dY_3~\theta(Y_0+Y_2)\theta(Y_0+Y_3)
~e^{\aleph(\nu_1)(Y_1-Y_0)+\aleph(\nu_2)(Y_0+Y_2)+\aleph(\nu_3)(Y_0+Y_3)}
\nonumber\\
&&\hspace{-1mm}
\times~\!\int\! {d^2x_4d^2x_5d^2x_6\over x_{45}^2x_{56}^2x_{46}^2}\Big({x_{45}^2\over x_{14}^2x_{15}^2}\Big)^{\half+i\nu_1}
\Big({x_{56}^2\over x_{25}^2x_{36}^2}\Big)^{\half+i\nu_2}
\Big({x_{46}^2\over x_{34}^2x_{36}^2}\Big)^{\half+i\nu_3}~+~O\big({1\over N_c}\big)
\label{3poms4}
\end{eqnarray}
where two-dimensional integrals go over transverse directions orthogonal to both $n_1$ and $n_2$. 
Combining Eqs. (\ref{3poms2}), (\ref{3poms3}), and (\ref{3poms4}) one obtains
\begin{eqnarray}
&&\hspace{-1mm}
\big\langle \calf^{j_1}_{n_1}\big(x_{1t}+{w_{1t}\over 2},x_{1t}-{w_{1t}\over 2}\big)
\calf^{j_2}_{n_2}\big(x_{2t}+{w_{2t}\over 2},x_{2t}-{w_{2t}\over 2}\big)
\calf^{j_3}_{n_2}\big(x_{3t}+{w_{3t}\over 2},x_{3t}-{w_{3t}\over 2}\big)
\big\rangle
\nonumber\\
&&\hspace{-1mm}
=~i{\alpha_s^5N_c^7\over 4\pi^{13}}
\!\int\! {d\nu_1 d\nu_2 d\nu_3\over\big({1\over 4}+\nu_2^2\big)^2\big({1\over 4}+\nu_3^2\big)^2}
{2\pi\delta(\omega_1-\omega_2-\omega_3)
\big({s\over m_\perp^2}\big)^{\omega_1+\omega_2+\omega_3\over 2}
\over [\omega_1-\aleph(\nu_1)] [\omega_2-\aleph(\nu_2)] [\omega_3-\aleph(\nu_3)]}
\nonumber\\
&&\hspace{3mm}
\times~(w_{1t}^2)^{-\half+i\nu_1}(w_{2t}^2)^{-\half+i\nu_2}(w_{3t}^2)^{-\half+i\nu_3}
\prod_{k=1,2,3}\nu_k
{ 2^{-4i\nu_k}\Gamma\big({3\over 2}+i\nu_k\big)\Gamma(1-i\nu_k)\over \Gamma\big({3\over 2}-i\nu_k\big)\Gamma(1+i\nu_k)}
\nonumber\\
&&\hspace{3mm}
\times~\!\int\! {d^2x_4d^2x_5d^2x_6\over x_{45}^2x_{56}^2x_{46}^2}\Big({x_{45}^2\over x_{14}^2x_{15}^2}\Big)^{\half+i\nu_1}
\Big({x_{56}^2\over x_{25}^2x_{36}^2}\Big)^{\half+i\nu_2}
\Big({x_{46}^2\over x_{34}^2x_{36}^2}\Big)^{\half+i\nu_3}
\label{3poms5}
\end{eqnarray}
Here $2\pi\delta(\omega_1-\omega_2-\omega_3)$ comes from the longitudinal integral 
\begin{eqnarray}
&&\hspace{-1mm}
\int_0^\infty\! {dl_1dl_2dl_3\over l_1l_2l_3}~l_1^{-\omega_1}l_2^{-\omega_2}l_3^{-\omega_3}
\!\int_{-\infty}^{Y_1}\! dY_0\!\int\! dY_2 dY_3~\theta(Y_0+Y_2)\theta(Y_0+Y_3)
\nonumber\\
&&\hspace{-1mm}
\times~e^{\aleph(\nu_1)(Y_1-Y_0)+\aleph(\nu_2)(Y_0+Y_2)+\aleph(\nu_3)(Y_0+Y_3)}
\nonumber\\
&&\hspace{-1mm}
=~{2\pi\delta(\omega_1-\omega_2-\omega_3)
\big({s\over m_\perp^2}\big)^{\omega_1+\omega_2+\omega_3\over 2}
\over [\omega_1-\aleph(\nu_1)] [\omega_2-\aleph(\nu_2)] [\omega_3-\aleph(\nu_3)]}
\!\int\! dY_0~e^{-i(\omega_1-\omega_2-\omega_3)Y_0}
\nonumber\\
&&\hspace{-1mm}
\rightarrow~{2\pi\delta(\omega_1-\omega_2-\omega_3)
\big({s\over m_\perp^2}\big)^{\omega_1+\omega_2+\omega_3\over 2}
\over [\omega_1-\aleph(\nu_1)] [\omega_2-\aleph(\nu_2)] [\omega_3-\aleph(\nu_3)]}
\label{3poms6}
\end{eqnarray}
Strictly speaking, the integral over $Y_0$ is divergent so we need some regularization to understand it. 
Following Ref.  \cite{Balitsky:2015tca}  
we take $n_2\neq n_3$ but $n_1\cdot n_2\simeq n_1\cdot n_3$.  
We can use our  formulas for $n_2=n_3$ case until longitudinal distances between frames
``2'' and ``3'' are smaller than typical transverse separation $x_{ij\perp}^2\sim m_\perp^{-2}$, i.e. when
$l_2l_3s_{23}\leq  m_\perp^{-2}$. In terms of rapidities $Y_2$ and $Y_3$
 this restriction means
$Y_2+Y_3~\leq~ \ln{s_{12}\over s_{23}}$ so instead of Eq. (\ref{3poms6})  we get
\begin{eqnarray}
&&\hspace{-1mm}
\!\int_{-\infty}^{Y_1}\! dY_0\!\int\! dY_2dY_3~\theta(Y_0+Y_2)\theta(Y_0+Y_3)\theta\big( \ln{s_{12}\over s_{23}}-Y_2-Y_3\big)
\nonumber\\
&&\hspace{11mm}
\times~e^{\aleph(\nu_1)(Y_1-Y_0)+\aleph(\nu_2)(Y_0+Y_2)+\aleph(\nu_3)(Y_0+Y_3)}
\nonumber\\
&&\hspace{-1mm}
=~{\big({s_{23}\over s_{12}}\big)^{\omega_2+\omega_3-\omega_1\over 2}
\over  (\omega_1-\omega_2-\omega_3)
(\omega_1-\aleph_1)\big(\omega_2-\aleph_2+{\omega_1-\omega_2-\omega_3\over 2}\big)
\big(\omega_3-\aleph_3+{\omega_1-\omega_2-\omega_3\over 2}\big)}
\nonumber\\
&&\hspace{11mm}
\stackrel{\omega_2+\omega_3\rightarrow\omega_1}{\rightarrow} 
~{\big({s_{23}\over s_{12}}\big)^{\omega_2+\omega_3-\omega_1\over 2}
\over (\omega_1-\omega_2-\omega_3)
(\omega_1-\aleph_1)(\omega_2-\aleph_2)(\omega_3-\aleph_3)}.  
\label{3poms7}
\end{eqnarray}
Thus, 
\begin{equation}
2\pi\delta(\omega_1-\omega_2-\omega_3)~\Leftrightarrow ~\lim_{n_3\rightarrow n_2}
{\big({n_2\cdot n_3\over n_1\cdot n_2}\big)^{\omega_2+\omega_3-\omega_1\over 2}\over (\omega_1-\omega_2-\omega_3)}
\label{3poms8}
\end{equation}
Let us emphasize that the divergence over $Y_0$ in r.h.s of eq. (\ref{3poms7}) leading to this $\delta$-function 
comes from boost invariance: at $n_2=n_3$ one can multiply $n_1$ by some $\lambda$ 
and $n_2$ by $\lambda^{-1}$ and the correlator (\ref{3poms2}) will not change. Thus, the singularity at 
$\omega_1=\omega_2+\omega_3$ is of general nature and should be present in a general formula (\ref{Cgeneral}).
Note, however, that for the correlator of 3 local ``forward'' operators these 
singularities seem to disappear, see the discussion in the end of Sect. \ref{sect:firstorder}.

To compare with the result (\ref{result}) for $n_2\neq n_3$ let us finish the calculation in this $n_2=n_3$ case.
The transverse integral was calculated in Ref. \cite{Korchemsky:1997fy} and the result is 
\begin{eqnarray}
&&\hspace{-1mm}
\!\int\! {d^2x_4d^2x_5d^2x_6\over x_{45}^2x_{56}^2x_{46}^2}\Big({x_{45}^2\over x_{14}^2x_{15}^2}\Big)^{\half+i\nu_1}
\Big({x_{56}^2\over x_{25}^2x_{36}^2}\Big)^{\half+i\nu_2}
\Big({x_{46}^2\over x_{34}^2x_{36}^2}\Big)^{\half+i\nu_3}
\nonumber\\
&&\hspace{-1mm}
=~{\pi^3\Omega\big(\half+i\nu_1,\half+i\nu_2,\half+i\nu_3\big)\over (x_{12}^2)^{\half+i\nu_1+i\nu_2-i\nu_3}
 (x_{13}^2)^{\half+i\nu_1+i\nu_3-i\nu_2} (x_{23}^2)^{\half+i\nu_2+i\nu_3-i\nu_1}}
 \label{3poms9}
\end{eqnarray}
where $\Omega$ is related to Meyer G-function, see the explicit expression in Ref. \cite{Korchemsky:1997fy} 
(for convenience, we extracted factor $\pi^3$ from the definition in Ref. \cite{Korchemsky:1997fy}).

Rewriting the intgral (\ref{3poms9}) in terms of $\gamma_i=2i\nu_i-1$ one obtains
\begin{eqnarray}
&&\hspace{-3mm}
\big\langle \calf^{j_1}_{n_1}\big(x_{1t}+{w_{1t}\over 2},x_{1t}-{w_{1t}\over 2}\big)
\calf^{j_2}_{n_2}\big(x_{2t}+{w_{2t}\over 2},x_{2t}-{w_{2t}\over 2}\big)
\calf^{j_3}_{n_2}\big(x_{3t}+{w_{3t}\over 2},x_{3t}-{w_{3t}\over 2}\big)
\big\rangle
\nonumber\\
&&\hspace{5mm}
=~-i{\alpha_s^5N_c^7\over 64\pi^{10}}{2\pi\delta(\omega_1-\omega_2-\omega_3)\over x_{12t}^2x_{13t}^2 x_{23t}^2}
\!\int_{-1-i\infty}^{-1+i\infty}\! {d\gamma_1d\gamma_2d\gamma_3\over \gamma_2^2\gamma_3^2(2+\gamma_2)^2(2+\gamma_3)^2}
\nonumber\\
&&\hspace{11mm}
\times~
\prod_{k=1}^3{(1+ \gamma_k)2^{-2\gamma_k}\Gamma\big(2+{\gamma_k\over 2}\big)\Gamma\big(\half-{\gamma_k\over 2}\big)
\over \Gamma\big(1-{\gamma_k\over 2}\big)\Gamma\big({3\over 2}+{\gamma_k\over 2}\big)}
{\Omega\big(1-{\gamma_1\over 2},1-{\gamma_2\over 2},1-{\gamma_3\over 2}\big)
\over 
[\omega_1-\taleph(\gamma_1)][\omega_2-\taleph(\gamma_2)][\omega_3-\taleph(\gamma_3)]}
\nonumber\\
&&\hspace{11mm}
\times~
\Big({w_{1t}^2x_{23t}^2\over x_{12t}^2x_{13t}^2}\Big)^{\gamma_1\over 2}
\Big({w_{2t}^2x_{13t}^2\over x_{12t}^2x_{23t}^2}\Big)^{\gamma_2\over 2}
\Big({w_{3t}^2x_{12t}^2\over x_{13t}^2x_{23t}^2}\Big)^{\gamma_3\over 2}\big({s\over m_\perp^2}\big)^{\omega_1+\omega_2+\omega_3\over 2}
\label{3poms10}
\end{eqnarray}
 where $\gamma_i=2i\nu_i-1$ similarly to Eq. (\ref{result4}).
 
Taking residues at $\gamma^\ast_i$ (roots of the equation (\ref{simproot}))  one obtains
\footnote{Similarly to the case of integral (\ref{result4}), the poles at $\gamma_i=0$ should cancel with
contributions of low-order diagrams without gluon ladder(s) in Fig. \ref{fig:3poms}.}
\begin{eqnarray}
&&\hspace{-3mm}
\big\langle \calf^{j_1}_{n_1}\big(x_{1t}+{w_{1t}\over 2},x_{1t}-{w_{1t}\over 2}\big)
\calf^{j_2}_{n_2}\big(x_{2t}+{w_{2t}\over 2},x_{2t}-{w_{2t}\over 2}\big)
\calf^{j_3}_{n_2}\big(x_{3t}+{w_{3t}\over 2},x_{3t}-{w_{3t}\over 2}\big)
\big\rangle
\nonumber\\
&&\hspace{-1mm}
=~8g^{10}{N_c^2\over \pi^2}2\pi\delta(\omega_1-\omega_2-\omega_3)
\big({s\over x_{12t}^2}\big)^{\omega_1+\omega_2-\omega_3\over 2}
\big({s\over x_{13t}^2}\big)^{\omega_1+\omega_3-\omega_2\over 2}
\big({s\over x_{23t}^2}\big)^{\omega_2+\omega_3-\omega_1\over 2}
\nonumber\\
&&\hspace{-1mm}
\times~
\Omega\big(1-{\gamma^\ast_1\over 2},1-{\gamma^\ast_2\over 2},1-{\gamma^\ast_3\over 2}\big){1\over x_{12t}^2x_{13t}^2 x_{23t}^2}
\Big({w_{1t}^2x_{23t}^2\over x_{12t}^2x_{13t}^2}\Big)^{\gamma^\ast_1\over 2}\Big({w_{2t}^2x_{13t}^2\over x_{12t}^2x_{23t}^2}\Big)^{\gamma^\ast_2\over 2}
\Big({w_{3t}^2x_{12t}^2\over x_{13t}^2x_{23t}^2}\Big)^{\gamma^\ast_3\over 2}
\nonumber\\
&&\hspace{-1mm}
\times~
{1\over {\gamma_2^\ast}^2{\gamma_3^\ast}^2\big(1+{\gamma^\ast_2\over 2}\big)^2\big(1+{\gamma_3^\ast\over 2}\big)^2}
\prod_{k=1}^3{(1+ \gamma^\ast_k)2^{-2\gamma^\ast_k}\Gamma\big(2+{\gamma^\ast_k\over 2}\big)\Gamma\big(\half-{\gamma^\ast_k\over 2}\big)
\over \Gamma\big(1-{\gamma^\ast_k\over 2}\big)\Gamma\big({3\over 2}+{\gamma^\ast_k\over 2}\big)\aleph'(\gamma^\ast_k)}
\label{3poms11}
\end{eqnarray}
Here again we replaced $\big({s\over m_\perp^2})^{\omega_1+\omega_2+\omega_3\over 2}$ 
by $\big({s\over x_{12t}^2}\big)^{\omega_1+\omega_2-\omega_3\over 2}
\big({s\over x_{13t}^2}\big)^{\omega_1+\omega_3-\omega_2\over 2}
\big({s\over x_{23t}^2}\big)^{\omega_2+\omega_3-\omega_1\over 2}$ which is within LLA accuracy. 
This result agrees with Eq. (30) from Ref. \cite{Balitsky:2015tca}. In terms of structure constant (\ref{Cgeneral}) we have
\begin{eqnarray}
&&\hspace{-1mm}
F\big[\gamma^\ast(\omega_1,g^2),\gamma^\ast(\omega_2,g^2),\gamma^\ast(\omega_3,g^2)\big]
\label{3poms12}\\
&&\hspace{-1mm}
\stackrel{\omega_1=\omega_2+\omega_3}{=}
8ig^{10}{N_c^2\over \pi^2}
{\Omega\big(1-{\gamma^\ast_1\over 2},1-{\gamma^\ast_2\over 2},1-{\gamma^\ast_3\over 2}\big)\over {\gamma_2^\ast}^2{\gamma_3^\ast}^2\big(1+{\gamma^\ast_2\over 2}\big)^2\big(1+{\gamma_3^\ast\over 2}\big)^2}
\prod_{k=1}^3{(1+ \gamma^\ast_k)2^{-2\gamma^\ast_k}\Gamma\big(2+{\gamma^\ast_k\over 2}\big)\Gamma\big(\half-{\gamma^\ast_k\over 2}\big)
\over \Gamma\big(1-{\gamma^\ast_k\over 2}\big)\Gamma\big({3\over 2}+{\gamma^\ast_k\over 2}\big)\aleph'(\gamma^\ast_k)}
\nonumber
\end{eqnarray}

It is instructive to compare with the result (\ref{result}) at small $\gamma^\ast_i \simeq -{8g^2\over\omega_i}\ll 1$. 
The estimate of the function $\Omega$ at small $\gamma_i$ reads \cite{Balitsky:2015tca,Balitsky:2015oux}
\begin{equation}
\hspace{-1mm}
\Omega\big(1-{\gamma_1\over 2},1-{\gamma_2\over 2},1-{\gamma_3\over 2}\big)
~=~{16\over\gamma_1\gamma_2\gamma_3}\big(1+{\gamma_1+\gamma_2\over\gamma_3}+{\gamma_1+\gamma_3\over\gamma_2}+{\gamma_2+\gamma_3\over\gamma_1}\big)
~+~O(\gamma_i)
\label{3poms13}
\end{equation}
so we get
\begin{eqnarray}
&&\hspace{-1mm}
\big\langle \calf^{j_1}_{n_1}\big(x_{1t}+{w_{1t}\over 2},x_{1t}-{w_{1t}\over 2}\big)
\calf^{j_2}_{n_2}\big(x_{2t}+{w_{2t}\over 2},x_{2t}-{w_{2t}\over 2}\big)
\calf^{j_3}_{n_2}\big(x_{3t}+{w_{3t}\over 2},x_{3t}-{w_{3t}\over 2}\big)
\big\rangle
\label{3poms14}\\
&&\hspace{-1mm}
=~-{g^2N_c^2\over 2\pi^2}\omega_1[2\pi\delta(\omega_1-\omega_2-\omega_3)]
\Big({w_{1t}^2x_{23t}^2\over x_{12t}^2x_{13t}^2}\Big)^{\gamma^\ast_1\over 2}\Big({w_{2t}^2x_{13t}^2\over x_{12t}^2x_{23t}^2}\Big)^{\gamma^\ast_2\over 2}
\Big({w_{3t}^2x_{12t}^2\over x_{13t}^2x_{23t}^2}\Big)^{\gamma^\ast_3\over 2}\big[1+O\big({g^2\over\omega}\big)\big]
\nonumber
\end{eqnarray}
which corresponds to structure constant (\ref{Cgeneral}) with
$$
F~=~-2i\pi g^2
$$
at $\omega_1=\omega_2+\omega_3$. This contribution to structure constant is imaginary in accordance with the fact that 
the physical amplitude in Fig. (\ref{fig:3poms}) is purely imaginary if left and right sides are symmetric. 
(The corresponding cross section describes diffractive scattering, see Refs. \cite{Balitsky:1997mk,Balitsky:2001gj}). 
Since the leading-order structure constant (\ref{Fotvet}) is real, it is natural to assume that Eq. (\ref{3poms12}) gives the leading contribution to the imaginary part of structure
constant at $\omega_1=\omega_2+\omega_3$ in the BFKL limit.

\subsection{Calculation of $\Lambda(\nu_1,\nu_2,\nu_3)$ \label{sect:lambda}}
\begin{figure}[htb]
\begin{center}
\includegraphics[width=77mm]{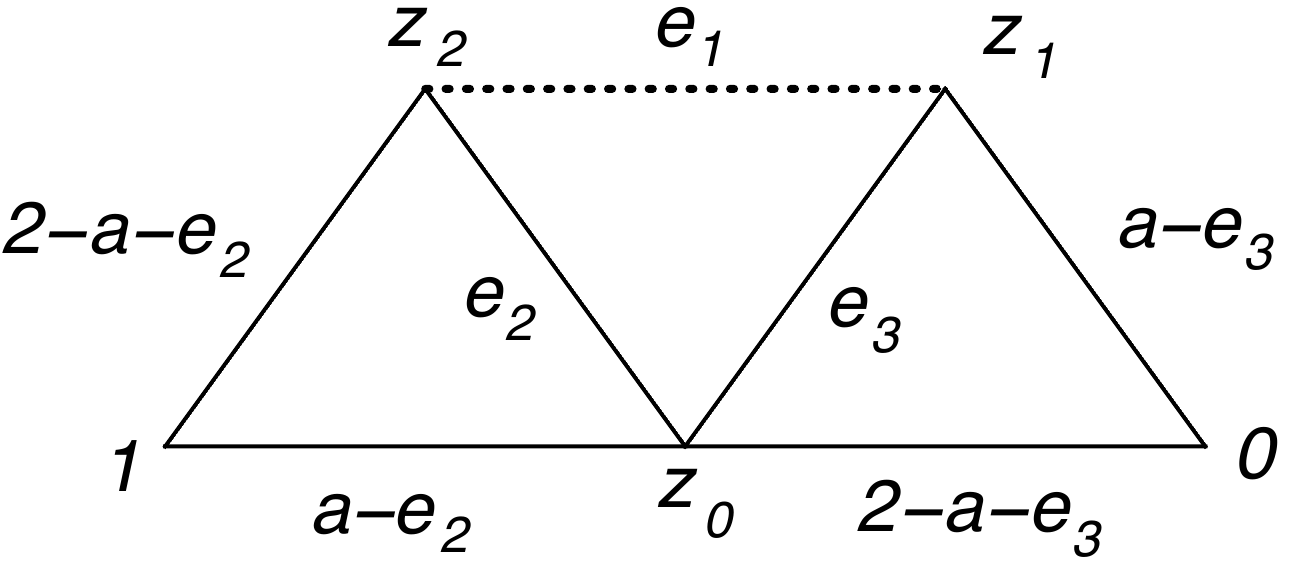}
\end{center}
\caption{Correlator defining the function $\Lambda(\epsilon_i)$ ($\equiv\Lambda(e_i)$) \label{fig:figinteg} in Eq. (\ref{lambda1}). Dotted line depicts the 
unusial
propagator $[z_{12x}^2+(z_{1y}+z_{2y})^2]^{-e_1}$ between points $z_1$ and $z_2$.}
\end{figure}

The function $\Lambda(\nu_1,\nu_2,\nu_3)$ is represented by the integral (\ref{lambda}). 
It is convenient to take $x=(1,0)$ and rewrite the integral as
\begin{eqnarray}
&&\hspace{-1mm}
\Lambda(\nu_1,\nu_2,\nu_3)~\equiv~\bLa(\epsilon_1,\epsilon_2,\epsilon_3)~=~
{1\over\pi^3}\!\int\! d^2z_1{d^2z_2\over (1-z_2)^4}{d^2z_0\over z_0^4}
\nonumber\\
&&\hspace{-1mm}
\times~[z_{12x}^2+(z_{1y}+z_{2y})^2]^{-\epsilon_1}
\Big[{z_{20}^2\over (1-z_2)^2(1-z_0)^2}\Big]^{-\epsilon_2}
\Big[{z_{10}^2\over z_1^2z_0^2}\Big]^{-\epsilon_3}
\label{lambda1}
\end{eqnarray}
where we denote $\epsilon_i\equiv\half-i\nu_i$ in a view of a later estimate at $\epsilon_i\rightarrow 0$. 
Unfortunately, the integral (\ref{lambda1}) diverges as $\epsilon_i\rightarrow 0$ so we need to define it as an analytic continuation of 
a convergent integral
\begin{eqnarray}
&&\hspace{-1mm}
\bLa^{a}(\epsilon_1,\epsilon_2,\epsilon_3)~\equiv~
{1\over\pi^3}\!\int\! d^2z_0d^2z_1d^2z_2
{1\over (|1-z_0|^2)^{a-\epsilon_2}(|1-z_2|^2)^{2-a-\epsilon_2}}
\nonumber\\
&&\hspace{-1mm}
\times~(\barz_2-z_1)^{-\epsilon_1}(z_2-\barz_1)^{-\epsilon_1}
{1\over |z_{20}^2|^{\epsilon_2}|z_1^2|^{a-\epsilon_3}|z_{10}^2|^{\epsilon_3}|z_0^2|^{2-a-\epsilon_3}}
\label{lambda2}
\end{eqnarray}
This integral is obviously convergent if $\epsilon_i>0$ and $|1-a|<\epsilon_i$.  (We will relax the condition $\epsilon_i>0$ later). 
The ``Feynman diagram'' integral is depicted in Fig. \ref{fig:figinteg}
the denominators being conventional 2-dim propagators (albeit with non-integer powers) except $[z_{12x}^2+(z_{1y}+z_{2y})^2]^{-\epsilon_1}$ denoted by a dotted line.

To calculate this integral, we will rewrite  $[z_{12x}^2+(z_{1y}+z_{2y})^2]$ in the denominator as $(\barz_2-z_1)(z_2-\barz_1)$ where
$z\equiv z_x+iz_y,~\barz\equiv z_x-iz_y$ and use the expansion
\begin{equation}
{1\over (\barz_2-z_1)^{\epsilon_1}(z_2-\barz_1)^{\epsilon_1}}~=~
\sum_{m,n=0}^\infty{\Gamma(\epsilon_1+m+n)\over\Gamma(\epsilon_1)m!n!}(1-\barz_2)^mz_1^n
\sum_{k,l=0}^\infty{\Gamma(\epsilon_1+k+l)\over\Gamma(a)k!l!}(1-z_2)^k\barz_1^l
\label{xpanshen}
\end{equation}
Next, we use the expansion (\ref{xpanshen}), calculate the integrals, reassemble the sum over $m,n,k,l$ and continue analytically to $a=0$
in the final result.

Using the integral 
\begin{eqnarray}
&&\hspace{-11mm}
\int\! {d^2z_1\over\pi}
{z_1^m\barz_1^n\over |z_2-z_1|^{2\alpha}|z_1|^{2\beta}}
~
=~{\Gamma(\alpha+\beta-n-1)\over \Gamma(\alpha)\Gamma(\beta-n)|z_2^2|^{\alpha+\beta-n-1}}z_2^{m-n}
{\Gamma(1-\alpha)\Gamma(1-\beta+m)\over\Gamma(2-\alpha-\beta+m)}
\label{integral1}
\end{eqnarray}
one obtains after some algebra
\begin{eqnarray}
&&\hspace{-1mm}
\bLa^{a}(\epsilon_1,\epsilon_2,\epsilon_3)~
\nonumber\\
&&\hspace{-1mm}=~{\sin\pi(a+\epsilon_2)\sin\pi\epsilon_2\sin\pi(\epsilon_3-a)\sin\pi\epsilon_3\over \pi^3\sin\pi(\epsilon_2+\epsilon_3)}
\!\!\sum_{m,n,k,l=0}^\infty{\Gamma(\epsilon_1+m+n)\over\Gamma(\epsilon_1)m!n!}
{\Gamma(\epsilon_1+k+l)\over\Gamma(\epsilon_1)k!l!}
\nonumber\\
&&\hspace{-1mm}
\times~(-1)^{m+l}{\Gamma(a+\epsilon_2-1+m)\Gamma(1-a-m)\over \Gamma(\epsilon_2)}
{\Gamma(1-\epsilon_2)\Gamma(\epsilon_2+a-1+k)\over\Gamma(a+k)}
\nonumber\\
&&\hspace{-1mm}
\times~
{\Gamma(\epsilon_3+l+1-a)\Gamma(a-l-1)\over \Gamma(\epsilon_3)}
{\Gamma(1-\epsilon_3)\Gamma(1+\epsilon_3-a+n)\over\Gamma(2-a+n)}
\nonumber\\
&&\hspace{-1mm}
\times~{\Gamma(\epsilon_2+m)\Gamma(\epsilon_3+l)\over \Gamma(\epsilon_2+\epsilon_3+m+l)}
{\Gamma(\epsilon_2+k)\Gamma(\epsilon_3+n)\over\Gamma(\epsilon_2+\epsilon_3+k+n)}
\nonumber\\
&&\hspace{-1mm}
=~
-{\sin\pi(a+\epsilon_2)\sin\pi\epsilon_2\sin\pi(\epsilon_3-a)\sin\pi\epsilon_3\over \pi^3\sin\pi(\epsilon_2+\epsilon_3)}
\sum_{m,n=0}^\infty{\Gamma(\epsilon_1+m+n)\over\Gamma(\epsilon_1)m!n!}
\sum_{k,l=0}^\infty{\Gamma(\epsilon_1+k+l)\over\Gamma(\epsilon_1)k!l!}
\nonumber\\
&&\hspace{-1mm}
\times~(-1)^{m+l}\!\int_0^1 \! du_1dv_1du_2dv_2dt_1dt_2~u_1^{a+\epsilon_2-2+m}(1-u_1)^{-a-m}(1-v_1)^{-\epsilon_2}v_1^{\epsilon_2+a-2+k}
u_2^{\epsilon_3-a+l}
\nonumber\\
&&\hspace{-1mm}
\times~
(1-u_2)^{a-2-l}(1-v_2)^{-\epsilon_3}v_2^{\epsilon_3-a+n}
t_1^{\epsilon_2-1+m}(1-t_1)^{\epsilon_3-1+l}t_2^{\epsilon_2-1+k}(1-t_2)^{\epsilon_3-1+n}
\end{eqnarray}
Now one can reassemble the sum (\ref{xpanshen}) and get
\begin{eqnarray}
&&\hspace{-1mm}
\bLa^a(\epsilon_1,\epsilon_2,\epsilon_3)~=~
-{\sin\pi(a+\epsilon_2)\sin\pi\epsilon_2\sin\pi(\epsilon_3-a)\sin\pi\epsilon_3\over \pi^3\sin\pi(\epsilon_2+\epsilon_3)}
\int_0^1 \! du_1dv_1du_2dv_2dt_1dt_2~
\nonumber\\
&&\hspace{-1mm}
\times~
u_1^{a+\epsilon_2-2}(1-u_1)^{\epsilon_1-a}(1-v_1)^{-\epsilon_2}v_1^{\epsilon_2+a-2}
u_2^{\epsilon_3-a}(1-u_2)^{\epsilon_1+a-2}(1-v_2)^{-\epsilon_3}v_2^{\epsilon_3-a}
\nonumber\\
&&\hspace{-1mm}
\times~t_1^{\epsilon_2-1}(1-t_1)^{\epsilon_3-1}t_2^{\epsilon_2-1}(1-t_2)^{\epsilon_3-1}
\big[(1-u_1)(1-v_2)+u_1t_1+( 1-u_1)v_2t_2\big]^{-\epsilon_1}
\nonumber\\
&&\hspace{11mm}
\times~
\big[(1-u_2)(1-v_1)+u_2(1-t_1)+( 1-u_2)v_1(1-t_2)\big]^{-\epsilon_1}
\label{lambda3}
\end{eqnarray}
Next, using Mellin-Barnes integral
\begin{eqnarray}
&&\hspace{-1mm}
{\Gamma(a)\over (A+B+C)^a}~=~{1\over (2\pi i)^2}\!\int\! ds_1ds_2~{\Gamma(a-s_1)\over A^{a-s_1}}{\Gamma(s_1-s_2)\over B^{s_1-s_2}}{\Gamma(s_2)\over C^{s_2}}
\label{MB}
\end{eqnarray}
one can rewrite Eq. (\ref{lambda3}) as follows
\begin{eqnarray}
&&\hspace{-1mm}
\bLa^a(\epsilon_1,\epsilon_2,\epsilon_3)~
=~-{\sin\pi(a+\epsilon_2)\sin\pi(\epsilon_2)\sin\pi(\epsilon_3-a)\sin\pi(\epsilon_3)\over \pi^3\sin\pi(\epsilon_2+\epsilon_3)\Gamma^2(\epsilon_1)}I^a(\epsilon_1,\epsilon_2,\epsilon_3)
\end{eqnarray}
where $I^a(\epsilon_1,\epsilon_2,\epsilon_3)=I^{a,a,a,a}(\epsilon_1,\epsilon_2,\epsilon_3)$ and 
\begin{eqnarray}
&&\hspace{-1mm}
I^{a_1,a_2,a_3,a_4}(\epsilon_1,\epsilon_2,\epsilon_3)~
\label{begin}\\
&&\hspace{-1mm}
=~\!\int_{\delta-i\infty}^{\delta+i\infty}\! {ds_1ds_2ds_3ds_4\over (2\pi i)^4}~
\Gamma(\epsilon_1-s_1-s_2)\Gamma(s_1)\Gamma(s_2)\Gamma(\epsilon_1-s_3-s_4)\Gamma(s_3)\Gamma(s_4)
\nonumber\\
&&\hspace{-1mm}
\times~{\Gamma(\epsilon_2-1+a_3-s_1)\Gamma(s_1-a_3+1)
\over\Gamma(\epsilon_2)}~
{\Gamma(1+\epsilon_3-a_1-s_3)\Gamma(a_1-1+s_3)
\over\Gamma(\epsilon_3)}~
\nonumber\\
&&\hspace{-1mm}
\times~
{\Gamma(\epsilon_2-1+a_2-s_4)\Gamma(1-\epsilon_1-\epsilon_2+s_3+s_4)\over\Gamma(a_2-\epsilon_1+s_3)}
\nonumber\\
&&\hspace{-1mm}
\times~{\Gamma(1+\epsilon_3-a_4-s_2)\Gamma(1-\epsilon_1-\epsilon_3+s_1+s_2)\over\Gamma(2-a_4-\epsilon_1+s_1)}
{\Gamma(\epsilon_2-s_1)\Gamma(\epsilon_3-s_3)\over \Gamma(\epsilon_2+\epsilon_3-s_1-s_3)}
{\Gamma(\epsilon_2-s_2)\Gamma(\epsilon_3-s_4)\over \Gamma(\epsilon_2+\epsilon_3-s_2-s_4)}
\nonumber
\end{eqnarray}
Here we assume $|1-a_i|<\delta<{\epsilon_i\over 2}$, then the MB integral is well-defined with all the ``left'' poles of the type $\Gamma(s_i+...)$ to the left of the contour of integration
over $s_i$ and ``right'' poles $\sim\Gamma(...-s_i)$ to the right of the contour. 

Next, we need to continue analytically to $a_i=0$. We will do this separately for each $a_i$ paying attention to 
the poles which intersect the contour of integration and taking residues in those poles as explained in the book \cite{Smirnov:2006ry}.
First, note that analytic continuation in $a_4$ is trivial: ``right'' pole at $s_2=1+\epsilon_3-a_4$ moves to the right and away from the contour. 
Thus, we set $a_4=0$ in what follows. At a next step,
we continue $a_1$ to $a_1=0$. There are two poles affected by that: pole at $s_3=1+\epsilon_3-a_1$ and pole at $s_3=1-a_1$. While the first 
pole is always to the right of the contour, the second pole intersects the contour so we need to take a residue at $s_3=1-a_1$. The integral 
$I^{a_1,a_2,a_3,0}$ at $a_1\leq1-\delta$ takes the form
\begin{eqnarray}
&&\hspace{-1mm}
I^{a_1,a_2,a_3,0}~
=~{\rm r.h.s.~of~Eq. (\ref{begin})}\Big|_{a_1=a_4=0}
\nonumber\\
&&\hspace{-1mm}
+~\Gamma(1-a_1)\!\int_{\delta-i\infty}^{\delta+i\infty}\!{ds_1ds_2ds_4\over (2\pi i)^3}
\Gamma(\epsilon_1-s_1-s_2)\Gamma(s_1)\Gamma(s_2)\Gamma(a_1+\epsilon_1-1-s_4)\Gamma(s_4)
\label{a1cont}\\
&&\hspace{-1mm}
\times~{\Gamma(\epsilon_2-1+a_3-s_1)\Gamma(s_1-a_3+1)
\over\Gamma(\epsilon_2)}{\Gamma(\epsilon_2-1+a_2-s_4)\Gamma(2-a_1-\epsilon_1-\epsilon_2+s_4)\over\Gamma(1-a_1+a_2-\epsilon_1)}
\nonumber\\
&&\hspace{-1mm}
\times~{\Gamma(1+\epsilon_3-s_2)\Gamma(1-\epsilon_1-\epsilon_3+s_1+s_2)\over\Gamma(2-\epsilon_1+s_1)}
{\Gamma(\epsilon_2-s_1)\Gamma(a_1+\epsilon_3-1)\over \Gamma(a_1+\epsilon_2+\epsilon_3-s_1-1)}
{\Gamma(\epsilon_2-s_2)\Gamma(\epsilon_3-s_4)\over \Gamma(\epsilon_2+\epsilon_3-s_2-s_4)}
\nonumber
\end{eqnarray}
Now we should continue from $a_1=1-\delta$ to $a_1=0$. In the first term in r.h.s. of Eq. (\ref{a1cont}) there are no more
crossings of the contour so we can just set $a_1=0$. In the second term, the pole at $s_4=a_1+\epsilon_1+\epsilon_2-2$ 
will always stay to the left of the cut but the pole at $s_4=a_1+\epsilon_1-1$ will move from $s_4=\epsilon_1-\delta$ to $s_4=\epsilon_1-1$ 
so it will cross the contour and we need to take the residue. The residue yields
\begin{eqnarray}
&&\hspace{-1mm}
\Gamma(1-a_1)\!\int_{\delta-i\infty}^{\delta+i\infty}\!{ds_1ds_2\over (2\pi i)^2}
\Gamma(\epsilon_1-s_1-s_2)\Gamma(s_1)\Gamma(s_2)\Gamma(a_1+\epsilon_1-1)
\label{a1conti}\\
&&\hspace{11mm}
\times~{\Gamma(\epsilon_2-1+a_3-s_1)\Gamma(s_1-a_3+1)
\over\Gamma(\epsilon_2)}{\Gamma(\epsilon_2-\epsilon_1+a_2-a_1)\Gamma(2-a_1-\epsilon_1-\epsilon_2+s_4)\over\Gamma(1-a_1+a_2-\epsilon_1)}
\nonumber\\
&&\hspace{11mm}
\times~{\Gamma(1+\epsilon_3-s_2)\Gamma(1-\epsilon_1-\epsilon_3+s_1+s_2)\over\Gamma(2-\epsilon_1+s_1)}
{\Gamma(\epsilon_2-s_1)\Gamma(a_1+\epsilon_3-1)\over \Gamma(a_1+\epsilon_2+\epsilon_3-s_1-1)}
\nonumber\\
&&\hspace{77mm}
\times~
{\Gamma(\epsilon_2-s_2)\Gamma(1-\epsilon_1-\epsilon_3-a_1)\over \Gamma(1+\epsilon_2+\epsilon_3-\epsilon_1-s_2-a_1)}
\nonumber
\end{eqnarray}
The continuation $a_1\rightarrow 0$ in Eq. (\ref{a1conti}) does not cross the integration contours so we can set 
$a_1=0$ and get for Eq. (\ref{begin})
\begin{eqnarray}
&&\hspace{-1mm}
I^{a_1\rightarrow 0,a_2,a_3,0}(\epsilon_1,\epsilon_2,\epsilon_3)~
\\
&&\hspace{-1mm}=~\!\int_{\delta-i\infty}^{\delta+i\infty}\! {ds_1ds_2ds_3ds_4\over (2\pi i)^4}~
\Gamma(\epsilon_1-s_1-s_2)\Gamma(s_1)\Gamma(s_2)\Gamma(\epsilon_1-s_3-s_4)\Gamma(s_3)\Gamma(s_4)
\nonumber\\
&&\hspace{-1mm}
\times~{\Gamma(\epsilon_2-1+a_3-s_1)\Gamma(s_1-a_3+1)
\over\Gamma(\epsilon_2)}~
{\Gamma(1+\epsilon_3-s_3)\Gamma(s_3-1)
\over\Gamma(\epsilon_3)}~
\nonumber\\
&&\hspace{-1mm}
\times~
{\Gamma(\epsilon_2-1+a_2-s_4)\Gamma(1-\epsilon_1-\epsilon_2+s_3+s_4)\over\Gamma(a_2-\epsilon_1+s_3)}
\nonumber\\
&&\hspace{-1mm}
\times~{\Gamma(1+\epsilon_3-s_2)\Gamma(1-\epsilon_1-\epsilon_3+s_1+s_2)
\over\Gamma(2-\epsilon_1+s_1)}
{\Gamma(\epsilon_2-s_1)\Gamma(\epsilon_3-s_3)\over \Gamma(\epsilon_2+\epsilon_3-s_1-s_3)}
{\Gamma(\epsilon_2-s_2)\Gamma(\epsilon_3-s_4)\over \Gamma(\epsilon_2+\epsilon_3-s_2-s_4)}
\nonumber\\
&&\hspace{-1mm}
+~\!\int_{\delta-i\infty}^{\delta+i\infty}\!{ds_1ds_2ds_4\over (2\pi i)^3}
\Gamma(\epsilon_1-s_1-s_2)\Gamma(s_1)\Gamma(s_2)\Gamma(\epsilon_1-1-s_4)\Gamma(s_4)
\nonumber\\
&&\hspace{-1mm}
\times~{\Gamma(\epsilon_2-1+a_3-s_1)\Gamma(s_1-a_3+1)
\over\Gamma(\epsilon_2)}{\Gamma(\epsilon_2-1+a_2-s_4)\Gamma(2-\epsilon_1-\epsilon_2+s_4)
\over\Gamma(1+a_2-\epsilon_1)}
\nonumber\\
&&\hspace{-1mm}
\times~{\Gamma(1+\epsilon_3-s_2)\Gamma(1-\epsilon_1-\epsilon_3+s_1+s_2)\over\Gamma(2-\epsilon_1+s_1)}
{\Gamma(\epsilon_2-s_1)\Gamma(\epsilon_3-1)\over \Gamma(\epsilon_2+\epsilon_3-s_1-1)}
{\Gamma(\epsilon_2-s_2)\Gamma(\epsilon_3-s_4)\over \Gamma(\epsilon_2+\epsilon_3-s_2-s_4)}
\nonumber\\
&&\hspace{-1mm}
+~\!\int_{\delta-i\infty}^{\delta+i\infty}\!{ds_1ds_2\over (2\pi i)^2}
\Gamma(\epsilon_1-s_1-s_2)\Gamma(s_1)\Gamma(s_2)\Gamma(\epsilon_1-1)
\nonumber\\
&&\hspace{-1mm}
\times~{\Gamma(\epsilon_2-1+a_3-s_1)\Gamma(s_1-a_3+1)
\over\Gamma(\epsilon_2)}{\Gamma(\epsilon_2-\epsilon_1+a_2)\Gamma(2-\epsilon_1-\epsilon_2+s_4)
\over\Gamma(1+a_2-\epsilon_1)}
\nonumber\\
&&\hspace{-1mm}
\times~{\Gamma(1+\epsilon_3-s_2)\Gamma(1-\epsilon_1-\epsilon_3+s_1+s_2)\over\Gamma(2-\epsilon_1+s_1)}
{\Gamma(\epsilon_2-s_1)\Gamma(\epsilon_3-1)\over \Gamma(\epsilon_2+\epsilon_3-s_1-1)}
{\Gamma(\epsilon_2-s_2)\Gamma(1-\epsilon_1-\epsilon_3)
\over \Gamma(1+\epsilon_2+\epsilon_3-\epsilon_1-s_2)}
\nonumber
\end{eqnarray}
Repeating this procedure for $a_2$ and $a_3$ one obtains after some algebra
\begin{eqnarray}
&&\hspace{-1mm}
\bLa(\epsilon_1,\epsilon_2,\epsilon_3)~=~-
{\sin^2\pi\epsilon_2\sin^2\pi\epsilon_3\over \pi^3\sin\pi(\epsilon_2+\epsilon_3)\Gamma^2(\epsilon_1)}
I^{a_1\rightarrow 0,a_2\rightarrow 0,a_3\rightarrow 0,0}(\epsilon_1,\epsilon_2,\epsilon_3)
\label{bLaresult}
\end{eqnarray}
\begin{eqnarray}
&&\hspace{-1mm}
I^{a_1\rightarrow 0,a_2\rightarrow 0,a_3\rightarrow 0,0}(\epsilon_1,\epsilon_2,\epsilon_3)
~=~J_1(\epsilon_i)+J_2(\epsilon_i)+...+J_{14}(\epsilon_i)
\label{sumj}
\end{eqnarray}
where 
\begin{eqnarray}
&&\hspace{-3mm}
J_1
~=~\!\int_{\delta-i\infty}^{\delta+i\infty}\! {ds_1ds_2ds_3ds_4\over (2\pi i)^4}~
\Gamma(\epsilon_1-s_1-s_2)\Gamma(s_1)\Gamma(s_2)\Gamma(\epsilon_1-s_3-s_4)\Gamma(s_3)\Gamma(s_4)
\nonumber\\
&&\hspace{-3mm}
\times~{\Gamma(\epsilon_2-1-s_1)\Gamma(s_1+1)
\over\Gamma(\epsilon_2)}~
{\Gamma(1+\epsilon_3-s_3)\Gamma(s_3-1)
\over\Gamma(\epsilon_3)}~
{\Gamma(\epsilon_2-1-s_4)\Gamma(1-\epsilon_1-\epsilon_2+s_3+s_4)\over\Gamma(-\epsilon_1+s_3)}
\nonumber\\
&&\hspace{-3mm}
\times~{\Gamma(1+\epsilon_3-s_2)\Gamma(1-\epsilon_1-\epsilon_3+s_1+s_2)\over\Gamma(2-\epsilon_1+s_1)}
{\Gamma(\epsilon_2-s_1)\Gamma(\epsilon_3-s_3)\over \Gamma(\epsilon_2+\epsilon_3-s_1-s_3)}
{\Gamma(\epsilon_2-s_2)\Gamma(\epsilon_3-s_4)\over \Gamma(\epsilon_2+\epsilon_3-s_2-s_4)}
\label{j1234}
\end{eqnarray}
%
\begin{eqnarray}
&&\hspace{-1mm}
J_2~=~{\Gamma(\epsilon_2-1)\Gamma(1+\epsilon_3-\epsilon_2)\over\Gamma(\epsilon_2)\Gamma(\epsilon_3)}
\!\int_{\delta-i\infty}^{\delta+i\infty}\! {ds_1ds_2ds_3\over(2\pi i)^3}~
\Gamma(\epsilon_1-s_1-s_2)\Gamma(s_1)\Gamma(s_2)
\nonumber\\
&&\hspace{-1mm}
\times~
\Gamma(1+\epsilon_1-\epsilon_2-s_3)\Gamma(s_3)
\Gamma(\epsilon_2-1-s_1)\Gamma(s_1+1)
\Gamma(1+\epsilon_3-s_3)\Gamma(s_3-1)
\label{j123}\\
&&\hspace{-1mm}
\times~{\Gamma(1+\epsilon_3-s_2)\Gamma(1-\epsilon_1-\epsilon_3+s_1+s_2)\over\Gamma(2-\epsilon_1+s_1)}
{\Gamma(\epsilon_2-s_1)\Gamma(\epsilon_3-s_3)\over \Gamma(\epsilon_2+\epsilon_3-s_1-s_3)}
{\Gamma(\epsilon_2-s_2)\over \Gamma(1+\epsilon_3-s_2)}
\nonumber
\end{eqnarray}
%
\begin{eqnarray}
&&\hspace{-1mm}
J_3~=~{\Gamma(\epsilon_3-1)\over\Gamma(\epsilon_2)}
\!\int_{\delta-i\infty}^{\delta+i\infty}\!  {ds_1ds_2ds_4\over(2\pi i)^3}~
\Gamma(\epsilon_1-s_1-s_2)\Gamma(s_1)\Gamma(s_2)\Gamma(\epsilon_1-1-s_4)\Gamma(s_4)
\nonumber\\
&&\hspace{-1mm}
\times~\Gamma(\epsilon_2-1-s_1)\Gamma(s_1+1)~
{\Gamma(\epsilon_2-1-s_4)\Gamma(2-\epsilon_1-\epsilon_2+s_4)
\over\Gamma(1-\epsilon_1)}
{\Gamma(1+\epsilon_3-s_2)\over\Gamma(2-\epsilon_1+s_1)}
\nonumber\\
&&\hspace{-1mm}
\times~\Gamma(1-\epsilon_1-\epsilon_3+s_1+s_2)
{\Gamma(\epsilon_2-s_1)\over \Gamma(\epsilon_2+\epsilon_3-s_1-1)}
{\Gamma(\epsilon_2-s_2)\Gamma(\epsilon_3-s_4)\over \Gamma(\epsilon_2+\epsilon_3-s_2-s_4)}
\label{j124}
\end{eqnarray}
%

\begin{eqnarray}
&&\hspace{-1mm}
J_4~=~{\Gamma(\epsilon_2-1)\over\Gamma(\epsilon_3)\Gamma(1-\epsilon_1+\epsilon_2)}
\!\int_{\delta-i\infty}^{\delta+i\infty}\! {ds_2ds_3ds_4\over(2\pi i)^3}~
\Gamma(1+\epsilon_1-\epsilon_2-s_2)
\Gamma(s_2)\Gamma(\epsilon_1-s_3-s_4)
\nonumber\\
&&\hspace{-1mm}
\times~\Gamma(s_3)\Gamma(s_4)
\Gamma(1+\epsilon_3-s_3)\Gamma(s_3-1)~
{\Gamma(\epsilon_2-1-s_4)\Gamma(1-\epsilon_1-\epsilon_2+s_3+s_4)
\over(\epsilon_3-s_3)\Gamma(-\epsilon_1+s_3)}
\nonumber\\
&&\hspace{-1mm}
\times~
\Gamma(1+\epsilon_3-s_2)\Gamma(\epsilon_2-\epsilon_1-\epsilon_3+s_2)
{\Gamma(\epsilon_2-s_2)\Gamma(\epsilon_3-s_4)\over \Gamma(\epsilon_2+\epsilon_3-s_2-s_4)}
\label{j234}
\end{eqnarray}
%
\begin{eqnarray}
&&\hspace{-1mm}
J_5~=~-{\Gamma(\epsilon_1-\epsilon_2)\over(1-\epsilon_2)}\Gamma(\epsilon_3-1)
\Gamma(1+\epsilon_3-\epsilon_2)
\!\int_{\delta-i\infty}^{\delta+i\infty}\!  {ds_1ds_2\over(2\pi i)^2}~
\Gamma(\epsilon_1-s_1-s_2)\Gamma(s_1)\Gamma(s_2)
\nonumber\\
&&\hspace{-1mm}
\times~\Gamma(\epsilon_2-1-s_1)\Gamma(s_1+1)
{\Gamma(1+\epsilon_3-s_2)\Gamma(1-\epsilon_1-\epsilon_3+s_1+s_2)
\Gamma(\epsilon_2-s_1)\Gamma(\epsilon_2-s_2)
\over\Gamma(2-\epsilon_1+s_1)
\Gamma(\epsilon_2+\epsilon_3-s_1-1)\Gamma(1+\epsilon_3-s_2)}
\label{j12}
\end{eqnarray}
%
\begin{eqnarray}
&&\hspace{-1mm}
J_6~=~{\Gamma(\epsilon_2-\epsilon_1)\Gamma(1-\epsilon_2)\Gamma(1+\epsilon_3-\epsilon_1)
\over\Gamma(\epsilon_2)\Gamma(1-\epsilon_1)}\Gamma(\epsilon_1-1)\Gamma(\epsilon_3-1)
\label{j12(3)}\\
&&\hspace{-1mm}
\times~
\!\int_{\delta-i\infty}^{\delta+i\infty}\!  {ds_1ds_2\over(2\pi i)^2}~\Gamma(\epsilon_1-s_1-s_2)
\Gamma(s_1)\Gamma(s_2)
\Gamma(\epsilon_2-1-s_1)\Gamma(s_1+1)
\nonumber\\
&&\hspace{-1mm}
\times~{\Gamma(1+\epsilon_3-s_2)\Gamma(1-\epsilon_1-\epsilon_3+s_1+s_2)
\Gamma(\epsilon_2-s_1)\Gamma(\epsilon_2-s_2)
\over\Gamma(2-\epsilon_1+s_1) \Gamma(\epsilon_2+\epsilon_3-s_1-1)
\Gamma(1+\epsilon_2+\epsilon_3-\epsilon_1-s_2)}
\nonumber
\end{eqnarray}
%
\begin{eqnarray}
&&\hspace{-1mm}
J_7~=~{\Gamma^2(\epsilon_2-1)\Gamma(1+\epsilon_3-\epsilon_2)
\over\Gamma(\epsilon_3)\Gamma(1+\epsilon_2-\epsilon_1)}
\!\int_{\delta-i\infty}^{\delta+i\infty}\! {ds_2ds_3\over(2\pi i)^2}~
\Gamma(1+\epsilon_1-\epsilon_2-s_2)
\label{j23}\\
&&\hspace{-1mm}
\times~\Gamma(s_2)\Gamma(1+\epsilon_1-\epsilon_2-s_3)\Gamma(s_3)
\Gamma(1+\epsilon_3-s_3)\Gamma(s_3-1)~
\nonumber\\
&&\hspace{-1mm}
\times~\Gamma(1+\epsilon_3-s_2)\Gamma(\epsilon_2-\epsilon_1-\epsilon_3+s_2)
{\Gamma(\epsilon_3-s_3)\over \Gamma(1+\epsilon_3-s_3)}
{\Gamma(\epsilon_2-s_2)\over \Gamma(1+\epsilon_3-s_2)}
\nonumber
\end{eqnarray}
%
\begin{eqnarray}
&&\hspace{-1mm}
J_8~=~-\!\int_{\delta-i\infty}^{\delta+i\infty}\!  {ds_2ds_4\over(2\pi i)^2}~
\Gamma(1+\epsilon_1-\epsilon_2-s_2)
\nonumber\\
&&\hspace{-1mm}
\times~\Gamma(\epsilon_2-1)\Gamma(s_2)\Gamma(\epsilon_1-1-s_4)\Gamma(s_4)
{\Gamma(\epsilon_2-1-s_4)\Gamma(2-\epsilon_1-\epsilon_2+s_4)
\over(1-\epsilon_3)\Gamma(1-\epsilon_1)}
\nonumber\\
&&\hspace{-1mm}
\times~
{\Gamma(1+\epsilon_3-s_2)\Gamma(\epsilon_2-\epsilon_1-\epsilon_3+s_2)\over\Gamma(2-\epsilon_1+s_1)}
{\Gamma(\epsilon_2-s_2)\Gamma(\epsilon_3-s_4)\over \Gamma(\epsilon_2+\epsilon_3-s_2-s_4)}
\label{j24}
\end{eqnarray}
%

\begin{eqnarray}
&&\hspace{-1mm}
J_9~=~
{\Gamma(\epsilon_2-1)\Gamma(\epsilon_1+\epsilon_3-\epsilon_2)\over\Gamma(\epsilon_3)}
\Gamma(2\epsilon_2-\epsilon_1-\epsilon_3)
\!\int_{\delta-i\infty}^{\delta+i\infty}\! {ds_3ds_4\over(2\pi i)^2}~
\Gamma(\epsilon_1-s_3-s_4)\Gamma(s_3)
\nonumber\\
&&\hspace{-1mm}
\times~\Gamma(s_4)\Gamma(\epsilon_3-s_3)\Gamma(s_3-1)
{\Gamma(\epsilon_2-1-s_4)\Gamma(1-\epsilon_1-\epsilon_2+s_3+s_4)\Gamma(\epsilon_3-s_4)
\over
\Gamma(-\epsilon_1+s_3) \Gamma(2\epsilon_2-\epsilon_1-s_4)}
\label{j34}
\end{eqnarray}
%

\begin{eqnarray}
&&\hspace{-1mm}
J_{10}~=~-{\Gamma^2(\epsilon_2-1)\Gamma(\epsilon_1-\epsilon_2)
\over (1-\epsilon_3)\Gamma(1+\epsilon_2-\epsilon_1)}\Gamma(1+\epsilon_3-\epsilon_2)
\!\int_{\delta-i\infty}^{\delta+i\infty}\!  {ds_2\over 2\pi i}~
\Gamma(1+\epsilon_1-\epsilon_2-s_2)\Gamma(s_2)
\nonumber\\
&&\hspace{-1mm}
\times~
\Gamma(1+\epsilon_3-s_2)\Gamma(\epsilon_2-\epsilon_1-\epsilon_3+s_2)
{\Gamma(\epsilon_2-s_2)\over\Gamma(1+\epsilon_3-s_2)}
\label{j2}
\end{eqnarray}
%
\begin{eqnarray}
&&\hspace{-1mm}
J_{11}~=~-\Gamma(\epsilon_1-1)\Gamma(\epsilon_2-1)
{\Gamma(\epsilon_2-\epsilon_1)\Gamma(1-\epsilon_2)\over(1-\epsilon_3)\Gamma(1-\epsilon_1)}
\!\int_{\delta-i\infty}^{\delta+i\infty}\!  {ds_2\over 2\pi i}~
\Gamma(1+\epsilon_1-\epsilon_2-s_2)\Gamma(s_2)
\nonumber\\
&&\hspace{-1mm}
\times~
{\Gamma(1+\epsilon_3-s_2)\Gamma(\epsilon_2-\epsilon_1-\epsilon_3+s_2)
\over\Gamma(1+\epsilon_2-\epsilon_1)}
{\Gamma(\epsilon_2-s_2)\Gamma(1+\epsilon_3-\epsilon_1)\over \Gamma(1+\epsilon_2+\epsilon_3-\epsilon_1-s_2)}
\label{j2(3)}
\end{eqnarray}
%
\begin{eqnarray}
&&\hspace{-1mm}
J_{12}~=~\Gamma^2(\epsilon_2-1)
\Gamma(\epsilon_1+\epsilon_3-\epsilon_2)
{\Gamma(2\epsilon_2-\epsilon_1-\epsilon_3)\Gamma(1+\epsilon_3-\epsilon_2)
\over \Gamma(\epsilon_3)\Gamma(1-\epsilon_1+\epsilon_2)}
\label{j3}\\
&&\hspace{-1mm}
\times~
\!\int_{\delta-i\infty}^{\delta+i\infty}\! {ds_3\over 2\pi i}~
\Gamma(1+\epsilon_1-\epsilon_2-s_3)\Gamma(s_3)
\Gamma(1+\epsilon_3-s_3)\Gamma(s_3-1)
{\Gamma(\epsilon_3-s_3)\over \Gamma(1+\epsilon_3-s_3)}
\nonumber
\end{eqnarray}
%
\begin{eqnarray}
&&\hspace{-1mm}
J_{13}~=~
-{\Gamma(1-\epsilon_3)\Gamma(\epsilon_2-1)\Gamma(\epsilon_1+\epsilon_3-\epsilon_2)
\over(1-\epsilon_3)\Gamma(1-\epsilon_1)}
\!\int_{\delta-i\infty}^{\delta+i\infty}\!  {ds_4\over 2\pi i}~
\Gamma(\epsilon_1-1-s_4)\Gamma(s_4)
\nonumber\\
&&\hspace{-1mm}
\times~
\Gamma(\epsilon_2-1-s_4)\Gamma(2-\epsilon_1-\epsilon_2+s_4)
{\Gamma(2\epsilon_2-\epsilon_1-\epsilon_3)\Gamma(\epsilon_3-s_4)
\over \Gamma(2\epsilon_2-\epsilon_1-s_4)}
\label{j4}
\end{eqnarray}
%
\begin{eqnarray}
&&\hspace{-1mm}
J_{14}~=~
-\Gamma(1-\epsilon_3)\Gamma^2(\epsilon_2-1)\Gamma(\epsilon_1+\epsilon_3-\epsilon_2)\Gamma(\epsilon_1-\epsilon_2)
{\Gamma(2\epsilon_2-\epsilon_1-\epsilon_3))\Gamma(1+\epsilon_3-\epsilon_2)\over
(1-\epsilon_3)\Gamma(1+\epsilon_2-\epsilon_1)}
\nonumber\\
\label{j}
\end{eqnarray}
%
\begin{eqnarray}
&&\hspace{-1mm}
J_{15}~=~-
\Gamma(1-\epsilon_3)\Gamma(\epsilon_2-1)\Gamma(\epsilon_1+\epsilon_3-\epsilon_2)\Gamma(\epsilon_1-1)
\nonumber\\
&&\hspace{-1mm}
\times~{\Gamma(\epsilon_2-\epsilon_1)\Gamma(1-\epsilon_2)\over\Gamma(1-\epsilon_1)}
{\Gamma(2\epsilon_2-\epsilon_1-\epsilon_3)\Gamma(1+\epsilon_3-\epsilon_1)\over (1-\epsilon_3)\Gamma(1+2\epsilon_2-2\epsilon_1)}
\end{eqnarray}
The combination of Eq. (\ref{bLaresult}) and (\ref{sumj}) is the final result for the function $\tiLa(\gamma_i)$ (our notation is $\gamma_i=-2\epsilon_i$).
Unfortunately, I was not able to find a representation of the sum (\ref{bLaresult}) which would be
explicitly symmetric in $\epsilon_1,\epsilon_2$, and $\epsilon_3$. However, the result (\ref{Lambdaresult})
in the limit $\epsilon_i\rightarrow 0$ obtained below is symmetric.

To get $\bLa(\epsilon_i)$ at small $\epsilon_i$ we need to estimate the behavior of the integrals $J_1$-$J_{13}$ as $\epsilon_i\rightarrow 0$. 
As an example, let us consider integral $J_9$ given by Eq. (\ref{j34}). The contours of integration over $s_3$ and $s_4$ are pinched between ``left'' and ``right'' poles as the separation between them vanishes in the limit $\epsilon_i\rightarrow 0$. 
Shifting contours of integration 
over $s_3$ and $s_4$ to the left of the real axis and taking residues at $s_3=0$ and $s_4=0$ one obtains
\begin{eqnarray}
&&\hspace{-1mm}
J_9~=~{\Gamma(\epsilon_2-1)\Gamma(\epsilon_1+\epsilon_3-\epsilon_2)\over\Gamma(\epsilon_3)}
\Gamma(2\epsilon_2-\epsilon_1-\epsilon_3)
\!\int_{-\delta-i\infty}^{-\delta+i\infty}\! {ds_3ds_4\over (2\pi i)^2}~
\Gamma(\epsilon_1-s_3-s_4)\Gamma(s_3)
\nonumber\\
&&\hspace{11mm}
\times~\Gamma(s_4)\Gamma(\epsilon_3-s_3)\Gamma(s_3-1)
{\Gamma(\epsilon_2-1-s_4)\Gamma(1-\epsilon_1-\epsilon_2+s_3+s_4)\Gamma(\epsilon_3-s_4)
\over
\Gamma(-\epsilon_1+s_3) \Gamma(2\epsilon_2-\epsilon_1-s_4)}
\nonumber\\
&&\hspace{-1mm}
+~{\Gamma^2(\epsilon_2-1)\Gamma(\epsilon_1+\epsilon_3-\epsilon_2)
\over\Gamma(2\epsilon_2-\epsilon_1)}
\Gamma(2\epsilon_2-\epsilon_1-\epsilon_3)
\nonumber\\
&&\hspace{11mm}
\times~\!\int_{-\delta-i\infty}^{-\delta+i\infty}\! {ds_3\over 2\pi i}~
\Gamma(\epsilon_1-s_3)\Gamma(s_3)
\Gamma(\epsilon_3-s_3)\Gamma(s_3-1)
{\Gamma(1-\epsilon_1-\epsilon_2+s_3)
\over
\Gamma(-\epsilon_1+s_3) }
\nonumber\\
&&\hspace{-1mm}
+~{\Gamma(\epsilon_2-1)\over\Gamma(-\epsilon_1)}\Gamma(2\epsilon_2-\epsilon_1-\epsilon_3)
\Gamma(\epsilon_1+\epsilon_3-\epsilon_2)
\nonumber\\
&&\hspace{11mm}
\times\!\int_{-\delta-i\infty}^{-\delta+i\infty}\! {ds_4\over 2\pi i}
\Gamma(\epsilon_1-s_4)\Gamma(s_4)
\Gamma(\epsilon_2-1-s_4)\Gamma(1-\epsilon_1-\epsilon_2+s_4)
{\Gamma(\epsilon_3-s_4)\over \Gamma(2\epsilon_2-\epsilon_1-s_4)}
\nonumber\\
&&\hspace{22mm}
\times~\big[\psi(\epsilon_1-s_4)+\psi(\epsilon_3)-\psi(1-\epsilon_1-\epsilon_2+s_4)+\psi(-\epsilon_1)
-2\psi(1)-1\big]
\nonumber\\
&&\hspace{-1mm}
+~{\Gamma^2(\epsilon_2-1)\over\Gamma(-\epsilon_1)}\Gamma(2\epsilon_2-\epsilon_1-\epsilon_3)
\Gamma(\epsilon_1+\epsilon_3-\epsilon_2)
\Gamma(\epsilon_1)
\Gamma(1-\epsilon_1-\epsilon_2)
{\Gamma(\epsilon_3)\over \Gamma(2\epsilon_2-\epsilon_1)}
\nonumber\\
&&\hspace{22mm}
\times~\big[\psi(\epsilon_1)+\psi(\epsilon_3)-\psi(1-\epsilon_1-\epsilon_2)+\psi(-\epsilon_1)
-2\psi(1)-1\big]
\label{jestimate}
\end{eqnarray}
where $\psi(x)\equiv\Gamma'(x)/\Gamma(x)$.
Now the integrals over $s_3$ and/or $s_4$ in the r.h.s. of Eq. (\ref{jestimate}) 
are not pinched so the only singularities at $\epsilon_i\rightarrow 0$ come from the 
explicit factors like $\Gamma(\epsilon_i)$ or $\psi(\epsilon_i)$. Actually, it is easy to see that the last non-integral term is the most singular so one obtains
\begin{eqnarray}
&&\hspace{-1mm}
J_9~\stackrel{\epsilon_i\rightarrow 0}{\simeq}~
{2\epsilon_2-\epsilon_1
\over\epsilon_2^2\epsilon_3^2(\epsilon_1+\epsilon_3-\epsilon_2)(2\epsilon_2-\epsilon_1-\epsilon_3)}
~+~O\big({1\over \epsilon^4}\big)
\label{j9estimate}
\end{eqnarray}
Similar estimates of remaining integrals yield
\begin{eqnarray}
&&\hspace{-1mm}
J_1\sim J_2\sim O\big({1\over \epsilon^4}\big),~~~J_3\simeq~{(\epsilon_2+\epsilon_3)^2\over \epsilon_1^2\epsilon_2^3\epsilon_3^2},~~~
J_4\simeq -{\epsilon_2+\epsilon_3\over\epsilon_2^3\epsilon_3^2(\epsilon_1+\epsilon_3-\epsilon_2)},
\nonumber\\
&&\hspace{-1mm}
J_5\simeq {\epsilon_2+\epsilon_3\over \epsilon_1\epsilon_2^3\epsilon_3(\epsilon_1-\epsilon_2)},~~~
J_6\simeq -{\epsilon_2+\epsilon_3\over\epsilon_1^2\epsilon_2^2\epsilon_3(\epsilon_1-\epsilon_2)},~~~
J_7\simeq {1\over\epsilon_2^3\epsilon_3(\epsilon_1+\epsilon_3-\epsilon_2)},
\nonumber\\
&&\hspace{-1mm}
J_8\simeq -{\epsilon_2+\epsilon_3\over \epsilon_1\epsilon_2^3\epsilon_3(\epsilon_1+\epsilon_3-\epsilon_2)},~~~
J_{10}\simeq {1\over\epsilon_2^3(\epsilon_1-\epsilon_2)(\epsilon_1+\epsilon_3-\epsilon_2)}
\nonumber\\
&&\hspace{-1mm}
J_{11}\simeq -{1\over\epsilon_1\epsilon_2^2(\epsilon_1-\epsilon_2)(\epsilon_1+\epsilon_3-\epsilon_2)},~~~
J_{12}\simeq {-1\over\epsilon_2^2\epsilon_3(\epsilon_1+\epsilon_3-\epsilon_2)(2\epsilon_2-\epsilon_1-\epsilon_3)}
\nonumber\\
&&\hspace{-1mm}
J_{13}\simeq
{2\epsilon_2-\epsilon_1
\over \epsilon_1\epsilon_2^2\epsilon_3(\epsilon_1+\epsilon_3-\epsilon_2)(2\epsilon_2-\epsilon_1-\epsilon_3)},~~~
J_{14}\simeq {-1\over\epsilon_2^2(2\epsilon_2-\epsilon_1-\epsilon_3)(\epsilon_1+\epsilon_3-\epsilon_2)(\epsilon_1-\epsilon_2)}
\nonumber\\
&&\hspace{-1mm}
J_{15}\simeq {1\over\epsilon_1\epsilon_2(2\epsilon_2-\epsilon_1-\epsilon_3)
(\epsilon_1-\epsilon_2)(\epsilon_1+\epsilon_3-\epsilon_2)}
\label{Jayz}
\end{eqnarray}
It is easy to see that 
$$
J_4+J_7+J_8+J_9+J_{10}+J_{11}+J_{12}+J_{13}+J_{14}+J_{15}~=~O\big({1\over \epsilon^4}\big)
$$
so we get
\begin{equation}
I^{a_1\rightarrow 0,a_2\rightarrow 0,a_3\rightarrow 0,0}(\epsilon_1,\epsilon_2,\epsilon_3)
~\simeq~J_3+J_5+J_6~\simeq~-{\epsilon_2+\epsilon_3\over\epsilon_1^2\epsilon_2^2\epsilon_3^2}
~+~O\big({1\over \epsilon^4}\big)
\end{equation}
and therefore
\begin{eqnarray}
&&\hspace{-1mm}
\bLa(\epsilon_1,\epsilon_2,\epsilon_3)~=~1+O(\epsilon_i)
\label{Lambdaresult}
\end{eqnarray}
which is quoted in Eq. (\ref{tila}) in terms of $\gamma_i=-2\epsilon_i$.
Note that the symmetric form of this result  for $\bLa(\epsilon_1,\epsilon_2,\epsilon_3)$ is a check for
for the calculation of the integral (\ref{lambda1}) which is not obviously symmetric in $\epsilon_1,\epsilon_2,\epsilon_3$.

\bibliography{triplebib}
\bibliographystyle{JHEP}

\end{document}